\documentclass[fleqn,usenatbib]{mnras}

\usepackage{newtxtext,newtxmath}
\usepackage[T1]{fontenc}
\usepackage{graphicx}
\usepackage{amsmath}
\usepackage{orcidlink}
\usepackage{booktabs}
\usepackage{bbm}
\usepackage{xcolor}
\usepackage{placeins}

\DeclareRobustCommand{\VAN}[3]{#2}
\let\VANthebibliography\thebibliography
\def\thebibliography{\DeclareRobustCommand{\VAN}[3]{##3}\VANthebibliography}

\graphicspath{{Plots/}}

\title[Surrogate Marginalisation of Beam Uncertainty]{Towards end-to-end Bayesian forward models in global 21-cm cosmology: surrogate modelling and marginalisation of beam uncertainty}

\author[J. L. Tutt et al.]{
Jacob L. Tutt\textsuperscript{\orcidlink{0009-0002-5358-4292}},$^{1,2}$\thanks{E-mail: jlt67@cam.ac.uk}
Dominic J. Anstey\textsuperscript{\orcidlink{0000-0003-1742-7417}},$^{1,2}$
John Cumner\textsuperscript{\orcidlink{0000-0001-8479-3746}},$^{1,2}$
Harry T. J. Bevins\textsuperscript{\orcidlink{0000-0002-4367-3550}},$^{1,2}$
Jean Cavillot\textsuperscript{\orcidlink{0000-0001-5731-1721}},$^{3}$
\and
Eloy de Lera Acedo\textsuperscript{\orcidlink{0000-0001-8530-6989}},$^{1,2}$ \\
$^{1}$Astrophysics Group, Cavendish Laboratory, J.\ J.\ Thomson Avenue, Cambridge, CB3 0HE, UK\\
$^{2}$Kavli Institute for Cosmology, Madingley Road, Cambridge, CB3 0HA, UK\\
$^{3}$Antenna Group, Universit\'e Catholique de Louvain, Louvain-la-Neuve, B-1348, Belgium
}

\date{Accepted XXX. Received YYY; in original form ZZZ}
\pubyear{\the\year{}}

\begin{document}
\label{firstpage}
\pagerange{\pageref{firstpage}--\pageref{lastpage}}
\maketitle

\begin{abstract}
Robust statistical inference in global 21-cm cosmology requires end-to-end uncertainty quantification that jointly handles the highly degenerate cosmological signal, foreground emission, and instrumental response. Although electromagnetic simulations capture physical antenna properties in a parametrised way, multi-hour runtimes make their integration within likelihood-based sampling frameworks infeasible. Most existing approaches therefore assume a single precomputed beam, a fragile assumption given our demonstration that realistic mismatches can severely bias the recovered cosmological and foreground parameters. To address this, we present an accelerated and differentiable Bayesian framework that incorporates an informed surrogate representation of chromatic beam uncertainty directly into a forward-modelling pipeline. Treating the physical antenna properties as nuisance quantities, we apply a two-stage decomposition directly to simulated directivity patterns, reducing the instrumental parameterisation by two orders of magnitude while retaining the angular and spectral structure required for accurate beam reconstruction. Exploiting the linearity of the resulting surrogate, we utilise analytically marginalisation to allow the continuous instrumental uncertainty to be propagated into the final posteriors and Bayesian evidence without directly sampling the beam space. Testing the framework against a suite of unseen beams and cosmological signals, we recover the true inputs at approximately the instrumental-noise level. We further show that, for the geometric and environmental uncertainty considered here, as few as 100 electromagnetic simulations are sufficient to construct an effective surrogate, substantially reducing the simulation burden for future analyses. This framework provides a scalable, statistically rigorous route towards hardware-accelerated uncertainty quantification in global 21-cm cosmology.
\end{abstract}

\begin{keywords}
methods: data analysis -- methods: statistical -- dark ages, reionization, first stars -- cosmology: observations
\end{keywords}

\section{Introduction}
\label{sec:introduction}

The constraints placed on fundamental physics by modern astrophysics and cosmology are predominantly anchored by two distinct observational pillars. The first of these, the Cosmic Microwave Background (CMB), provides a high-precision snapshot of the Universe at the epoch of recombination ($z\sim1100$), revealing a Universe of remarkable homogeneity \citep{COBE, WMAP, Planck2014, Planck2016, Planck2020}. The second, in stark contrast, is the richly structured distribution of matter observed in the local Universe, which has been traced in increasing detail, from the large-scale cosmic web mapped by galaxy surveys such as BOSS \citep{Boss} and DESI \citep{DESI}, through to the positions, motions and chemical abundances of individual stars within the Milky Way, measured by surveys such as Gaia \citep{GAIAOG}, 4MOST \citep{4MOST} and WEAVE \citep{WEAVE}.

Separating these two regimes lies a comparatively unexplored period, commonly divided into three epochs: the Cosmic Dark Ages (DA, $z \sim 1100$--$30$), before the formation of the first luminous sources; the Cosmic Dawn (CD, $z \sim 30$--$10$), when the first stars and galaxies began to form; and the Epoch of Reionisation (EoR, $z \sim 10$--$6$), during which radiation from these early sources progressively ionised the neutral intergalactic medium (IGM). The recent revelation of an unexpectedly large population of massive galaxies at $z>10$ by the James Webb Space Telescope \citep[JWST; see e.g.,][]{mismatchuv, jwstsfe, JWSTIMF, jwststochasticstar} has underscored the critical need to directly constrain the rich processes governing this high-redshift era.

While many probes promise insights into this period, including secondary CMB anisotropies \citep{McQuinn_2005, Reichardt_2021, iliev_2024} and quasar absorption spectra \citep{Gunn_1965, Fan_2006, Becker_2015, Eilers_2018, Qin_2021, Bosman_2022}, the 21-cm signal, originating from the hyperfine transition within the neutral hydrogen that makes up the IGM, acts as a direct tracer of both the spatial morphology of reionisation and the thermal history of the gas \citep[for comprehensive reviews, see][]{Furlanetto_2006, Pritchard_2012, Barkana_2016, Mesinger_2019}. It therefore provides a window into the physics of the first luminous sources, placing constraints on their formation efficiency, spectral emissivity and initial mass function \citep{gesseyjones2022,Schauer2019} as well as contributions from their associated X-ray binaries \citep{Sartorio_2023} and other potential exotic processes \citep{Mittal_2022,Barkana_2018,gesseyjones2024}.

Current 21-cm experiments can be broadly divided into three observational approaches. The first comprises those that seek to measure the monopole (sky-averaged) component through isolated radiometers such as EDGES \citep{Bowman_2018}, PRIZM \citep{Philip_2018}, SARAS \citep{Singh_2018}, MIST \citep{Monsalve_2024}, RHINO \citep{Bull_2025} and REACH \citep{Acedo_2022}. The second includes interferometric arrays targeting the statistical spatial fluctuations in the 21-cm brightness field through its power spectrum, including HERA \citep{HERA}, LOFAR \citep{LOFAR}, NenuFAR \citep{nenufar}, and the MWA \citep{MWA}. The third, as promised by next-generation instruments like the upcoming SKA-Low \citep{SKA}, aims to go beyond statistical measurements to achieve direct tomographic imaging. Despite differing in their observational strategies, all three methodologies share similar hurdles regarding the unprecedented accuracy with which the sources of systematics must be characterised and understood.

A major challenge common to both is the presence of bright Galactic and extragalactic foregrounds, which exceed the expected 21-cm signal by three to four orders of magnitude \citep{Shaver_1999}. Furthermore, signal extraction is complicated by the foreground emissions being modulated by both a spatially evolving and frequency-dependent instrumental response, introducing spectral artefacts that are degenerate with those of the underlying cosmological signature. In interferometric measurements, this effect contributes to foreground mode mixing and wedge-like contamination of otherwise cleaner Fourier modes \citep[see][]{Liu_2014a, Liu_2014b, Thyagarajan_2016, Neben_2016, Fagnoni_2020, O_Hara_2025}. In global experiments, the same physical coupling appears as chromatic distortions in the antenna temperature \citep{Vedantham_2014, Anstey_2021}. In this work, we focus on the latter; however, similar approaches to the treatment of instrumental uncertainty have been explored in the context of interferometers \citep{kern_2025,wilensky_2024,Wilensky_2025}.

Methodologically, existing global 21-cm analysis frameworks handle this inference challenge through one of four primary avenues. Firstly, some experiments rely on instrumental designs engineered to be minimally chromatic \citep{Singh_2018, Singh_2022}, which in principle allows the data to be modeled using simple, smooth functional forms \citep{Bevins_2021}. While avoiding conditioning on a specific foreground or electromagnetic beam model is advantageous, achieving sufficient achromaticity across a wide bandwidth to avoid biasing the inference is exceptionally difficult in practice. Beyond these, there are further informed approaches that explicitly account for the chromatic foreground and beam structure. These include methods that use a beam-factor-based chromaticity correction (BFCC) to account for the impact of the antenna before modelling the foregrounds and signal \citep{Monsalve_2017, Bowman_2018,sims_2023,Sims_2025b}, pipelines that use training sets of simulated beam-weighted foreground spectra to define basis functions from which the data can be reconstructed \citep{Tauscher_2018a, Tauscher_2020b, Rapetti_2020, Hibbard_2020, Bassett_2021b, Saxena_2023}, and frameworks that use physically motivated Bayesian forward models to evaluate the convolution of the beam with a parameterised sky model \citep{Anstey_2021, Anstey_2022, Pattison_2023, pattison_2025a, Pagano_2023, Shen_2021, Shen_2022,Mittal_2024, Robins_2026}. While the latter three approaches explicitly account for the beam in the analysis, they do so under the assumption that it is known exactly.

In practice, however, the fidelity of any beam model is fundamentally limited by physical discrepancies between simulation and reality, as well as numerical effects such as finite mesh resolution. Any electromagnetic model assumes an idealised geometry, yet real-world antennas suffer from finite manufacturing and assembly tolerances as well as time-dependent environmental changes. Errors in this description propagate directly into the beam-weighted foreground model and can therefore bias the recovered 21-cm posterior. This sensitivity has been demonstrated across a series of increasingly realistic analyses. Early work by \citet{Hibbard_2020} showed that the foreground basis depends strongly on the assumed beam model, with beam mismatch driving foreground residuals from the millikelvin level to kelvin-scale amplitudes. To account for this, subsequent studies introduced idealized beam variations into their training sets, including the simple frequency-dependent Gaussian perturbations of \citet{Tauscher_2021}, while \citet{Bassett_2021a} demonstrated that insufficient coverage of this instrumental parameter space can severely bias signal recovery and underestimate posterior uncertainties.

More recently, \citet{Hibbard_2024} developed the \texttt{MEDEA} emulator to interpolate complex electromagnetic simulations across physical antenna and environmental parameters, providing an efficient way to represent realistic beam variations; however, their propagation into 21-cm cosmological inference was not demonstrated. Complementary studies have quantified the resulting requirements and consequences, showing that even minute deviations in the antenna pattern \citep{Cumner_2024}, physically plausible structural mismatches \citep{Pattison_2026}, or dynamic environmental conditions \citep{pattison_2025b} can directly bias recovered cosmological parameters.

Therefore, achieving robust, statistically principled 21-cm inference requires a framework capable of simultaneously accounting for realistic end-to-end uncertainties in both the foreground sky and instrumental response. Neglecting these effects risks biased cosmological constraints and artificially narrow posteriors. Motivated by these limitations, this work presents, to our knowledge, the first joint inference framework for global 21-cm cosmology that propagates continuous instrumental beam uncertainty, learned from electromagnetic beam simulations, alongside foreground and cosmological uncertainties within a physically motivated Bayesian forward model. We achieve this by extending the hardware-accelerated, differentiable pipeline developed for the Radio Experiment for the Analysis of Cosmic Hydrogen \citep[REACH;][]{Tutt_2026}, ensuring the inference is no longer conditioned on a single, fixed beam template.

The remainder of this paper is structured as follows. In Section~\ref{sec:beam}, we introduce the simulated beam ensemble and describe the two-stage basis construction used to compress the chromatic beam uncertainty into a tractable parameterisation. Section~\ref{sec:bayesian} describes the simulated radiometer data, the extension of the REACH forward model to include beam uncertainty, and the Bayesian marginalisation framework used for inference. The performance of the method is then demonstrated in Section~\ref{sec:results}, and finally conclusions drawn in Section~\ref{sec:conclusions}.

\section{Beam Parameterisation}
\label{sec:beam}

\begin{figure}
    \centering
    \includegraphics[width=\linewidth]{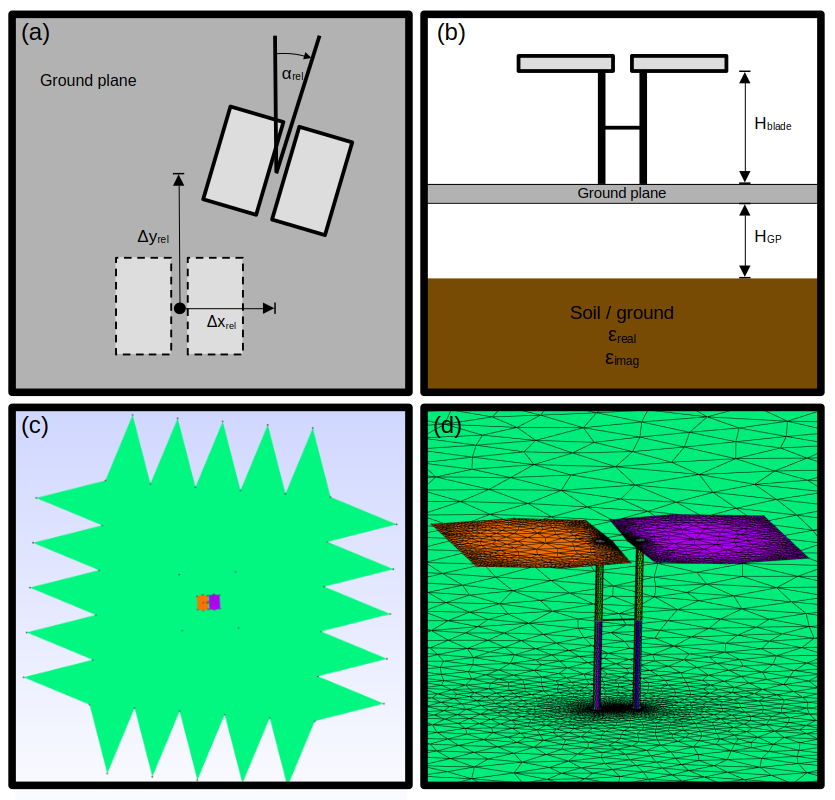}
    \caption{(a) Parameterization of the displacement of the dipole blades position relative to centre. Note that the entire system is subsequently realigned to maintain the centre line of the blades being upon the y axis. (b) The two heights varied, between ground level and the ground plane, and the ground plane and the dipole blades. Along with the two soil permittivity parameters. (c) a zoomed out view of the antenna model. (d) a close view of the central dipole and balun in addition, triangulated with the solver mesh.}
    \label{fig:mesh and dimensions}
\end{figure}

\begin{figure*}
    \centering
    \includegraphics[width=\textwidth]{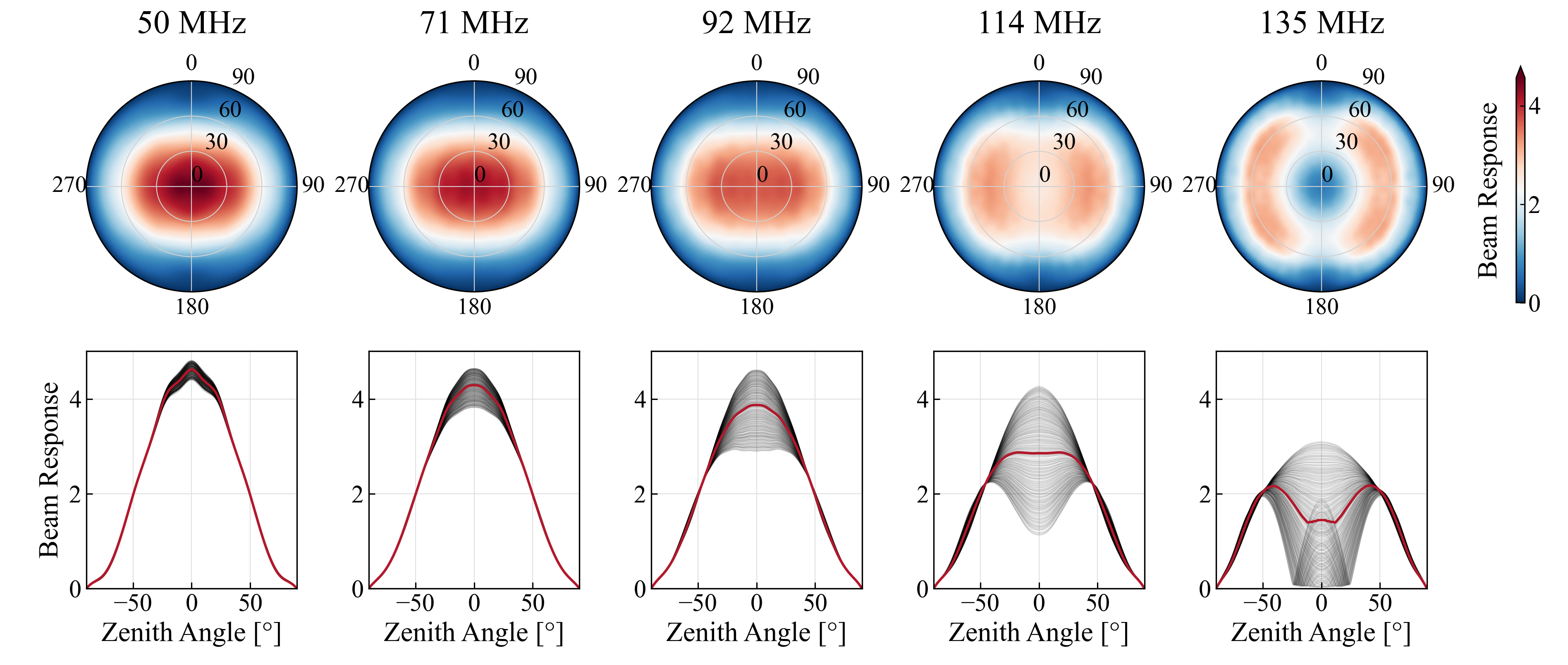}
    \caption{Illustration of the variation in chromatic beam structure across the simulated REACH antenna realisations drawn from the seven-dimensional physical parameter space. Top row: polar projections of the normalised directivity, $D(\nu,\theta,\phi)$, at a range of frequencies across the observing band for one representative beam, shown in local angular coordinates above the horizon. Bottom row: azimuthal slices of the directivity for all 400 training set beams, illustrating the frequency-dependent structure introduced by the physical perturbations.}
    \label{fig:beam_chromaticity}
\end{figure*}

In this section, we describe the Computational Electromagnetic (CEM) simulations used to generate a physically motivated ensemble of realistic beams (Section~\ref{sec:beam_simulations}). We then outline how these beams can then be used to learn the underlying structure of the chromatic variations (Section~\ref{sec:basis_functions}) and the novel method this work presents for efficient parameterisation for use within statistical inference (Section~\ref{sec:beam_parameterisation}).

Although we specifically focus our analysis on the instrumental uncertainties associated with the hexagonal blade dipole antenna currently deployed in the Karoo desert as part of the REACH experiment \citep{Acedo_2022, Cumner_2022}, the methodology presented can be readily applied to any antenna geometry. 

\subsection{Beam Simulations}
\label{sec:beam_simulations}

We characterise each simulated beam using its normalised antenna power pattern, or directivity, $D(\nu,\theta,\phi)$, where $\theta$ and $\phi$ denote the zenith and azimuthal angles, respectively. Following the convention defined by \citet{Cumner_2024}, this is calculated directly from the far-field electric field components via
\begin{equation}
D(\nu,\theta,\phi)
=
4\pi
\frac{
|E_{\theta}(\nu,\theta,\phi)|^2
+
|E_{\phi}(\nu,\theta,\phi)|^2
}{
\int_{4\pi}
\left[
|E_{\theta}(\nu,\theta,\phi)|^2
+
|E_{\phi}(\nu,\theta,\phi)|^2
\right]
\,\mathrm{d}\Omega
}.
\label{eq:beam_directivity}
\end{equation}
Here, $E_{\theta}(\nu,\theta,\phi)$ and $E_{\phi}(\nu,\theta,\phi)$ represent the orthogonal components of the far-field electric field vector. In this work, we isolate the above-horizon component of the spatial response by enforcing
\begin{equation}
D(\nu,\theta>\pi/2,\phi)=0,
\label{eq:horizon_cut}
\end{equation}
before re-normalising the pattern at each simulated frequency channel, $\nu_i$, such that
\begin{equation}
\int_{4\pi} D(\nu_i,\Omega)\,\mathrm{d}\Omega = 4\pi.
\label{eq:beam_normalisation}
\end{equation}
Consequently, the antenna radiation efficiency, $\eta(\nu)$, is assumed constant, however as $\eta(\nu)$ enters the antenna temperature calculation simply as a frequency-dependent multiplicative scalar, it can be integrated into future expansions without any changes to the methodology.

To sample the variations in these patterns, we simulate 500 distinct antenna configurations. We use Latin-hypercube sampling to distribute the configurations across a seven-dimensional physical parameter space, detailed in Table~\ref{tab:beam_simulation_parameters}, which captures two primary channels of instrumental uncertainty. The first being geometric perturbations in the antenna system arising from manufacturing and installation tolerances. Specifically, we vary the ground-plane height ($H_{\mathrm{GP}}$) as well as the dipole's blade height ($H_{\mathrm{blade}}$), rotation ($\alpha_{\mathrm{rel}}$), and lateral $x$ and $y$ displacements ($\Delta x_{\mathrm{rel}}$, $\Delta y_{\mathrm{rel}}$) relative to the ground plane. The coordinate centre is maintained directly between the blades with the y axis aligned with the gap between the dipole blades through a coordinate transformation after any blades shifts take place. The second channel comprises variations in the complex dielectric relative permittivity of the soil ($\epsilon_{\mathrm{real}}$ and $\epsilon_{\mathrm{imag}}$), which account for environmental fluctuations such as changing soil moisture content and temperature. Schematic representation of these parameters is shown in Figure~\ref{fig:mesh and dimensions}, along with images of the overall antenna model including ground plane and the meshed central dipole and balun element.

The antenna simulations were carried out using a in-house Method-of-Moments (MoM) solver. The MoM \citep{Harrington1993} is used here to solve the Electric Field Integral Equations (EFIE). The MoM linear system of equations can be written as:
\begin{equation}
    \mathbf{Z} \  \mathbf{I} = \mathbf{V},
\end{equation}
with $\textbf{Z} $ the MoM matrix, $\textbf{V}$ the excitation vector and  $\textbf{I}$ the vector of current coefficients to be obtained. The classical delta-gap excitation is selected here to determine the excitation vector $\textbf{V}$. The antenna is simulated in the presence of a semi-infinite soil of a given permittivity. The MoM matrix is obtained here by summing two matrices:
\begin{equation}
    \mathbf{Z}  = \mathbf{Z_{FS}} + \mathbf{Z_{soil}}
\end{equation}
the first contribution $\mathbf{Z_{FS}}$ is the homogeneous free-space MoM matrix which does not depend on the presence of the soil. The second term $\mathbf{Z_{soil}}$ accounts for the presence of the soil and is efficiently calculated with the inhomogeneous plane waves method \citep{Hu1999,Alkhalifeh2016}. Only this second matrix needs to be recomputed when tuning the soil permittivity. Once both matrices are calculated, the current coefficients $\mathbf{I}$ can be obtained using direct Gaussian inversion. From these currents, the radiation pattern of the antenna can be computed while accounting for the reflection by the soil as done in \citet{Cavillot2020}. The ratio of power penetrating into the soil to the power delivered at the feed of the antenna is then computed using the method described in \citet{Cavillot2024}.

In reality, the physical parameters of the deployed instrument are expected to be known to a much higher precision than the bounds specified in Table~\ref{tab:beam_simulation_parameters}. However, we deliberately inflate these ranges to ensure that the resulting ensemble of simulated beams encompasses a conservative upper limit on the instrumental uncertainty, providing a rigorous stress test for our downstream data analysis pipeline. While this ensemble provides a robust proof of concept for end-to-end beam uncertainty propagation, the simulation suite does not represent an exhaustive description of all instrumental systematics. Complex deformations of the ground plane mesh \citep{Pattison_2026} and numerical uncertainties in the antenna solver are left for future work with more extensive simulation budgets.

For the analysis presented below, the simulated ensemble is randomly split into a training set of 400 beams, from which the structure of the beam variations is learned, and an independent validation set of 100 beams. The effect of these physical perturbations on the chromatic beam structure is illustrated in Figure~\ref{fig:beam_chromaticity}. 

Throughout this work, we benchmark the methodology using a representative subset of 11 beams from the validation set, chosen to span increasing distances from the centre of the sampled parameter volume, from typical to increasingly extreme realisations. The distance of each beam $b$ from the centre of the parameter space is quantified using the Euclidean distance after standardising each physical parameter by its distribution across the ensemble,
\begin{equation}
d_b
=
\left[
\sum_{i=1}^{7}
\left(
\frac{p_{b,i}-\bar{p}_i}{s_i}
\right)^2
\right]^{1/2},
\label{eqn:normalised_beam_parameter_distance}
\end{equation}
\noindent where $p_{b,i}$ is the value of the $i$th physical parameter for beam $b$, and $\bar{p}_i$ and $s_i$ are the corresponding ensemble mean and sample standard deviation. After ranking the validation beams by $d_b$, we select the realisations at every 10th percentile, with the parameters of the selected beams listed in Table~\ref{tab:validation_beam_parameters}.

\begin{table}
    \centering
    \caption{The parameter configurations of the validation beams used to benchmark the framework, comprising one realisation selected at each 10th percentile of the normalised parameter-distance distribution and listed in order of increasing distance from the centre of the ensemble. Geometrical quantities are quoted in mm, rotation in degrees, and soil-permittivity components are dimensionless.}
    \label{tab:beam_simulation_parameters}
    \begin{tabular}{llr@{--}ll}
        \toprule
        Symbol & Description & \multicolumn{2}{c}{Range} & Units \\
        \midrule
        $H_{\mathrm{blade}}$ & Dipole blade height & $800$ & $1200$ & mm \\
        $H_{\mathrm{GP}}$ & Ground-plane height & $1000$ & $1500$ & mm \\
        $\alpha_{\mathrm{rel}}$ & Dipole rotation & $-10$ & $10$ & $\deg$ \\
        $\Delta x_{\mathrm{rel}}$ & Dipole $x$ displacement& $-200$ & $200$ & mm \\
        $\Delta y_{\mathrm{rel}}$ & Dipole $y$ displacement& $-200$ & $200$ & mm \\
        $\epsilon_{\mathrm{real}}$ & Soil relative permittivity (real) & $1$ & $10$ & -- \\
        $\epsilon_{\mathrm{imag}}$ & Soil relative permittivity (imaginary) & $-2$ & $0$ & -- \\
        \bottomrule
    \end{tabular}
\end{table}

\begin{table}
    \centering
    \caption{Parameter configurations of the validation beams used to generate the benchmark datasets, listed in order of increasing normalised distance from the centre of the prior volume. Geometrical quantities are quoted in mm, rotations in degrees, and soil-permittivity components are dimensionless.}
    \label{tab:validation_beam_parameters}
    \resizebox{\columnwidth}{!}{%
    \begin{tabular}{@{}lrrrrrrr@{}}
        \toprule
        Beam & $H_{\mathrm{blade}}$ & $H_{\mathrm{GP}}$ & $\alpha_{\mathrm{rel}}$ & $\Delta x_{\mathrm{rel}}$ & $\Delta y_{\mathrm{rel}}$ & $\epsilon_{\mathrm{real}}$ & $\epsilon_{\mathrm{imag}}$ \\
        \midrule
        0 & 904 & 1335 & 1.6 & 31 & 36 & 3.34 & -1.20 \\
        1 & 1080 & 1272 & 9.8 & 18 & -91 & 4.24 & -0.68 \\
        2 & 861 & 1428 & -6.0 & 72 & 86 & 6.00 & -1.26 \\
        3 & 1074 & 1000 & -1.0 & 87 & 4 & 7.58 & -0.44 \\
        4 & 973 & 1310 & 0.8 & 197 & -11 & 8.89 & -0.35 \\
        5 & 1035 & 1239 & -7.8 & -110 & 19 & 1.72 & -0.21 \\
        6 & 927 & 1024 & -0.3 & 138 & 161 & 6.62 & -0.51 \\
        7 & 824 & 1136 & -3.8 & -61 & 168 & 4.83 & -1.86 \\
        8 & 1188 & 1440 & -3.3 & -111 & -105 & 7.47 & -0.22 \\
        9 & 1100 & 1403 & 10.0 & -177 & -3 & 6.15 & -0.03 \\
        10 & 1173 & 1461 & 8.9 & -149 & -174 & 6.27 & -1.80 \\
        \bottomrule
    \end{tabular}
    }
\end{table}

\subsection{Basis Functions}
\label{sec:basis_functions}

\begin{figure*}
    \centering
    \includegraphics[width=\textwidth]{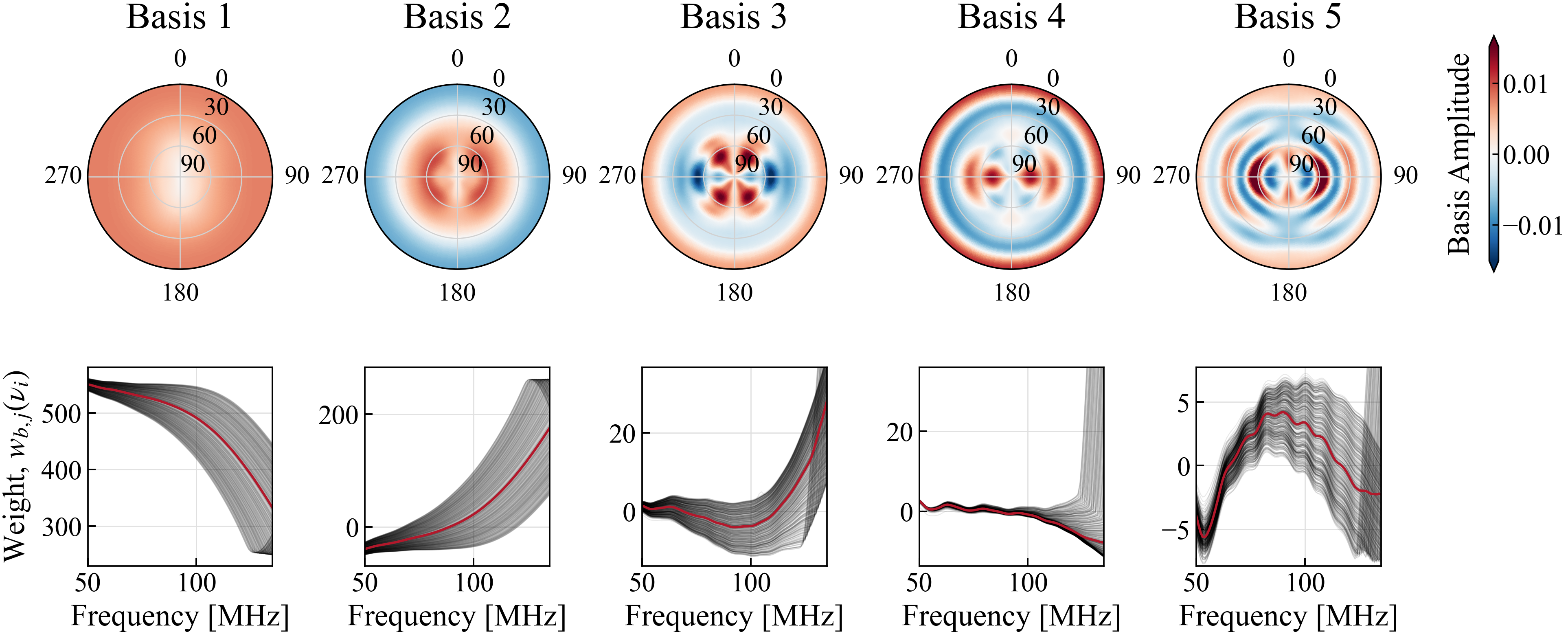}
    \caption{First five SVD beam basis functions and their corresponding frequency-dependent weight distributions across the 400-beam training set. Top row: angular basis functions $B_j(\theta,\phi)$ obtained from the row-stacked beam-frequency decomposition. Bottom row: associated weights $w_{b,j}(\nu_i)$ for all training beams, showing how the contribution of each angular mode varies coherently across frequency and across the simulated beam ensemble.}
    \label{fig:svd_basis_weights}
\end{figure*}

Given this training set of beams, we seek to learn the underlying variations in the beam's angular and spectral geometry such that arbitrary beams of coherent structure can be represented to the level of precision required for global 21-cm cosmology. A singular value decomposition (SVD) is particularly well suited to this task because it constructs an orthogonal basis ordered by decreasing contribution to the structure of the beam ensemble, allowing the dominant variations to be retained with the fewest possible modes. Motivated by the need to span this space as efficiently as possible, we adopt the approach of \citet{Pieterse_2024}, in which the frequency and angular information are treated jointly within a single decomposition, rather than performing separate decompositions at each frequency. This yields a unified set of basis functions, allowing the dominant beam variations to be captured with far fewer modes.

To do this, for each beam $b$ and frequency channel $\nu_i$, the directivity pattern $D_b(\nu_i,\theta,\phi)$ is flattened into a single angular vector and stored as a row of the matrix $\mathbf{M}$. The resulting matrix therefore has dimensions $(N_{\mathrm{train}} \cdot N_{\nu}) \times N_{\Omega}$, where $N_{\Omega}$ is the number of angular samples and $N_{\nu}$ is the number of frequency channels. Writing the combined beam-frequency row index as $\alpha \equiv (b,i)$, the SVD
of this matrix is
\begin{equation}
M_{\alpha p}
=
\sum_{j=1}^{N_{\mathrm{mode}}}
U_{\alpha j}\,\sigma_j\,(V^{\dagger})_{jp},
\label{eq:beam_svd_components}
\end{equation}
\noindent where $p$ indexes the angular samples, $j$ indexes the SVD modes, $N_{\mathrm{mode}}=\min(N_{\mathrm{train}}N_{\nu},N_{\Omega})$, and $\sigma_j$ are the singular values ordered by decreasing variance. The rows of $\mathbf{V}^{\dagger}$ define an orthogonal set of angular basis functions, $\{B_j(\theta,\phi)\}$, while the corresponding beam- and frequency-dependent behaviour is encoded by the left-singular vectors, with components $U_{(b,i),j}$. In the convention adopted throughout this work, the singular values are absorbed into these coefficients, such that the weight of mode $j$ for beam $b$ at frequency $\nu_i$ is defined as
\begin{equation}
w_{b,j}(\nu_i)
\equiv
\sigma_j\,U_{(b,i),j}.
\label{eq:beam_svd_weights}
\end{equation}

\noindent The first five resultant angular basis functions and their corresponding weight distributions across all 400 training beams are shown in Figure~\ref{fig:svd_basis_weights}. It can be seen that Basis 1 and Basis 2 capture the dominant beam structure, with Basis 1 contributing most strongly at lower frequencies and Basis 2 becoming increasingly important towards the upper end of the band. This is consistent with the chromatic evolution visible in Figure~\ref{fig:beam_chromaticity}. In contrast, the higher-order basis functions encode progressively finer angular structure and increasingly complex spectral behaviour, illustrating the additional structure that must be represented in order to describe the beam to the level of precision required.

Each beam can therefore be reconstructed as a weighted sum of angular basis functions,
\begin{equation}
\tilde{D}_b^{(R)}(\nu_i,\theta,\phi)
=
\sum_{j=1}^{R}
w_{b,j}(\nu_i)\,
B_j(\theta,\phi),
\label{eq:beam_svd_reconstruction}
\end{equation}
\noindent where $R$ is the number of retained SVD modes from the full set of $N_{\mathrm{mode}}=\min(N_{\mathrm{train}}N_{\nu},N_{\Omega})$ modes.

A critical question is determining the exact truncation rank, $R$, necessary to balance accuracy against the computational overhead of downstream statistical inference. Previous work by \citet{Cumner_2024} quantified beam reconstruction accuracy using the mean directivity error
\begin{equation}
\Delta D_R
=
\left\langle
\left\langle
\left|
D_b(\nu,\theta,\phi)
-
\tilde{D}_b^{(R)}(\nu,\theta,\phi)
\right|
\right\rangle_{\theta,\phi}
\right\rangle_{\nu},
\label{eq:pattern_truncation_metric}
\end{equation}
\noindent and showed that, for robust global-signal recovery, the beam error should be controlled to better than $-35$ dB in mean pattern difference.

In this work, we instead use a complementary metric by exploiting the hardware-accelerated forward model, described in more detail in Section~\ref{sec:bayesian}, to propagate the truncation error directly through to the associated error in the antenna temperature. Specifically, for each of the 400 beams in the training set, we compare the antenna temperature obtained from a truncated SVD representation with that from the full-fidelity beam. To provide a conservative upper limit on the truncation error, we choose an observational snapshot in which the Galactic Centre is overhead, thereby maximising both the beam-induced chromatic distortion and the overall antenna temperature. This snapshot is evaluated at an apparent local sidereal time of $\mathrm{LST}=20.01$~h at the REACH observing location ($\mathrm{lat}=-30.83875^\circ$, $\mathrm{lon}=21.37492^\circ$), corresponding to 2019-07-01 at 00:00 UTC. The corresponding antenna-temperature residual is
\begin{equation}
\Delta T_{A,b}^{(R)}(\nu)
=
T_{A,b}^{(N_{\mathrm{ref}})}(\nu)
-
T_{A,b}^{(R)}(\nu),
\label{eq:svd_temperature_residual}
\end{equation}
\noindent where $T_{A,b}^{(X)}(\nu)$ is the antenna temperature for beam realisation $b$ reconstructed with $X$ SVD modes, with $R$ denoting the truncated reconstruction and $N_{\mathrm{ref}}$ the reference full reconstruction.

\begin{figure}
    \centering
    \includegraphics[width=\columnwidth]{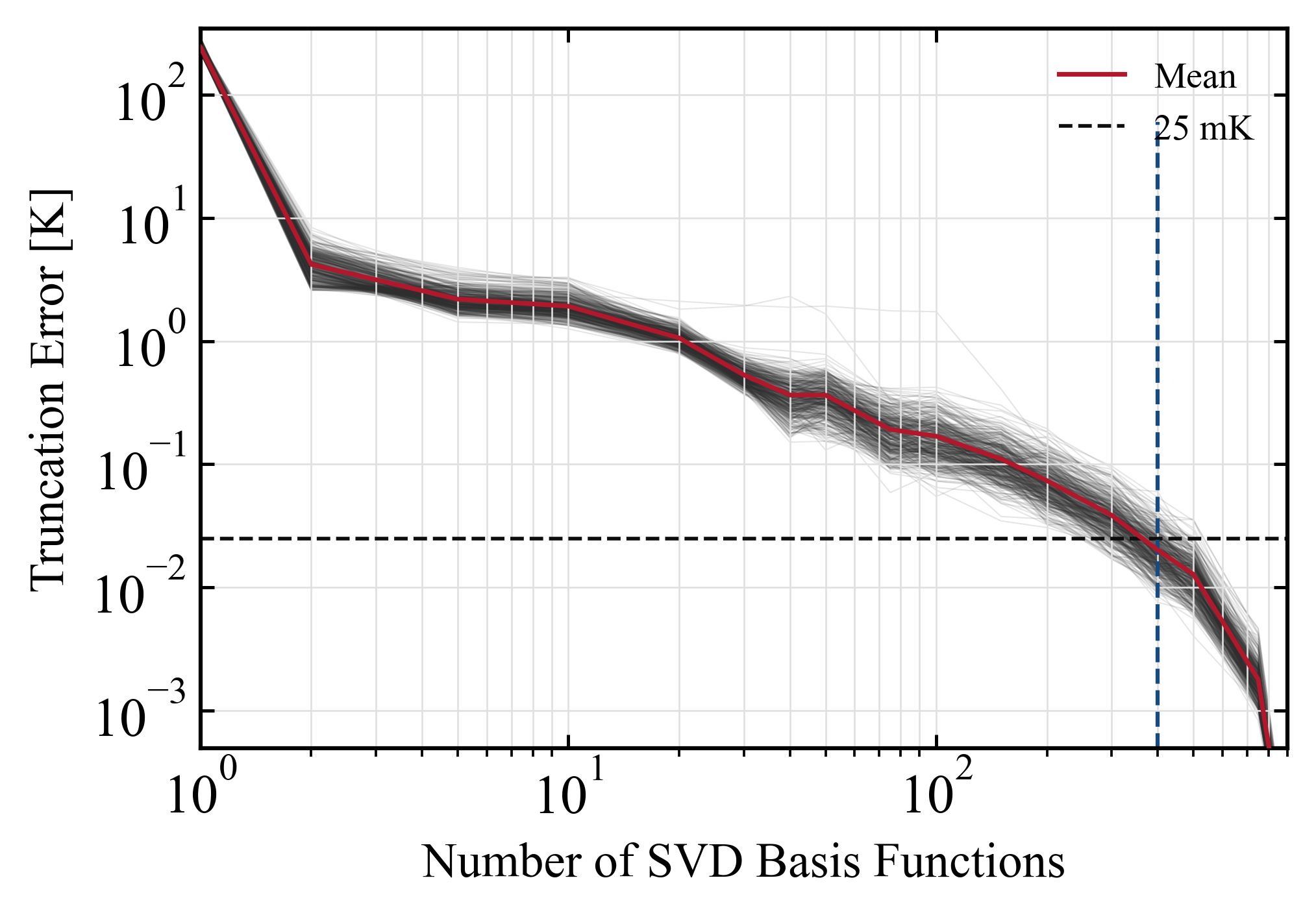}
    \caption{Antenna-temperature truncation error, $\Delta T_{A,b}^{(R)}(\nu)$, as a function of the number of retained SVD basis functions. Grey curves show the RMSE for individual training beams, evaluated by propagating each truncated beam reconstruction through the foreground forward model at the conservative Galactic-centre-overhead snapshot. The red curve shows the mean across the training set. The horizontal dashed line marks the adopted 25~mK thermal-noise requirement, while the vertical dashed line indicates the 400 SVD modes retained throughout the remainder of this work.}
    \label{fig:svd_truncation}
\end{figure}

We choose the truncation criterion adopted throughout this work to require that the mean root-mean-squared error (RMSE) across the full observing band, evaluated over the beam ensemble, remains below the expected thermal noise level of the instrument, taken here to be 25~mK \citep{Acedo_2022}. Using this metric, Figure~\ref{fig:svd_truncation} shows that retaining the first 400 SVD basis functions is sufficient, yielding a mean RMSE of 20.09 mK across the training set and hence is the number of basis functions used throughout the remainder of this work. At this stage, given that each basis function has a weight at every frequency channel across the observing band, a direct parameterisation of the beam in this space would require $400 \times N_{\nu} = 34{,}000$ parameters. This raises two distinct challenges: one of computational tractability, and one of statistical conditioning for cosmological inference.

From a computational perspective, exploring a posterior over such a high-dimensional space is demanding, but not fundamentally prohibitive. Sophisticated sampling frameworks can be adapted to these high-dimensional regimes, notably through gradient-based approaches \citep{Hoffman_2014} or block-conditional Gibbs-style sampling schemes \citep{wilensky_2024, Wilensky_2025}. The greater challenge arises because access to the Bayesian evidence is pivotal for statistically principled and validated signal recovery in the low signal-to-noise regime \citep[see][]{Sims_2025}. However, its exact evaluation using methods such as Nested Sampling \citep{Skilling_2006,yallup2025nested} becomes computationally prohibitive in spaces of this dimensionality. In this case, however, the linear nature of the beam coefficients is particularly advantageous. Analytical marginalisation provides a route to integrating over this large subspace exactly, and while such approaches are often limited by the cost of large matrix inversions, the hardware-accelerated pipeline used in this work naturally accommodates these operations through distributed linear algebra solvers \citep{Wiersema_2026}, making this regime tractable in practice. Approximate alternatives also exist, including Laplace-collapsed methods \citep{lovick_2026}, nested sampling within Gibbs \citep{yallup2026nestedsamplingslicewithingibbsefficient}, variational inference \citep{Gunapati_2022}, and deep-learning-driven approaches \citep{Jeffrey_2024, leeney_2026, Polanska_2025}.

However, setting aside the computational challenge, directly parameterising the beam in this space also hinders our ability to recover the 21-cm signal. As shown in Section~\ref{sec:data_generation}, the variations in antenna temperature induced by beam uncertainty are orders of magnitude larger than the expected cosmological signal. Separating the cosmological signal from beam uncertainty therefore requires the inference to be physically informed by how the SVD weight functions evolve coherently across the band and relative to one another. A direct SVD-weight parameterisation does not encode this structure efficiently, and its lack of success is demonstrated explicitly in Appendix~\ref{app:posterior_grids}. This motivates a second compression stage, not of the variations of the beam themselves, but of the variation within the 34,000-dimensional SVD-weight space.

\subsection{Efficient Beam Parameterisation}
\label{sec:beam_parameterisation}

\begin{figure*}

    \centering
    \includegraphics[width=\textwidth]{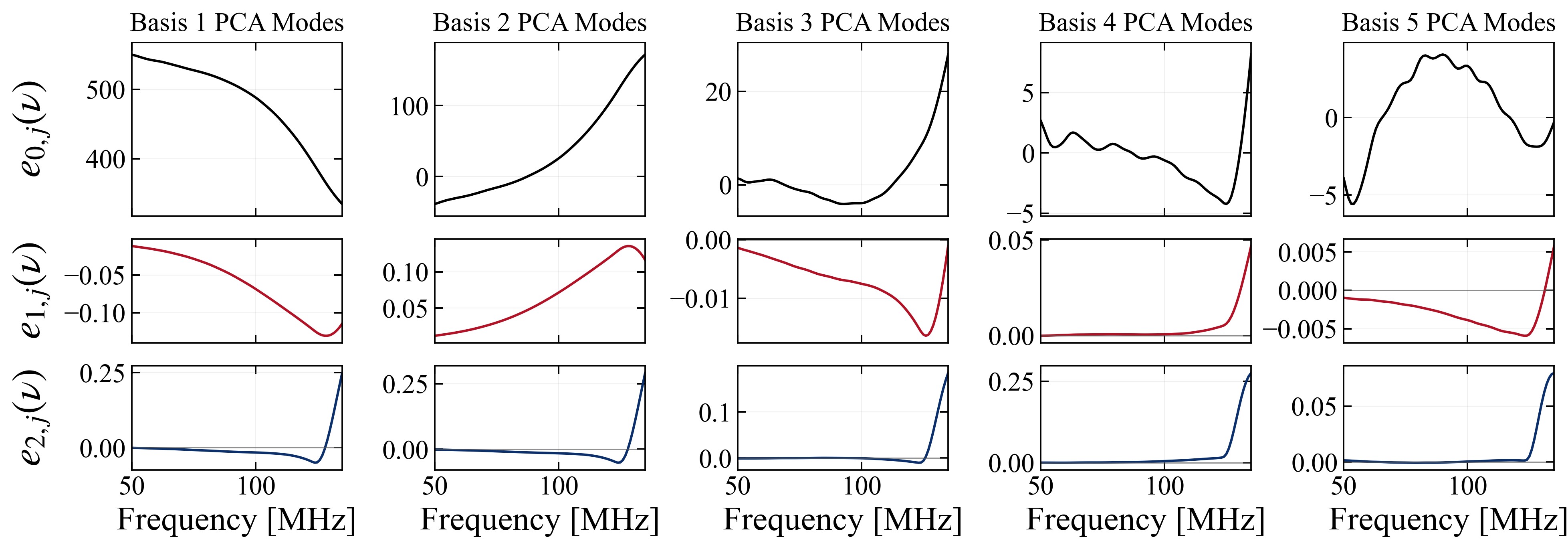}
    \caption{Mean SVD-weight functions and the first two PCA correction modes learned from the 400-beam training set. Columns correspond to the first five SVD angular basis functions. The top row shows the mean weight function $e_{0,j}(\nu_i)$, while the lower rows show the first two coherent corrections, $e_{1,j}(\nu_i)$ and $e_{2,j}(\nu_i)$, illustrating how the second-stage decomposition captures correlated spectral structure across both frequency and SVD mode. For a given beam, a coefficient-weighted combination of these PCA modes reconstructs each frequency-dependent SVD weight, $w_{b,j}(\nu_i)$. These weights are subsequently combined with the corresponding angular basis functions shown in Figure~\ref{fig:svd_basis_weights} to reconstruct the beam.}
    \label{fig:pca_weight_modes}
\end{figure*}

To address both of these challenges, we perform a second decomposition directly on the SVD weight functions. The first SVD described above compresses each individual beam into angular basis functions and their associated frequency-dependent weights. Here, we instead ask how those weight functions vary across the beam ensemble to seek a lower-dimensional representation that captures the physically motivated correlations between basis functions and across the observing band. Specifically, after subtracting the mean stacked weight vector over the training ensemble, we perform a singular value decomposition on the residual beam-to-beam variations. This is equivalent to carrying out a principal component analysis (PCA) in the SVD-weight space, and yields a compact set of correlated correction modes.

For each beam realisation $b$, the first 400 retained SVD weight functions are stacked into a single vector,
\begin{equation}
\mathbf{w}_b
=
\left[
w_{b,1}(\nu_1),\ldots,w_{b,400}(\nu_1),
\ldots,
w_{b,1}(\nu_{N_{\nu}}),\ldots,w_{b,400}(\nu_{N_{\nu}})
\right],
\label{eq:pca_weight_vector}
\end{equation}
\noindent such that each beam is described by a vector of length $400N_{\nu}$. Writing the mean stacked weight vector over the training ensemble as $\mathbf{e}_0$, the residual beam-to-beam variations are then described by a new orthogonal set of correction modes, $\{\mathbf{e}_m\}$. In component form, this gives
\begin{equation}
\tilde{w}_{b,j}^{(M)}(\nu_i)
=
e_{0,j}(\nu_i)
+
\sum_{m=1}^{M}
a_{b,m}\,e_{m,j}(\nu_i),
\label{eq:pca_weight_reconstruction_components}
\end{equation}
\noindent where $a_{b,m}$ is the PCA coefficient of beam $b$ along mode $m$, $M$ is the number of retained PCA modes, $e_{0,j}(\nu_i)$ is the mean SVD weight function, and $e_{m,j}(\nu_i)$ is the corresponding correction function for SVD basis mode $j$. Substituting this directly into the SVD beam expansion gives
\begin{equation}
\tilde{D}_b^{(M,R)}(\nu_i,\theta,\phi)
=
\sum_{j=1}^{R}
\left[
e_{0,j}(\nu_i)
+
\sum_{m=1}^{M}
a_{b,m}\,e_{m,j}(\nu_i)
\right]
B_j(\theta,\phi),
\label{eq:pca_beam_reconstruction}
\end{equation}
\noindent where, throughout this work, $R=400$ is fixed by the SVD truncation criterion derived above.

\begin{figure}
    \centering
    \includegraphics[width=\columnwidth]{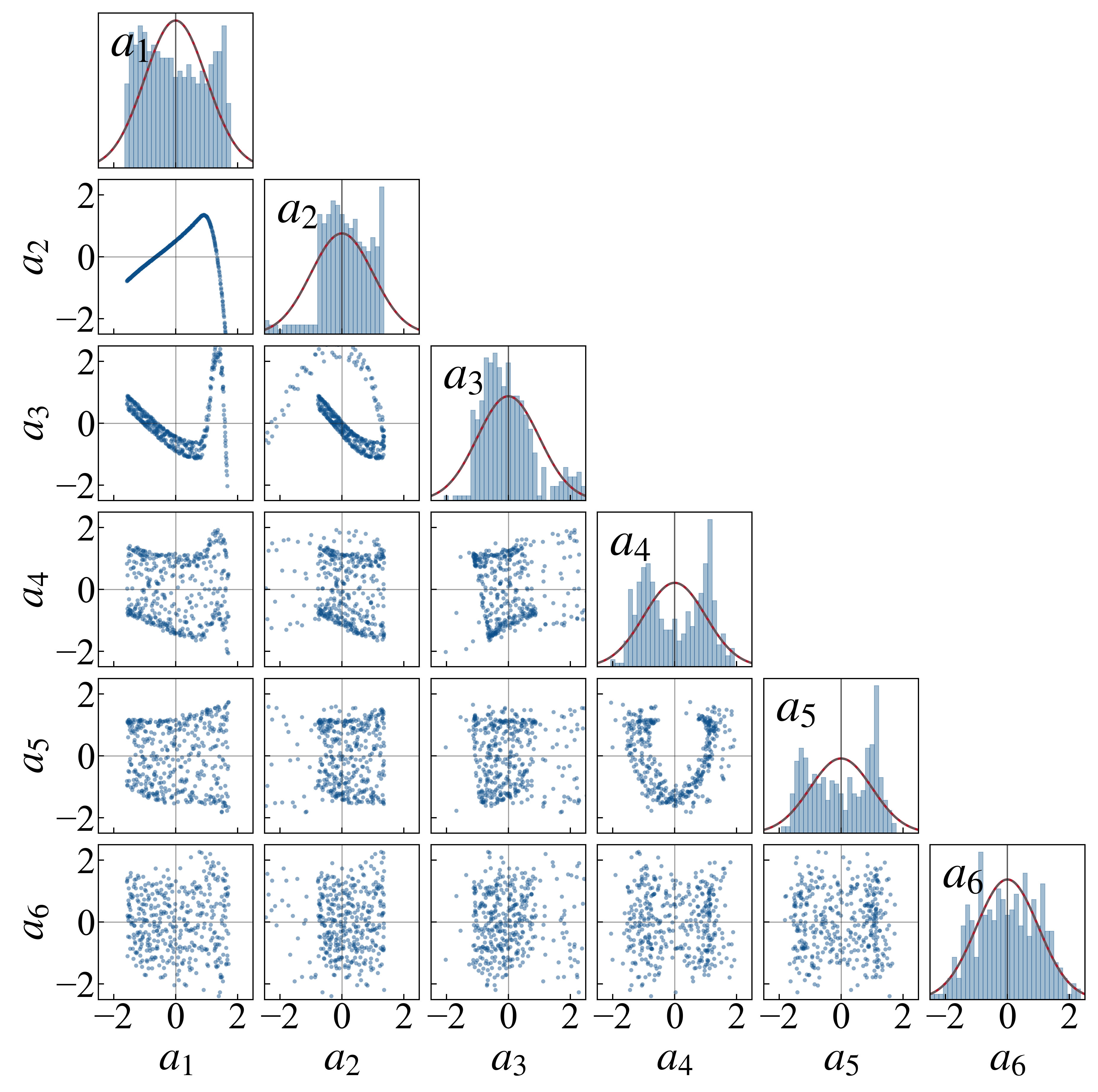}
    \caption{Distribution of the first six PCA beam coefficients across the 400-beam training set. Off-diagonal panels show the pairwise structure between coefficients, while diagonal panels show the one-dimensional coefficient distributions. The red curves show the Gaussian approximations adopted as the coefficient priors used to marginalise over beam uncertainty in the inference analyses.}
    \label{fig:pca_corner}
\end{figure}

The structure learned by this second-stage parameterisation is illustrated in Figure~\ref{fig:pca_weight_modes}. In comparison to the original SVD weight distributions shown in Figure~\ref{fig:svd_basis_weights}, we see that the mean weight functions, $e_{0,j}(\nu_i)$, capture much of the broad low-order spectral behaviour, while the subsequent functions, $e_{m,j}(\nu_i)$, encode coherent departures from this mean. This correlated structure is further reflected in Figure~\ref{fig:pca_corner}, which shows the highly structured distribution of the corresponding coefficients across the training set. Together, these figures help motivate this approach to parameterisation: it is sufficiently flexible to capture the physically realisable beam variations, while remaining maximally conditioned on their correlated structure. This is essential if beam uncertainty is to be propagated through to the antenna temperature without compromising cosmological recovery by being overly flexible.

\begin{figure}
    \centering
    \includegraphics[width=\columnwidth]{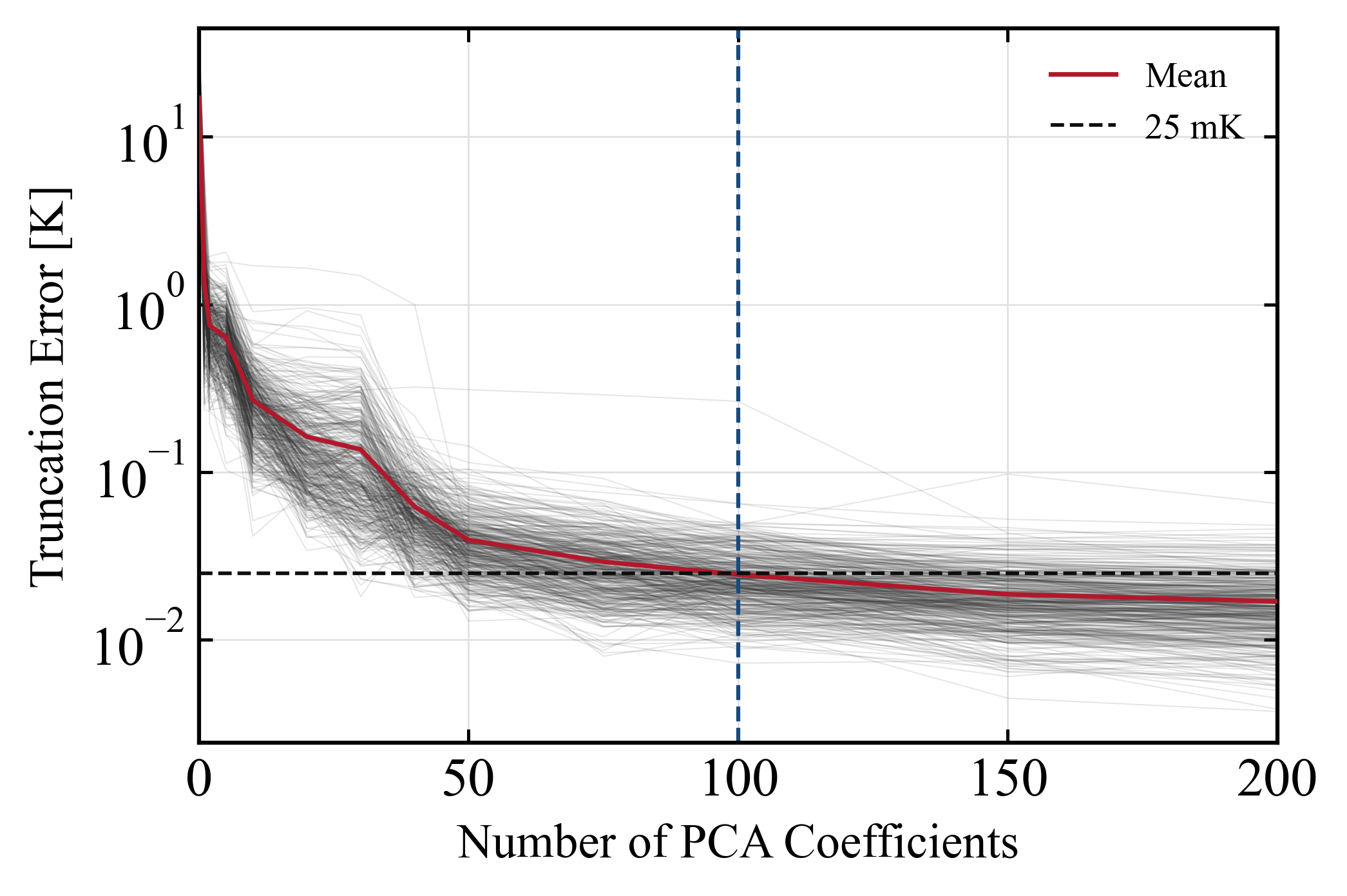}
    \caption{Antenna-temperature truncation error, $\Delta T_{A,b}^{(R,M)}(\nu)$, as a function of the number of retained PCA-weight functions. Grey curves show the RMSE for individual training beams, evaluated by propagating each truncated beam reconstruction through the foreground forward model at the conservative Galactic-centre-overhead snapshot. The red curve shows the mean across the training set. The horizontal dashed line marks the adopted 25~mK thermal-noise requirement, while the vertical dashed line indicates the 100 PCA modes retained throughout the remainder of this work.}
    \label{fig:pca_trunctation}
\end{figure}

In direct analogy to Section~\ref{sec:basis_functions}, after performing the second-stage decomposition we must determine how many PCA modes are required to reconstruct the beam with sufficient accuracy. We therefore use the same antenna-temperature truncation criterion to evaluate the error between the original beam and the PCA-truncated representation. Since the PCA coefficients are propagated back through the first-stage SVD bases before the beam is reconstructed, its ultimate accuracy is bounded by that of the retained SVD representation, which as previously stated is 20.09~mK.

Figure~\ref{fig:pca_trunctation} shows that the PCA reconstruction converges rapidly under this metric. Retaining the first 100 PCA coefficients yields a mean antenna-temperature RMSE of 24.47~mK, and is therefore sufficient to satisfy the adopted accuracy requirement. We have thus compressed the beam parameterisation by a factor of 340, while retaining the ability to reconstruct the beam to the precision required for downstream inference.

While the hierarchical construction adopted here, namely performing a second decomposition on the SVD weight space, may at first appear to be a convoluted approach to encoding the joint spatial and chromatic structure of the beam, it addresses a fundamental bottleneck exposed in recent literature. A seemingly more natural alternative would be to enforce this joint structure directly in the beam space by constructing the training matrix, $\mathbf{M}$, such that the frequency and angular axes are stacked vertically, thereby forcing the primary basis functions to span both angle and frequency simultaneously, corresponding to a matrix of dimensions $(N_{\Omega} \cdot N_{\nu}) \times N_{\mathrm{train}}$. However, this is precisely the alternative approach \citet{Pieterse_2024} explored, and showed that direct stacked-frequency construction converges much more slowly and suffers a substantial loss in reconstruction accuracy relative to the method presented in Section~\ref{sec:basis_functions}.

The two-stage framework presented addresses this trade-off by exploiting the fact that SVD weights occupy a highly structured and physically correlated space. The result is a parameterisation that retains the reconstruction accuracy, compresses the beam uncertainty to a dimesionality that is tractable for Bayesian evidence evaluation and conditions the allowed variations on the coherent spectral evolution of physically realisable beams across the band.

\section{Inference Framework}
\label{sec:bayesian}

In this section, we describe our approach for simulating high-fidelity antenna-temperature data (Section~\ref{sec:data_generation}), outline the mathematical formalism used to extend the REACH forward model so that beam uncertainty is propagated through to the antenna temperature (Section~\ref{sec:forward_model}), and finally describe how low signal-to-noise cosmological recovery can be carried out within a Bayesian framework.

\subsection{Radiometer Data Generation}
\label{sec:data_generation}

In order to fairly evaluate the performance of inference pipelines under realistic conditions, it is essential to generate data that accurately reflects the extent of the chromatic distortions introduced through the coupling of spatially and spectrally varying foregrounds to chromatic instrumental beams. Following \citealt{Anstey_2021}, for a given antenna position and observational window, this continuous sky model is built in the antenna's local Alt-Az $(\theta,\phi)$ coordinate frame by comparing the 2008 Global Sky Model \citep[GSM][]{deOliveiraCosta2008} evaluated at 408~MHz ($T_{408}$) and 230~MHz ($T_{230}$) to calculate a spatially varying spectral index field, $\beta(\theta,\phi,t)$. Here $T_{\mathrm{CMB}}=2.725~\mathrm{K}$ denotes the CMB monopole temperature, which is subtracted before spectral extrapolation and restored after scaling to the observing band:
\begin{equation} 
\label{eqn:spectralindexmap}
\beta(\theta, \phi, t)=-
\frac{
\log\!\left[(T_{230}(\theta,\phi, t) - T_{\mathrm{CMB}}) / (T_{408}(\theta,\phi, t) - T_{\mathrm{CMB}})\right]
}{
\log\!\left(230/408\right)
}.
\end{equation}
This formulation allows for the extrapolation of a given sky map at a reference frequency $\nu_0$, taken to be 230~MHz, to all frequencies within our band (50--135~MHz):
\begin{equation}
\label{eqn:skyapproximation}
T_{\mathrm{sky}}(\theta, \phi, \nu, t)
=
\left[T_{230}(\theta,\phi,t) - T_{\mathrm{CMB}}\right]
\left(\frac{\nu}{\nu_0}\right)^{-\beta(\theta,\phi,t)}
+ T_{\mathrm{CMB}}.
\end{equation}

\noindent This simulated sky temperature is then convolved with an antenna beam, described in Section~\ref{sec:beam_simulations}, to produce a time-dependent foreground contribution to the antenna spectra. Additional thermal noise realisations ($\hat{\sigma}$) can then be added to reflect true radiometer data, taken here to be homoscedastic Gaussian distributed with a standard deviation of 25~mK. The total simulated antenna temperature is therefore given by:
\begin{equation}
\label{eqn:simantenna}
T_{\mathrm{data}}(\nu, t)
=
\frac{1}{4\pi}
\int_{0}^{4\pi}
D(\theta,\phi,\nu)\,
T_{\mathrm{sky}}(\theta,\phi,\nu,t)\,
\mathrm{d}\Omega
+ \hat{\sigma}.
\end{equation}

\noindent Throughout this work, we focus on a 3-hour observational window where the Galactic centre remains below the horizon, representing a typical period where cosmological recovery would be targeted. For consistency with previous analyses, the observing window begins at an apparent local sidereal time of $\mathrm{LST}=2.05$~h at the REACH observing location, corresponding to 00:00:00 UTC on 2019-10-01.  While in reality this field is continuous, it is discretised into 1~MHz frequency channels and 5-minute time bins.

Finally, as we are considering the monopole component of the 21-cm signal, cosmological contributions to the antenna temperature are treated as an additional emission and absorption component. A Gaussian absorption profile is used to represent this:
\begin{equation}
\label{eqn:t21model}
T_{21}(\nu)
=
-A_{21}\,
\exp\!\left[
-\frac{(\nu - \nu_{21})^{2}}{2\sigma_{21}^{2}}
\right],
\end{equation}
\noindent where the baseline mock signal used in this work has an amplitude $A_{21} = 0.155~\mathrm{K}$, central frequency $\nu_{21} = 85~\mathrm{MHz}$, and width $\sigma_{21} = 15~\mathrm{MHz}$. To test the robustness of the inference pipeline under varying signal profiles, Section~\ref{sec:beam_marginalisation_results} considers three independent one-dimensional sweeps through the signal-parameter space, varying $A_{21} \in \{100,155,250,500\}~\mathrm{mK}$, $\nu_{21} \in \{75,85,95,105\}~\mathrm{MHz}$, and $\sigma_{21} \in \{5,10,15,20\}~\mathrm{MHz}$ about the baseline profile.

\begin{figure}
    \centering
    \includegraphics[width=\columnwidth]{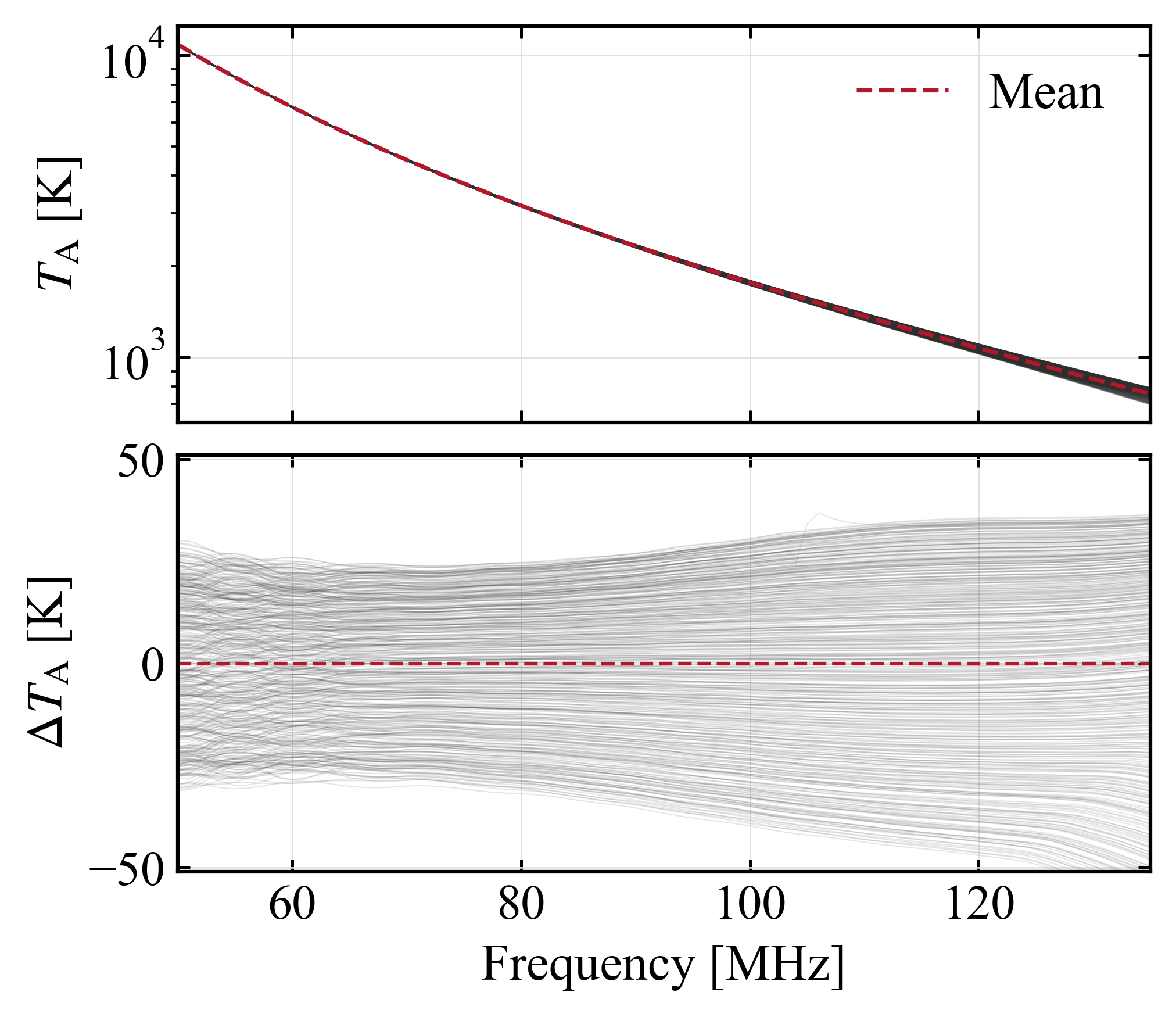}
    \caption{Generated antenna-temperature for all 500 simulated beams of the REACH dipole antenna in the Karoo Desert, South Africa, at 00:00:00 UTC on 2019-10-01. Top panel: foreground antenna temperatures, $T_{\mathrm{A}}(\nu)$, with the dashed curve showing the mean across the simulation suite. Bottom panel: deviations from this mean, $\Delta T_{\mathrm{A}}(\nu)$, demonstrating that physically plausible beam perturbations induce chromatic antenna-temperature structure at the level of tens-of-kelvin.}
    \label{fig:beam_temperature_variation}
\end{figure}

To illustrate how beam variations propagate into the data, Figure~\ref{fig:beam_temperature_variation} shows the foreground contribution to the antenna temperature for all 500 simulated beams. Despite originating from small physical perturbations around a fixed design, the resulting spectra vary significantly in amplitude and structure, with differences reaching tens of Kelvin across the band. With these variations being roughly two orders of magnitude larger than what theory predicts for the expected cosmological contribution, it further demonstrates that even modest errors in the chromatic beam response introduce structured residuals large enough to obscure cosmological signal inference.

\subsection{Forward Model}
\label{sec:forward_model}

Prior to this work, significant work has gone into developing and validating the phyiscally motivated data analysis pipeline first presented by the REACH collaboration \citep{Anstey_2021}. While for clarity we will focus on the joint inference of underlying cosmology, diffuse foreground emission and beam uncertainty, the mathematical framework remains modular and can be easily made to include further factors such as additional foreground contributions \citep{Mittal_2024, Robins_2026}, radio frequnecy interference \citep{Leeney_2023,Anstey_2024}, the ionosphere \citep{Shen_2021, Shen_2022}, and environmental effects \citep{Pattison_2023, pattison_2025a}. 

The REACH parameterisation of the diffuse foreground sky divides the sky into $N_{\beta}$ regions, each assigned a uniform spectral index $\beta_j$, and independently into $N_{\alpha}$ regions, each assigned a multiplicative amplitude correction $\alpha_i$ relative to a reference sky map (here taken to be the GSM at 230 MHz). These are defined by the binary masks $M_{\beta,j}$ and $M_{\alpha,i}$, which specify the fixed membership of each pixel $(\theta, \phi, t)$ in the corresponding regions. Overall the observed sky model for a given time and frequency, conditioned on the parameter set ${\alpha, \beta}$ is then modelled as a sum over these regions, expressed mathematically as:

\begin{equation}
\label{eqn:forward_model}
\begin{split}
T_{\mathrm{sky}}^{\mathrm{model }} (\theta&, \phi, \nu, t ) = \sum_{i=1}^{N_{\alpha}} \sum_{j=1}^{N_\beta}  M_{\alpha,i}(\theta, \phi, t) M_{\beta,j}(\theta, \phi, t) \\
& \times \left[ \alpha_i \left( T_{230}(\theta, \phi, t) - T_{\mathrm{CMB}} \right) \right] \left( \frac{\nu}{\nu_0} \right)^{-\beta_j} + T_{\mathrm{CMB}},
\end{split}
\end{equation}

\noindent While historically this modelled sky would then be convolved with a fixed beam $D(\theta, \phi, \nu)$ to determine the observed antenna temperature, in this work the beam is itself promoted to a model component, $\tilde{D}^{(M,R)}(\nu,\theta,\phi)$, parameterised by the PCA coefficients $\{a_m\}$ introduced in Section~\ref{sec:beam_parameterisation}. The corresponding model antenna temperature is then given by
\begin{equation}
\label{eqn:antenna_forward_model}
T_{\mathrm{A}}^{\mathrm{model}}(\nu, t)
=
\frac{1}{4\pi}
\int_{4\pi}
\tilde{D}^{(M,R)}(\nu,\theta,\phi)\,
T_{\mathrm{sky}}^{\mathrm{model}}(\theta,\phi,\nu,t)\,
\mathrm{d}\Omega.
\end{equation}

\noindent In direct analogy with the chromatic response functions $\mathcal{K}_{i,j}(\nu, t)$ introduced in the convention written in \citet{Tutt_2026}, the linearity of the beam model allows the coupling between the regional sky model and the beam bases to be pre-computed. Rather than evaluating the angular integral during inference, we define the basis response functions
\begin{equation}
\label{eqn:basis_response_functions}
\begin{split}
\mathcal{B}_{i,j,m}(\nu,t)
=
&
\frac{1}{4\pi}
\int_{4\pi}
M_{\alpha,i}(\theta,\phi,t)\,
M_{\beta,j}(\theta,\phi,t)\,
 \\\times
& \left[
T_{230}(\theta,\phi,t)-T_{\mathrm{CMB}}
\right] 
\sum_{k=1}^{R}
e_{m,k}(\nu)\,
B_k(\theta,\phi)\,
\mathrm{d}\Omega,
\end{split}
\end{equation}
\noindent where $B_k(\theta,\phi)$ are the retained SVD angular basis functions defined in Section~\ref{sec:basis_functions}, and $e_{m,k}(\nu)$ are the PCA-space mean ($m=0$) and correction functions ($m\geq 1$) defined in Section~\ref{sec:beam_parameterisation}.

From these, the foreground contribution to the antenna temperature can be calculated through a series of matrix operations on the pre-computed response functions, yielding a forward model well suited to hardware-accelerated inference. The resulting foreground model is therefore given by
\begin{equation}
\label{eqn:fast_forward_model}
T_{\mathrm{FG}}^{\mathrm{model}}(\nu,t)
=
\sum_{m=0}^{M} {a}_m
\sum_{i=1}^{N_{\alpha}}
\sum_{j=1}^{N_{\beta}}
\alpha_i \mathcal{B}_{i,j,m}(\nu,t)
\left( \frac{\nu}{\nu_0} \right)^{-\beta_j}
+
T_{\mathrm{CMB}},
\end{equation}
where $a_0 = 1$ as it corresponds to the mean PCA weight function. 

In the analyses presented in Section~\ref{sec:results}, we adopt a fixed foreground partition of 30 spectral-index regions ($N_{\beta}=30$). For the purposes of this work, the foreground-amplitude corrections are held fixed at $\alpha_i = 1$. The forward model is therefore conditioned on 30 foreground spectral-index parameters, $\boldsymbol{\beta}$, 100 retained beam coefficients, $\boldsymbol{a}$, and the three cosmological parameters, $\boldsymbol{\theta_{21}}$, of the Gaussian absorption profile, giving 133 free parameters in total. These naturally separate into parameters of scientific interest and nuisance parameters. The former comprise both the cosmological and foreground parameters, $\boldsymbol{\beta}$, since constraining the low-frequency radio sky remains an important scientific objective in its own right \citep{Carter_2025, Robins_2026}. The nuisance parameters are the beam coefficients, which enter only to propagate instrumental uncertainty through the inference.

\subsection{Bayesian Marginalisation}
\label{sec:beam_marginalisation}

To evaluate the uncertainties and degeneracies of the model parameters in light of the data, we employ Bayes theorem. This allows us to update our prior state of belief over the parameters, $\pi(\boldsymbol{\theta}_{\mathbf{M}} \mid \mathbf{M})$, using the conditional probability of the data given the model, $\mathcal{L}(\mathbf{D} \mid \boldsymbol{\theta}_{\mathbf{M}}, \mathbf{M})$, to yield the joint posterior distribution:
\begin{equation}
\label{eqn:bayes_theorem}
P(\boldsymbol{\theta}_{\mathbf{M}} \mid \mathbf{D}, \mathbf{M}) =
\frac{
\mathcal{L}(\mathbf{D} \mid \boldsymbol{\theta}_{\mathbf{M}}, \mathbf{M})
\,
\pi(\boldsymbol{\theta}_{\mathbf{M}} \mid \mathbf{M})
}{
\mathcal{Z}(\mathbf{D} \mid \mathbf{M})
},
\end{equation}
where $\mathcal{Z}(\mathbf{D} \mid \mathbf{M})$ is the Bayesian evidence (or marginal likelihood).

To draw samples from this posterior distribution while simultaneously evaluating the evidence, we employ the parallelised Nested Slice Sampling (NSS) algorithm implemented in \texttt{BlackJAX} \citep{yallup2025nested, cabezas2024blackjax}. Although the PCA beam parameterisation developed in Section~\ref{sec:beam_parameterisation} reduces the instrumental dimensionality to a level tractable for evidence-based inference \citep[see][]{lovick_2025}, the computational efficiency of NSS scaling remains highly sensitive to the total number of sampled parameters, typically scaling as $\sim \mathcal{O}(N_{\mathrm{Dim}}^3)$ \citep{polychord}. Fortuitously, because the forward model depends linearly on the beam coefficients, we can analytically marginalise over this subspace allowing us to rigorously incorporate instrumental uncertainty without explicitly sampling them.

To achieve this, we first isolate the linear beam correction modes from the astrophysical and cosmological components. Conditioned on the foreground and cosmological parameters, we define the mean basis model vector as
\begin{equation}
\label{eqn:mean_model}
\begin{split}
\mu(\nu,t \mid \boldsymbol{\alpha}, \boldsymbol{\beta}, \boldsymbol{\theta}_{21})
\equiv
\sum_{i=1}^{N_{\alpha}}
\sum_{j=1}^{N_{\beta}}
\alpha_i \mathcal{B}_{i,j,0} & (\nu,t)
\left( \frac{\nu}{\nu_0} \right)^{-\beta_j} \\ &
+
T_{\mathrm{CMB}}
+
T_{21}(\nu \mid  \boldsymbol{\theta}_{21}).
\end{split}
\end{equation}

\noindent The basis correction vectors are then, 
\begin{equation}
\label{eqn:design_matrix_columns}
h_{m}(\nu,t \mid \boldsymbol{\alpha}, \boldsymbol{\beta})
\equiv
\sum_{i=1}^{N_{\alpha}}
\sum_{j=1}^{N_{\beta}}
\alpha_i \mathcal{B}_{i,j,m}(\nu,t)
\left( \frac{\nu}{\nu_0} \right)^{-\beta_j}.
\end{equation}

\noindent Here $\mu(\nu,t \mid \boldsymbol{\alpha}, \boldsymbol{\beta}, \boldsymbol{\theta}_{21})$ and $h_m(\nu,t \mid \boldsymbol{\alpha}, \boldsymbol{\beta})$ are functions of frequency and time. In the inference, these are evaluated on the full frequency-time grid and flattened to form the vector $\boldsymbol{\mu}$ and the columns $\mathbf{h}_m$ of: 

\begin{equation}
\label{eqn:design_matrix}
\mathbf{H}(\boldsymbol{\alpha}, \boldsymbol{\beta})
\equiv
\left[
\mathbf{h}_{1}\;
\mathbf{h}_{2}\;
\cdots\;
\mathbf{h}_{M}
\right].
\end{equation}

\noindent such that the forward modelled data vector can therefore be written, conditional on the astrophysical parameters, as
\begin{equation}
\label{eqn:linear_beam_model}
\mathbf{M}
=
\boldsymbol{\mu}
+
\mathbf{H}\boldsymbol{a},
\end{equation}
such that the beam dependence is now explicitly linear in the pca weights $\boldsymbol{a} = (a_1,\ldots,a_M)^{\mathrm{T}}$.

Assuming homoscedastic Gaussian noise parametrised by a covariance matrix $\mathbf{C}_n$, the unmarginalised joint log-likelihood for both the astrophysical parameters, $\boldsymbol{\phi} \equiv (\boldsymbol{\beta}, \boldsymbol{\theta}_{21})$, and the beam parameters, $\boldsymbol{a}$, is expressed as
\begin{equation}
\label{eqn:original_likelihood}
\begin{split}
\ln \mathcal{L}(\mathbf{d}\mid\boldsymbol{\phi},\boldsymbol{a})
=
-\frac{1}{2}
\Big[
\left(
\mathbf{d}-\boldsymbol{\mu}-\mathbf{H}\boldsymbol{a}
\right)& ^{\mathrm{T}} 
\mathbf{C}_{n}^{-1}
\left(
\mathbf{d}-\boldsymbol{\mu}-\mathbf{H}\boldsymbol{a}
\right)
\\
&+
\ln \det \mathbf{C}_{n}
+
N_{\mathrm{data}}\ln(2\pi)
\Big].
\end{split}
\end{equation}

\noindent Given a Gaussian prior on the beam coefficients,
\begin{equation}
\label{eqn:beam_prior}
\boldsymbol{a}
\sim
\mathcal{N}
\left(
\boldsymbol{0},
\mathbf{\Sigma}_{a}
\right),
\end{equation}
\noindent where $\mathbf{\Sigma}_{a}$ is set to be the diagonal covariance matrix of the PCA coefficients across the training set. Some departure from Gaussianity is visible among the first few coefficient distributions in Figure~\ref{fig:pca_corner}, while the higher-order modes are increasingly well described by the adopted approximation. Importantly, its adequacy is tested empirically through the held-out recovery analyses in Section~\ref{sec:beam_marginalisation_results}, where it does not prevent robust signal recovery.

Since both the prior and the conditional likelihood are Gaussian, we can exploit conjugate identities to analytically integrate out $\boldsymbol{a}$ from the joint probability distribution. This marginalisation effectively inflates the instrumental noise covariance to form an effective data covariance matrix, $\mathbf{C}_{\mathrm{eff}}$:
\begin{equation}
\label{eqn:effective_covariance}
\mathbf{C}_{\mathrm{eff}}
=
\mathbf{C}_{n}
+
\mathbf{H}\,
\mathbf{\Sigma}_{a}\,
\mathbf{H}^{\mathrm{T}}.
\end{equation}
\noindent Substituting this expression into Equation~\ref{eqn:original_likelihood} yields the final marginalised log-likelihood function, which depends solely on the core astrophysical parameters:
\begin{equation}
\label{eqn:marginalised_likelihood}
\begin{split}
\ln \mathcal{L}_{\mathrm{marg}}(\mathbf{d}\mid\boldsymbol{\phi})
=
-\frac{1}{2}
\Big[
\left(
\mathbf{d}-\boldsymbol{\mu}
\right)^{\mathrm{T}} &
\mathbf{C}_{\mathrm{eff}}^{-1} 
\left(
\mathbf{d}-\boldsymbol{\mu}
\right) \\
& +
\ln \det \mathbf{C}_{\mathrm{eff}}
+
N_{\mathrm{data}}\ln(2\pi)
\Big].
\end{split}
\end{equation}

\noindent Overall, the sampled parameter space is collapsed from 134 dimensions down to just 34, significantly accelerating inference while still rigurously propagating full instrumental beam uncertainties into our final cosmological posteriors and evidence evaluation. The priors used for the sampled astrophysical and noise parameters, together with the Gaussian prior used for the marginalised beam coefficients, are summarised in Table~\ref{tab:priors}.

\begin{table}
\centering
\caption{Prior distributions used in the Bayesian inference. The beam coefficients are not sampled directly in the marginalised likelihood, but their Gaussian prior defines the beam-uncertainty covariance propagated through Equation~\ref{eqn:effective_covariance}.}
\label{tab:priors}
\begin{tabular}{llll}
\toprule
Parameter & Prior & Range & Units \\
\midrule
\multicolumn{4}{l}{\textit{Statistical noise}} \\
Noise amplitude, $\sigma_n$ & Log-uniform & $[10^{-4},10^{-1}]$ & K \\
\midrule
\multicolumn{4}{l}{\textit{Regional foregrounds}} \\
Spectral index, $\beta_j$ & Uniform & $[2.458,3.146]$ & -- \\
\midrule
\multicolumn{4}{l}{\textit{Global 21-cm signal}} \\
Amplitude, $A_{21}$ & Uniform & $[0.05,0.6]$ & K \\
Centre frequency, $\nu_{21}$ & Uniform & $[50,200]$ & MHz \\
Width, $\sigma_{21}$ & Uniform & $[2,40]$ & MHz \\
\midrule
\multicolumn{4}{l}{\textit{Beam uncertainty}} \\
Beam coefficients, $a_m$ & Gaussian & $\mathcal{N}(0,\Sigma_{a,mm})$ & -- \\
\bottomrule
\end{tabular}
\end{table}

\section{Results}
\label{sec:results}

In this section, we present the performance of the inference framework described in Section~\ref{sec:bayesian}. We begin by illustrating the failure of the conventional fixed-beam approach under incorrect beam conditioning (Section~\ref{sec:fixed_beam_inference}). We then demonstrate the performance of the beam-marginalisation framework across the suite of validation beams and cosmological realisations (Section~\ref{sec:beam_marginalisation_results}), before finally investigating the robustness of the methodology under different training-set sizes and distributions in Section~\ref{sec:changing_training_set} to establish practical requirements for future analyses.

\subsection{Fixed Beam Inference}
\label{sec:fixed_beam_inference}
\begin{figure}
    \centering
    \includegraphics[width=\columnwidth]{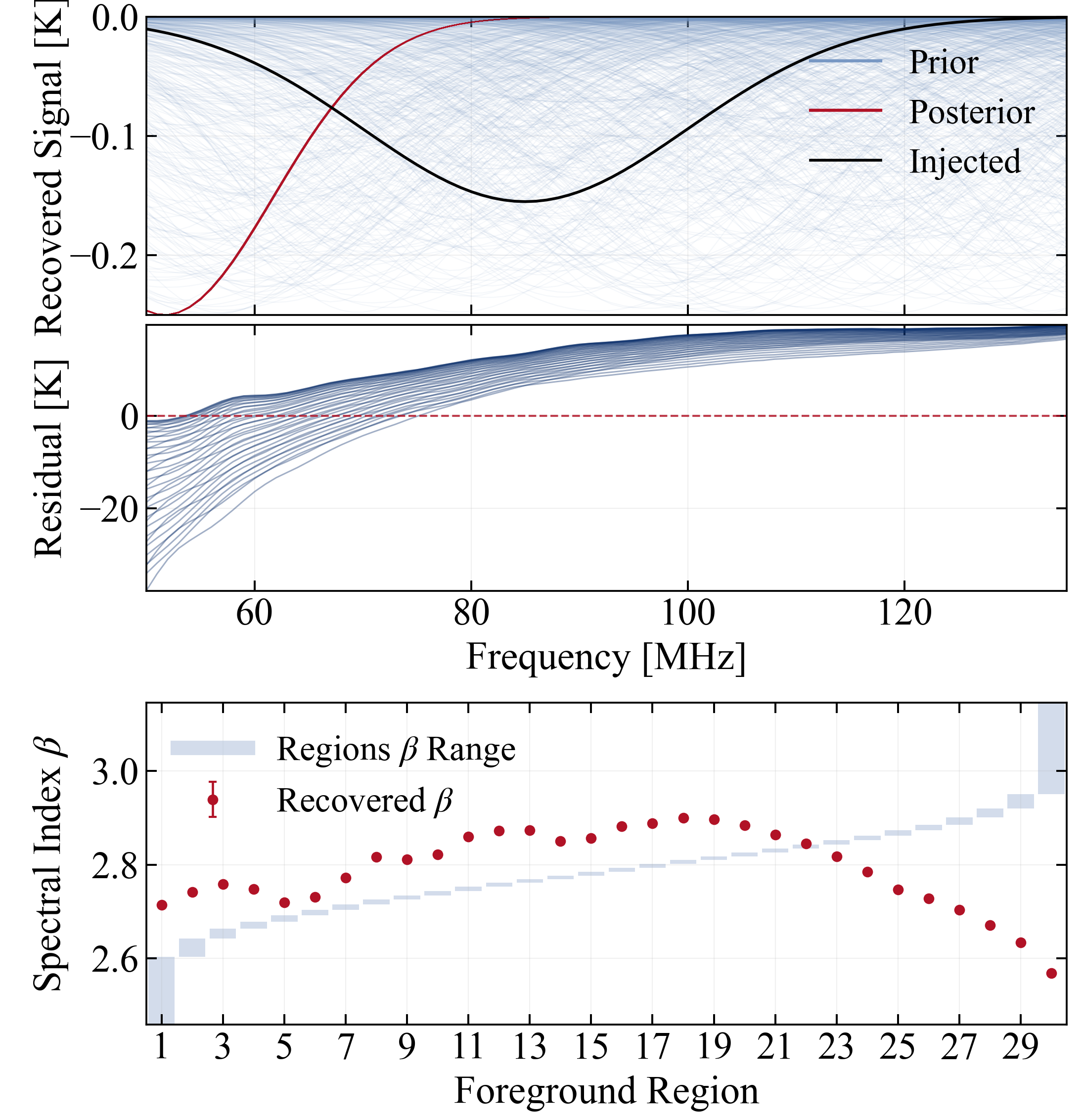}
    \caption{Representative failure mode of fixed-beam inference under beam mismatch. Top panel: recovered 21-cm signal posterior compared to the injected absorption profile. Middle panel: full-model residuals, showing unmodelled chromatic structure at the tens-of-kelvin level. Bottom panel: recovered foreground spectral-index parameters compared to the range of the spectral-index field $\beta(\nu, \theta, \phi)$ in that region used to generate the data.}
    \label{fig:fixed_beam_failure}
\end{figure}

To provide an end-to-end benchmark against which to compare the methodology presented in this work, we first consider the conventional approach in which the beam is treated as a known quantity. In this case, the analysis reduces to the simplified forward model described in previous REACH analyses \citep[for the exact formalism, see][]{Tutt_2026}, where the chromatic response functions are precomputed for a single assumed beam and held fixed throughout the inference.

Specifically, we consider data generated following the procedure outlined in Section~\ref{sec:data_generation} using Beam 0 from the validation set, and then perform the inference while incorrectly conditioning the forward model on Beam 1. Both beams lie at relatively small global distances ($\leq10\%$) from the centre of the sampled parameter space under the normalised distance metric defined in Equation~\ref{eqn:normalised_beam_parameter_distance}. We therefore use this comparison as a representative example of beam misspecification within the simulated uncertainty volume.

\begin{figure*}
    \centering
    \includegraphics[width=\textwidth]{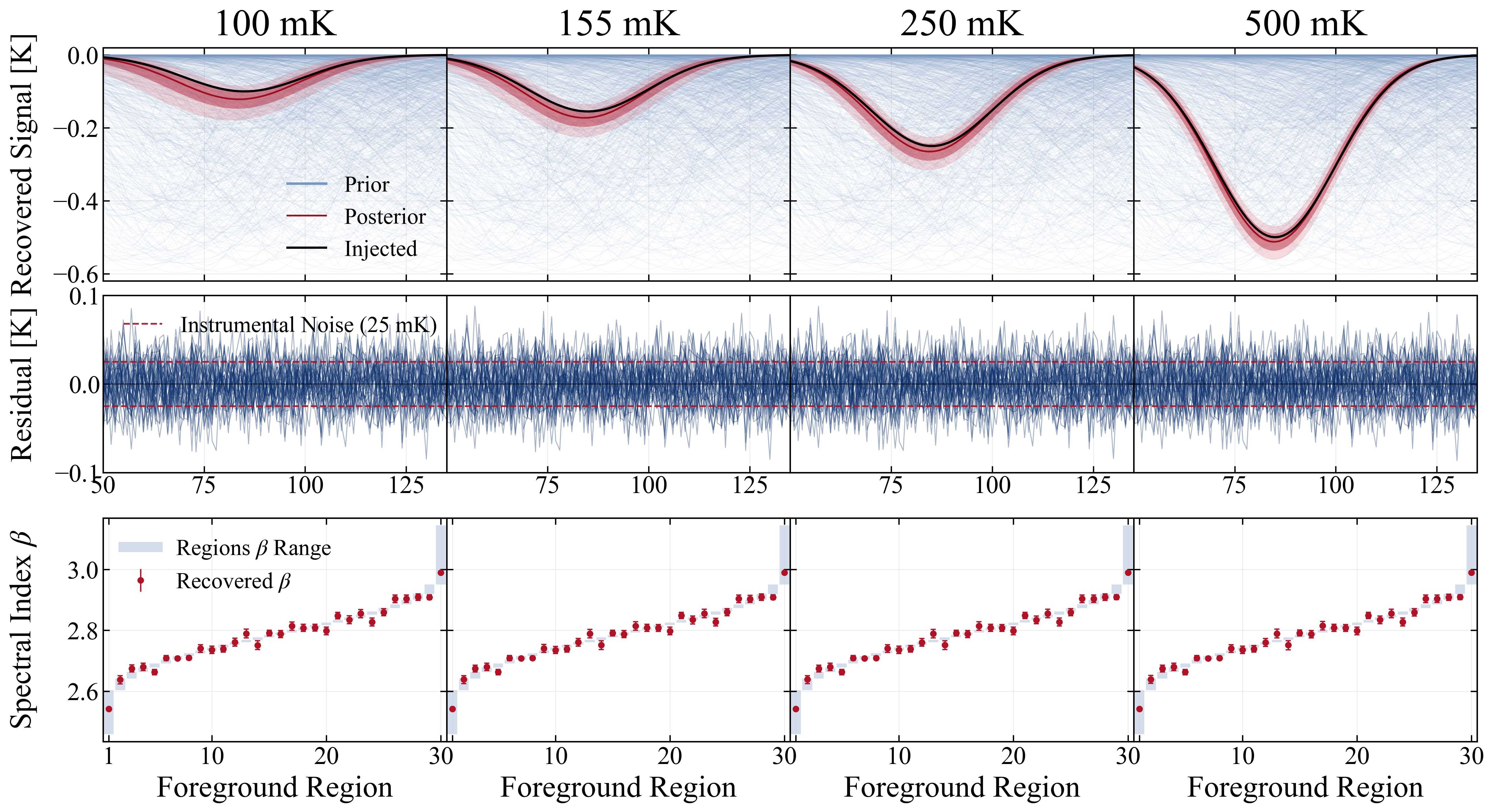}
    \caption{Beam-marginalised inference results for data generated with Beam~1 and four injected 21-cm absorption profiles with amplitudes $A_{21}=\{100,155,250,500\}$~mK, at fixed central frequency $\nu_{21}=85$~MHz and width $\sigma_{21}=15$~MHz. Top row: recovered signal posterior, with the 68 and 95 per cent credible regions shaded in red, the central red curve showing the posterior-median signal and the black curve showing the injected signal. Middle row: full-model residuals at the conditional maximum-posterior beam parameters; dashed red lines indicate the $\pm25$~mK instrumental-noise level. Bottom row: recovered foreground spectral-index parameters compared with the regional $\beta$ ranges used to generate the data.}
    \label{fig:400_signal_recovery_plot}
\end{figure*}

The resulting inference is shown in Figure~\ref{fig:fixed_beam_failure}. The beam mismatch leaves residuals at the level of tens of kelvin, many orders of magnitude larger than the expected cosmological signal. The inference therefore attempts to absorb unmodelled instrumental structure into the astrophysical components of the model, leading to a strongly biased recovery of the injected 21-cm absorption profile. Although foreground recovery is often assumed to require less stringent modelling accuracy than cosmological inference, the same beam mismatch also biases the recovered foreground spectral-index parameters, $\boldsymbol{\beta}$. Overall, this demonstrates that conditioning the analysis on an incorrect beam can lead to catastrophic failure in both foreground and cosmological inference.

\subsection{Beam Marginalised Inference}
\label{sec:beam_marginalisation_results}

\begin{figure*}
    \centering
    \includegraphics[width=\textwidth]{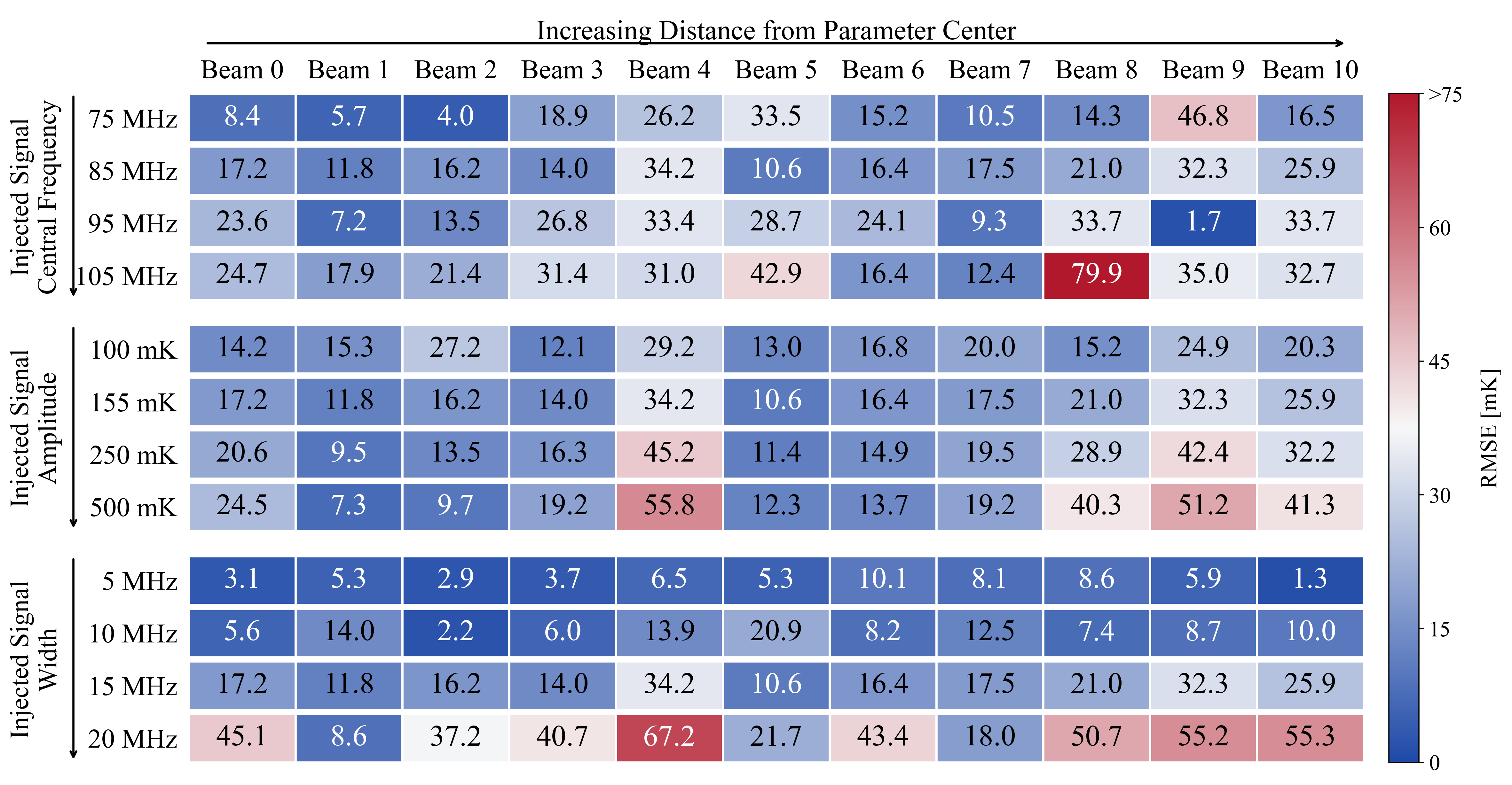}
    \caption{Signal-recovery RMSE across the validation grids using the 400-beam PCA prior. Columns show the 11 validation beams ordered by increasing normalised distance from the centre of the sampled physical parameter space. From top to bottom, the three row blocks independently vary the injected signal central frequency, amplitude and width, with the remaining signal parameters fixed at their baseline values. Consequently, the baseline injection, defined by $\nu_{21}=85$~MHz, $A_{21}=155$~mK and $\sigma_{21}=15$~MHz, is repeated once within each block. Each cell reports the RMSE between the injected signal and the posterior-mean recovered signal. The colour scale is capped at 75~mK to highlight cases above the simulated 25~mK instrumental-noise level.}
    \label{fig:400_signal_recovery_table}
\end{figure*}

We finally turn to the performance of the full inference framework across the complete suite of simulated observations. This contains all 11 validation beams listed in Table~\ref{tab:validation_beam_parameters}, each analysed across the three independent signal-parameter sweeps in amplitude, central frequency and width defined in Section~\ref{sec:data_generation}. Together, these comprise 110 unique inference runs, allowing us to test the robustness of the recovery to both the form of the cosmological signal within the observing band and increasing distance from the centre of the sampled beam-parameter space.

The exclusion of these beams from the training dataset allows us to ensure that their specific chromatic structure is not used to construct the angular basis functions, $B_k(\theta,\phi)$, in Section~\ref{sec:basis_functions}, the PCA basis functions, $e_{m,k}(\nu)$, in Section~\ref{sec:beam_parameterisation}, or the PCA coefficient covariance matrix, $\mathbf{\Sigma}_a$, in Equation~\ref{eqn:beam_prior}. In other words, the held-out beams are able to test whether the inference can generalise to unseen realisations drawn from the same physical uncertainty space.

For comparison with the fixed-beam failure mode shown in Figure~\ref{fig:fixed_beam_failure}, Figure~\ref{fig:400_signal_recovery_plot} shows the recovered signal posteriors, foreground spectral indices, and the associated residuals for the four signal-amplitude injections generated using Beam~1. The corresponding recoveries across the central-frequency and width sweeps are presented in Appendix~\ref{app:signal_shape_recovery}. Due to the marginalisation of the beam coefficients, the residuals are less trivial to define. We therefore reconstruct the conditional maximum-posterior beam coefficients at the maximum posterior of the sampled astrophysical parameters. Specifically, for $\hat{\boldsymbol{\phi}} = (\hat{\boldsymbol{\beta}},\hat{\boldsymbol{\theta}}_{21}, \hat{\sigma}_{n})$, we evaluate $\boldsymbol{\mu}$, $\mathbf{H}$, and $\mathbf{C}_{\mathrm{eff}}$ and compute
\begin{equation}
\hat{\boldsymbol{a}}
=
\mathbf{\Sigma}_a
\mathbf{H}^{\mathrm{T}}
\mathbf{C}_{\mathrm{eff}}^{-1}
\left(
\mathbf{d}
-
\boldsymbol{\mu}
\right).
\label{eqn:conditional_beam_map}
\end{equation}
\noindent The residuals shown in Figure~\ref{fig:400_signal_recovery_plot} are then computed as
\begin{equation}
\mathbf{r}
=
\mathbf{d}
-
\left[
\boldsymbol{\mu}
+
\mathbf{H}\hat{\boldsymbol{a}}
\right].
\label{eqn:conditional_residual}
\end{equation}

\noindent Across the four Beam 1 cases, the signal posteriors are consistent with the injected profiles. The recovered spectral-index parameters also lie within the corresponding regional ranges of the continuous input map, $\beta(\theta,\phi,t)$, used to generate the data. Furthermore, the corresponding residuals are reduced to the level of the simulated instrumental noise (25~mK), demonstrating that overall we are able to model the data to noise-limited levels, in stark contrast to Section~\ref{sec:fixed_beam_inference}.

Focusing on cosmological recovery across the complete suite, Figure~\ref{fig:400_signal_recovery_table} shows the RMSE between the injected signal and the posterior-mean signal for all beam and signal configurations. In the vast majority of cases, the recovered posterior is consistent with the injected signal at approximately the level of the instrumental noise.

Comparatively, it can be seen that, while signal recovery remains consistent with the injected profiles, both throughout the analyses presented here and those in Section~\ref{sec:changing_training_set}, Beam 4 typically yields larger RMSE values than other validation beams at a similar distance from the parameter-space centre. Through inspection of Table~\ref{tab:validation_beam_parameters}, and the choice of how distance is defined in Equation~\ref{eqn:normalised_beam_parameter_distance}, we can see that while Beam 4 is globally relatively close to the centre of the parameter space (the 40th percentile), it has the most extreme displacement from the ground-plane centre in the $x$ direction, $\Delta x_{\mathrm{rel}} = 197$~mm, and therefore sits on the edge of this particular physical parameter axis. Similarly, while still successfully recovered, Beams 9 and 10 also yield comparatively larger RMSE values than other beams, as by construction they lie close to the global edge of the sampled parameter space.

Considering each of the signal-parameter sweeps in turn, the central-frequency results remain broadly consistent with the injected profiles. The only notable exceptions are three cases with an RMSE greater than 40~mK, each of which occurs for an injection where the deepest part of the absorption profile lies close to the edge of the observing band. In these cases, the reduced spectral coverage on one side of the profile increases its degeneracy with the foreground model and consequently complicates signal recovery. A similarly stable recovery is observed across the signal-amplitude sweep. The comparatively larger RMSE values again tend to be associated with Beams 4, 9 and 10, with no clear systematic degradation as the injected amplitude is varied.

In contrast, a clear trend is observed as the width of the injected signal increases, with recovery degrading for broader absorption profiles. As the signal broadens, its evolution across the observing band becomes increasingly smooth and therefore more degenerate with both the foreground emission and the chromatic structure introduced through the beam--foreground coupling. While these tests do not directly compare antenna bandwidths, this behaviour reinforces the motivation for wide-band antenna designs that encompass the absorption profile as fully as possible. Although restricting the observing bandwidth may reduce instrumental chromaticity, it also limits the spectral leverage available to distinguish broad cosmological features from smooth foreground and instrumental structure. The framework presented here provides a route to retain broad frequency coverage while explicitly propagating uncertainty in the chromatic beam response.

Overall, the method shows robust performance across the grid of beams and cosmological injections, demonstrating that the framework is able to generalise to unseen realisations drawn from the same physical uncertainty space. 

\subsection{Sensitivity to the Beam Training Set}
\label{sec:changing_training_set}

\begin{figure*}
    \centering
    \includegraphics[width=\textwidth]{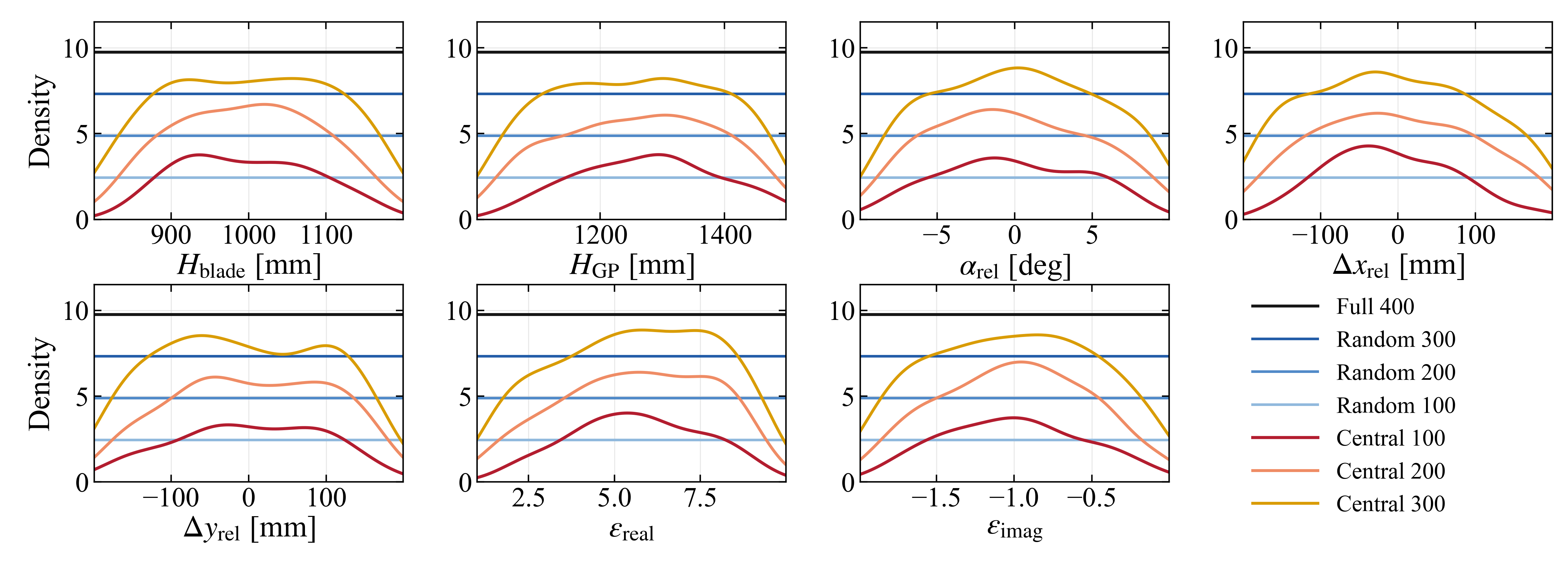}
    \caption{Physical-parameter distributions for the training sets used to construct the beam-uncertainty parameterisations. Each panel shows one of the seven simulation parameters varied in the electromagnetic beam suite. The distributions compare the full 400-beam training set with the reduced random and central training subsets, illustrating how the alternative training sets change both the density and coverage of the sampled physical uncertainty space.}
    \label{fig:training_parameter_distributions}
\end{figure*}

Given the demonstrated success of the methodology, we now explore the sensitivity of the inference to the size and distribution of the training dataset. A central requirement of the method presented is access to a simulation suite that sufficiently describes the beam uncertainty such that an efficient parameterisation can be learned and subsequently marginalised over. As the method is extended to larger numbers of physical uncertainty parameters, it is therefore important to understand how densely the training set must sample the physical parameter space (Section~\ref{sec:changing_training_set_sizes}). In addition, while a large simulation suite covering the full range of physically plausible perturbations is ideal, in practice it is unlikely that the full uncertainty volume will be perfectly known or sampled. We therefore also test the robustness of the method when the validation beams require extrapolation beyond the training-set support (Section~\ref{sec:changing_training_set_distributions}).

To investigate the effect of training-set size, we construct three additional training sets of 100, 200 and 300 beams by randomly sampling from the original training set, as visualised in Figure~\ref{fig:training_parameter_distributions}. The lower limit of 100 beams is set by the PCA trucation criterion derived in Section~\ref{sec:beam_parameterisation}, since the number of PCA modes that can be constructed is bounded by the number of SVD-weight vectors, $\mathbf{w}_b$, in the training ensemble.

To investigate the effect of training-set distribution, we construct a second set of reduced training ensembles by selecting the 100, 200 and 300 beams closest to the centre of the physical parameter space, using the distance metric defined in Equation~\ref{eqn:normalised_beam_parameter_distance}. These central training sets are also shown in Figure~\ref{fig:training_parameter_distributions}, and are designed to test how well the method performs when increasingly distant validation beams lie outside the region spanned by the training prior.

For each of these six reduced training sets, we repeat the full basis-construction procedure from the beginning. That is, each training set has unique angular basis functions, second-stage PCA functions, and PCA coefficient covariance matrix. The resulting six inference frameworks therefore marginalise over distinct learned beam-uncertainty spaces. We then apply each framework to simulated data from the same 11 beams, using the baseline injected cosmological signal with $A_{21} =155~\mathrm{mK}, \nu_{21}= 85~\mathrm{MHz}$ and $\sigma_{21} = 15~\mathrm{MHz}$.

\subsubsection{Training Set Size}
\label{sec:changing_training_set_sizes}

\begin{figure*}
    \centering
    \includegraphics[width=\textwidth]{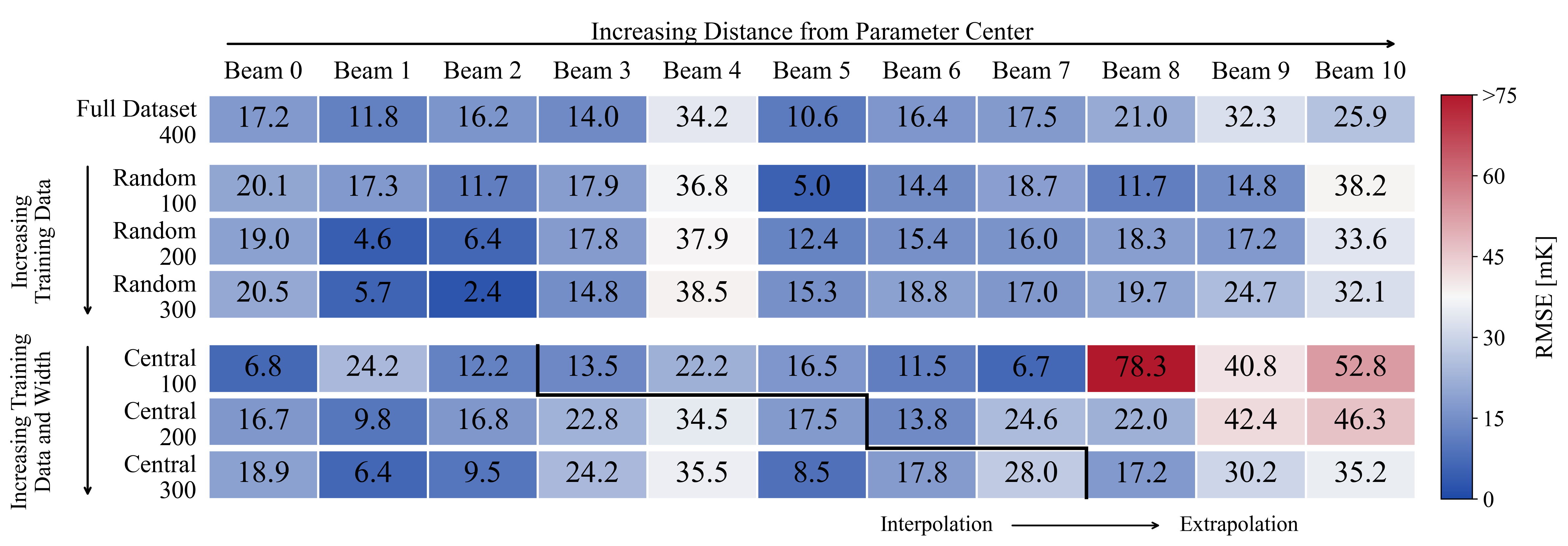}
    \caption{Signal-recovery RMSE for the validation beams under different beam uncertainty models built from different beam subsets. Columns show the 11 validation beams ordered by increasing distance from the centre of the sampled physical parameter space. Rows compare the full 400-beam training set, reduced random training sets, and reduced central training sets. The black line separates interpolation-like cases, where validation beams lie within the training-set support, from extrapolation-like cases where the beam prior must generalise beyond the reduced training volume.}
    \label{fig:signal_rmse_table}
\end{figure*}

We first present the effect of randomly thinning the training set, with the resulting RMSE between the posterior-mean recovered signal and the injected signal shown in Figure~\ref{fig:signal_rmse_table}. A priori, reducing the density with which the physical parameter space is sampled might be expected to degrade the ability of the learned beam prior to generalise to held-out realisations. However, no clear trend is observed as the training set is reduced from 400 to 100 beams. With the exception of the same higher-RMSE cases discussed above, namely Beams 4, 9 and 10, the recovered signals remain consistent with the simulated instrumental-noise level.

This indicates that, for the uncertainty volume considered here, the limiting factor is not the number of CEM simulations required to learn a generalisable beam-uncertainty parameterisation, but rather the reconstruction accuracy set by the number of retained PCA modes. This is encouraging for future applications of the method, as it suggests that end-to-end beam marginalisation may be significantly less simulation-intensive than expected, reducing the computational cost of implementation and allowing the approach to scale to larger numbers of physical uncertainty parameters.

\subsubsection{Training Set Coverage}
\label{sec:changing_training_set_distributions}

We next consider the effect of restricting the training set to the central 25\%, 50\% and 75\% regions of the physical parameter space. In doing so, the held-out validation beams can be classified into two categories, separated by the black stepped line in Figure~\ref{fig:signal_rmse_table}. The first contains beams that lie within the support of the training prior, and therefore represent interpolation-like cases. The second contains beams that lie outside the support of the training prior, and therefore represent extrapolation-like cases. 

Focusing on the extrapolation cases, we find that the method is able to extrapolate beyond the initial training support to a reasonable extent. For example, for the central 25\% training set, corresponding to 100 beams, the data generated using Beams 3--7 are all classified as out-of-sample, yet the recovered signals remain consistent with the injected profile at the level of the instrumental noise. As expected, this degrades for the most extreme cases, Beams 8--10, where the RMSE rises above 40~mK. This shows that the method cannot arbitrarily extrapolate to any beam outside the training support, as expected from both the structure encoded in the basis functions and the constraints imposed by the PCA prior.

Additionally, as expected, increasing the size of the central training set improves the recovery of the beams that previously failed. In particular, Beams 8--10 move closer to the extrapolation boundary as the training support is expanded, and their RMSE correspondingly decreases.

\section{Conclusions}
\label{sec:conclusions}

In this work, we have presented a significant advancement towards robust, end-to-end uncertainty propagation for global 21-cm cosmology. A confirmed detection of the sky-averaged 21-cm signal will require analysis frameworks that can accurately model the bright foreground sky and chromatic instrumental response, while also accounting for their degeneracies with the underlying cosmological signal. 

Despite this, analysis frameworks across the field have typically been conditioned on a single assumed beam model, preventing instrumental uncertainty from being propagated into the recovered signal posterior and Bayesian evidence. The method presented here addresses this limitation by learning the uncertainty in the beam chromaticity from a suite of electromagnetic simulations, and marginalising over these variations within a physically motivated Bayesian forward-modelling framework.

The methodology consists of three main components:

\begin{itemize}
\item \textbf{Simulating the beam uncertainty:} Using a suite of CEM realisations for the REACH hexagonal blade dipole antenna, spanning perturbations in antenna geometry and soil permittivity, we characterise the coherent angular and spectral changes induced by plausible instrumental differences.

\item \textbf{Efficient surrogate parameterisation:} We introduce a two-stage decomposition in which an initial SVD captures the dominant angular beam modes, while a second decomposition of the SVD weights learns the correlated spectral evolution of those modes. This compresses the instrumental parameterisation by two orders of magnitude while retaining antenna-temperature reconstruction accuracy at the expected thermal-noise level.

\item \textbf{Analytical beam marginalisation:} By exploiting the linearity of the beam model to analytically marginalise over the beam-uncertainty parameters, we allow instrumental uncertainty to be rigorously propagated into the recovered astrophysical and cosmological posteriors while collapsing the sampled parameter space to accelerate statistical inference.
\end{itemize}

Using this framework, we first demonstrated that the conventional fixed-beam approach can fail catastrophically under a representative beam mismatch drawn from the simulated uncertainty volume, producing residuals at the tens-of-kelvin level and leading to biased recovery of both the cosmological signal and the foreground spectral-index parameters.

When the instrumental variation is instead marginalised over, the recovery improves substantially. Across 44 held-out datasets, spanning 11 unseen beam realisations and four injected signal central frequencies, the recovered signal posteriors are consistent with the injected profiles at approximately the simulated noise level in the vast majority of cases, with only three cases exceeding an RMSE of 40~mK. Overall, the results demonstrate that the learned beam parameterisation is able to generalise to unseen realisations drawn from the same physical uncertainty space, while being sufficiently conditioned on the spectral and angular dependence of the beam to allow robust cosmological signal recovery.

Finally, we investigated the sensitivity of the method to the size and coverage of the simulation suite used to define the instrumental prior. Randomly reducing the training set from 400 to 100 beams produced no clear degradation in signal recovery, suggesting that, for the uncertainty volume considered here, reconstruction accuracy is limited more by the retained mode count than by the raw number of simulations. Restricting the training set to the central region of parameter space showed that moderate extrapolation beyond the training support is possible, but that performance degrades for the most extreme validation beams. This indicates that future applications need not require prohibitively large simulation suites, but they must still span the physically relevant directions of instrumental variation.

A natural direction for future work is to extend the simulation suite towards higher-fidelity representations of the REACH antennas, incorporating additional manufacturing tolerances, ground-plane deformation, radiation efficiency, coupling to the receiver signal chain and differences between electromagnetic solvers. The two-stage decomposition presented here is not the only possible approach to representing these variations, and alternative beam-modelling methods should also be explored, including physical-parameter interpolation with frameworks such as \texttt{MEDEA} \citep{Hibbard_2024} and neural-network emulators (Bevins et al., in prep.). Additionally, as further antenna geometries are deployed, joint analyses that independently marginalise over each antenna's beam uncertainty could exploit their complementary instrumental responses to break degeneracies and provide tighter astrophysical constraints on the 21-cm signal.

The framework presented here, combined with the hardware-accelerated, differentiable REACH pipeline, represents a significant advance towards the end-to-end analysis required for a robust global 21-cm detection: jointly constraining the cosmological signal, foreground sky and uncertain instrumental response within a single statistically principled model. By making continuous beam uncertainty part of the inference rather than a fixed assumption, this work establishes a practical foundation for robust, evidence-based detection of the sky-averaged 21-cm signal.

\section*{Acknowledgements}

The authors acknowledge the use of resources provided by the Isambard-AI National AI Research Resource (AIRR). Isambard-AI \citep{Isamabrd_2024} is operated by the University of Bristol and is funded by the UK Government’s Department for Science, Innovation and Technology (DSIT) via UK Research and Innovation; and the Science and Technology Facilities Council [ST/AIRR/I-A-I/1023]. This work used the DiRAC Data Intensive service (CSD3, project number APP52749) at the University of Cambridge, managed by the University of Cambridge University Information Services on behalf of the STFC DiRAC HPC Facility (www.dirac.ac.uk). The DiRAC component of CSD3 at Cambridge was funded by BEIS, UKRI and STFC capital funding and STFC operations grants. DiRAC is part of the UKRI Digital Research Infrastructure. JT is supported by the Harding Distinguished Postgraduate Scholars Programme (HDPSP) and the Science and Technology Facilities Council (STFC) DTP Studentship. JMC and DJA were supported by the Science and Technology Facilities Council. HTJB acknowledges support from the Kavli Institute for Cosmology Cambridge and the Kavli Foundation. We would also like to thank the Kavli Foundation for their support of REACH. EdLA acknowledges the support of STFC via an Ernest Rutherford Fellowship. This work was supported by UK Research and Innovation (UKRI) through the Horizon Europe Guarantee under grant EP/Y02916X/1 (REACH 21).

\section*{Data Availability}

The data that supports the findings of this study are available from the first author upon reasonable request.

\bibliographystyle{mnras}
\bibliography{references}

@article{Bevins_2023_margarine,
    author = {Bevins, Harry T. J. and Handley, William J. and Lemos, Pablo and Sims, Peter H. and de Lera Acedo, Eloy and Fialkov, Anastasia and Alsing, Justin},
    title = {Marginal post-processing of Bayesian inference products with normalizing flows and kernel density estimators},
    journal = {Monthly Notices of the Royal Astronomical Society},
    volume = {526},
    number = {3},
    pages = {4613--4626},
    year = {2023},
    doi = {10.1093/mnras/stad3028},
    eprint = {2205.12841},
    archivePrefix = {arXiv},
    primaryClass = {astro-ph.IM}
}

@misc{cabezas2024blackjax,
      title={BlackJAX: Composable {B}ayesian inference in {JAX}},
      author={Alberto Cabezas and Adrien Corenflos and Junpeng Lao and Rémi Louf},
      year={2024},
      eprint={2402.10797},
      archivePrefix={arXiv},
      primaryClass={cs.MS}
}

@article{yallup2025nested,
      title={Nested Slice Sampling: Vectorized Nested Sampling for {GPU}-Accelerated Inference},
      author={Yallup, David and Kroupa, Namu and Handley, Will},
      journal={Transactions on Machine Learning Research},
      year={2026},
      month=may,
      url={https://openreview.net/forum?id=5mF2eRl3gt}
}

@misc{yallup2026nestedsamplingslicewithingibbsefficient,
      title={Nested Sampling with Slice-within-Gibbs: Efficient Evidence Calculation for Hierarchical Bayesian Models},
      author={David Yallup},
      year={2026},
      eprint={2602.17414},
      archivePrefix={arXiv},
      primaryClass={stat.CO},
      url={https://arxiv.org/abs/2602.17414}
}

@ARTICLE{COBE,
       author = {{Smoot}, G.~F. and {Bennett}, C.~L. and {Kogut}, A. and {Wright}, E.~L. and {Aymon}, J. and {Boggess}, N.~W. and {Cheng}, E.~S. and {de Amici}, G. and {Gulkis}, S. and {Hauser}, M.~G. and {Hinshaw}, G. and {Jackson}, P.~D. and {Janssen}, M. and {Kaita}, E. and {Kelsall}, T. and {Keegstra}, P. and {Lineweaver}, C. and {Loewenstein}, K. and {Lubin}, P. and {Mather}, J. and {Meyer}, S.~S. and {Moseley}, S.~H. and {Murdock}, T. and {Rokke}, L. and {Silverberg}, R.~F. and {Tenorio}, L. and {Weiss}, R. and {Wilkinson}, D.~T.},
        title = "{Structure in the COBE Differential Microwave Radiometer First-Year Maps}",
      journal = {\apjl},
         year = 1992,
        month = sep,
       volume = {396},
        pages = {L1},
          doi = {10.1086/186504},
       adsurl = {https://ui.adsabs.harvard.edu/abs/1992ApJ...396L...1S}
}

@article{WMAP,
doi = {10.1086/377252},
url = {https://doi.org/10.1086/377252},
year = {2003},
month = {sep},
publisher = {},
volume = {148},
number = {1},
pages = {97},
author = {Bennett, C. L. and Hill, R. S. and Hinshaw, G. and Nolta, M. R. and Odegard, N. and Page, L. and Spergel, D. N. and Weiland, J. L. and Wright, E. L. and Halpern, M. and Jarosik, N. and Kogut, A. and Limon, M. and Meyer, S. S. and Tucker, G. S. and Wollack, E.},
title = {First-Year Wilkinson Microwave Anisotropy Probe
(WMAP)* Observations:
Foreground Emission},
journal = {The Astrophysical Journal Supplement Series}
}

@ARTICLE{Planck2014,
       author = {{Planck Collaboration} and {Ade}, P.~A.~R. and {Aghanim}, N. and {Armitage-Caplan}, C. and {Arnaud}, M. and {Ashdown}, M. and {Atrio-Barandela}, F. and {Aumont}, J. and {Baccigalupi}, C. and {Banday}, A.~J. and {Barreiro}, R.~B. and {Bartlett}, J.~G. and {Battaner}, E. and {Benabed}, K. and {Beno{\^\i}t}, A. and {Benoit-L{\'e}vy}, A. and {Bernard}, J.-P. and {Bersanelli}, M. and {Bielewicz}, P. and {Bobin}, J. and {Bock}, J.~J. and {Bonaldi}, A. and {Bond}, J.~R. and {Borrill}, J. and {Bouchet}, F.~R. and {Bridges}, M. and {Bucher}, M. and {Burigana}, C. and {Butler}, R.~C. and {Calabrese}, E. and {Cappellini}, B. and {Cardoso}, J.-F. and {Catalano}, A. and {Challinor}, A. and {Chamballu}, A. and {Chary}, R.-R. and {Chen}, X. and {Chiang}, H.~C. and {Chiang}, L.-Y. and {Christensen}, P.~R. and {Church}, S. and {Clements}, D.~L. and {Colombi}, S. and {Colombo}, L.~P.~L. and {Couchot}, F. and {Coulais}, A. and {Crill}, B.~P. and {Curto}, A. and {Cuttaia}, F. and {Danese}, L. and {Davies}, R.~D. and {Davis}, R.~J. and {de Bernardis}, P. and {de Rosa}, A. and {de Zotti}, G. and {Delabrouille}, J. and {Delouis}, J.-M. and {D{\'e}sert}, F.-X. and {Dickinson}, C. and {Diego}, J.~M. and {Dolag}, K. and {Dole}, H. and {Donzelli}, S. and {Dor{\'e}}, O. and {Douspis}, M. and {Dunkley}, J. and {Dupac}, X. and {Efstathiou}, G. and {Elsner}, F. and {En{\ss}lin}, T.~A. and {Eriksen}, H.~K. and {Finelli}, F. and {Forni}, O. and {Frailis}, M. and {Fraisse}, A.~A. and {Franceschi}, E. and {Gaier}, T.~C. and {Galeotta}, S. and {Galli}, S. and {Ganga}, K. and {Giard}, M. and {Giardino}, G. and {Giraud-H{\'e}raud}, Y. and {Gjerl{\o}w}, E. and {Gonz{\'a}lez-Nuevo}, J. and {G{\'o}rski}, K.~M. and {Gratton}, S. and {Gregorio}, A. and {Gruppuso}, A. and {Gudmundsson}, J.~E. and {Haissinski}, J. and {Hamann}, J. and {Hansen}, F.~K. and {Hanson}, D. and {Harrison}, D. and {Henrot-Versill{\'e}}, S. and {Hern{\'a}ndez-Monteagudo}, C. and {Herranz}, D. and {Hildebrandt}, S.~R. and {Hivon}, E. and {Hobson}, M. and {Holmes}, W.~A. and {Hornstrup}, A. and {Hou}, Z. and {Hovest}, W. and {Huffenberger}, K.~M. and {Jaffe}, A.~H. and {Jaffe}, T.~R. and {Jewell}, J. and {Jones}, W.~C. and {Juvela}, M. and {Keih{\"a}nen}, E. and {Keskitalo}, R. and {Kisner}, T.~S. and {Kneissl}, R. and {Knoche}, J. and {Knox}, L. and {Kunz}, M. and {Kurki-Suonio}, H. and {Lagache}, G. and {L{\"a}hteenm{\"a}ki}, A. and {Lamarre}, J.-M. and {Lasenby}, A. and {Lattanzi}, M. and {Laureijs}, R.~J. and {Lawrence}, C.~R. and {Leach}, S. and {Leahy}, J.~P. and {Leonardi}, R. and {Le{\'o}n-Tavares}, J. and {Lesgourgues}, J. and {Lewis}, A. and {Liguori}, M. and {Lilje}, P.~B. and {Linden-V{\o}rnle}, M. and {L{\'o}pez-Caniego}, M. and {Lubin}, P.~M. and {Mac{\'\i}as-P{\'e}rez}, J.~F. and {Maffei}, B. and {Maino}, D. and {Mandolesi}, N. and {Maris}, M. and {Marshall}, D.~J. and {Martin}, P.~G. and {Mart{\'\i}nez-Gonz{\'a}lez}, E. and {Masi}, S. and {Massardi}, M. and {Matarrese}, S. and {Matthai}, F. and {Mazzotta}, P. and {Meinhold}, P.~R. and {Melchiorri}, A. and {Melin}, J.-B. and {Mendes}, L. and {Menegoni}, E. and {Mennella}, A. and {Migliaccio}, M. and {Millea}, M. and {Mitra}, S. and {Miville-Desch{\^e}nes}, M.-A. and {Moneti}, A. and {Montier}, L. and {Morgante}, G. and {Mortlock}, D. and {Moss}, A. and {Munshi}, D. and {Murphy}, J.~A. and {Naselsky}, P. and {Nati}, F. and {Natoli}, P. and {Netterfield}, C.~B. and {N{\o}rgaard-Nielsen}, H.~U. and {Noviello}, F. and {Novikov}, D. and {Novikov}, I. and {O'Dwyer}, I.~J. and {Osborne}, S. and {Oxborrow}, C.~A. and {Paci}, F. and {Pagano}, L. and {Pajot}, F. and {Paladini}, R. and {Paoletti}, D. and {Partridge}, B. and {Pasian}, F. and {Patanchon}, G. and {Pearson}, D. and {Pearson}, T.~J. and {Peiris}, H.~V. and {Perdereau}, O. and {Perotto}, L. and {Perrotta}, F. and {Pettorino}, V. and {Piacentini}, F. and {Piat}, M. and {Pierpaoli}, E. and {Pietrobon}, D. and {Plaszczynski}, S. and {Platania}, P. and {Pointecouteau}, E.},
        title = "{Planck 2013 results. XVI. Cosmological parameters}",
      journal = {\aap},
         year = 2014,
        month = nov,
       volume = {571},
          eid = {A16},
        pages = {A16},
          doi = {10.1051/0004-6361/201321591},
archivePrefix = {arXiv},
       eprint = {1303.5076},
 primaryClass = {astro-ph.CO},
       adsurl = {https://ui.adsabs.harvard.edu/abs/2014A&A...571A..16P}
}

@ARTICLE{Planck2016,
       author = {{Planck Collaboration} and {Ade}, P.~A.~R. and {Aghanim}, N. and {Arnaud}, M. and {Ashdown}, M. and {Aumont}, J. and {Baccigalupi}, C. and {Banday}, A.~J. and {Barreiro}, R.~B. and {Bartlett}, J.~G. and {Bartolo}, N. and {Battaner}, E. and {Battye}, R. and {Benabed}, K. and {Beno{\^\i}t}, A. and {Benoit-L{\'e}vy}, A. and {Bernard}, J.-P. and {Bersanelli}, M. and {Bielewicz}, P. and {Bock}, J.~J. and {Bonaldi}, A. and {Bonavera}, L. and {Bond}, J.~R. and {Borrill}, J. and {Bouchet}, F.~R. and {Boulanger}, F. and {Bucher}, M. and {Burigana}, C. and {Butler}, R.~C. and {Calabrese}, E. and {Cardoso}, J.-F. and {Catalano}, A. and {Challinor}, A. and {Chamballu}, A. and {Chary}, R.-R. and {Chiang}, H.~C. and {Chluba}, J. and {Christensen}, P.~R. and {Church}, S. and {Clements}, D.~L. and {Colombi}, S. and {Colombo}, L.~P.~L. and {Combet}, C. and {Coulais}, A. and {Crill}, B.~P. and {Curto}, A. and {Cuttaia}, F. and {Danese}, L. and {Davies}, R.~D. and {Davis}, R.~J. and {de Bernardis}, P. and {de Rosa}, A. and {de Zotti}, G. and {Delabrouille}, J. and {D{\'e}sert}, F.-X. and {Di Valentino}, E. and {Dickinson}, C. and {Diego}, J.~M. and {Dolag}, K. and {Dole}, H. and {Donzelli}, S. and {Dor{\'e}}, O. and {Douspis}, M. and {Ducout}, A. and {Dunkley}, J. and {Dupac}, X. and {Efstathiou}, G. and {Elsner}, F. and {En{\ss}lin}, T.~A. and {Eriksen}, H.~K. and {Farhang}, M. and {Fergusson}, J. and {Finelli}, F. and {Forni}, O. and {Frailis}, M. and {Fraisse}, A.~A. and {Franceschi}, E. and {Frejsel}, A. and {Galeotta}, S. and {Galli}, S. and {Ganga}, K. and {Gauthier}, C. and {Gerbino}, M. and {Ghosh}, T. and {Giard}, M. and {Giraud-H{\'e}raud}, Y. and {Giusarma}, E. and {Gjerl{\o}w}, E. and {Gonz{\'a}lez-Nuevo}, J. and {G{\'o}rski}, K.~M. and {Gratton}, S. and {Gregorio}, A. and {Gruppuso}, A. and {Gudmundsson}, J.~E. and {Hamann}, J. and {Hansen}, F.~K. and {Hanson}, D. and {Harrison}, D.~L. and {Helou}, G. and {Henrot-Versill{\'e}}, S. and {Hern{\'a}ndez-Monteagudo}, C. and {Herranz}, D. and {Hildebrandt}, S.~R. and {Hivon}, E. and {Hobson}, M. and {Holmes}, W.~A. and {Hornstrup}, A. and {Hovest}, W. and {Huang}, Z. and {Huffenberger}, K.~M. and {Hurier}, G. and {Jaffe}, A.~H. and {Jaffe}, T.~R. and {Jones}, W.~C. and {Juvela}, M. and {Keih{\"a}nen}, E. and {Keskitalo}, R. and {Kisner}, T.~S. and {Kneissl}, R. and {Knoche}, J. and {Knox}, L. and {Kunz}, M. and {Kurki-Suonio}, H. and {Lagache}, G. and {L{\"a}hteenm{\"a}ki}, A. and {Lamarre}, J.-M. and {Lasenby}, A. and {Lattanzi}, M. and {Lawrence}, C.~R. and {Leahy}, J.~P. and {Leonardi}, R. and {Lesgourgues}, J. and {Levrier}, F. and {Lewis}, A. and {Liguori}, M. and {Lilje}, P.~B. and {Linden-V{\o}rnle}, M. and {L{\'o}pez-Caniego}, M. and {Lubin}, P.~M. and {Mac{\'\i}as-P{\'e}rez}, J.~F. and {Maggio}, G. and {Maino}, D. and {Mandolesi}, N. and {Mangilli}, A. and {Marchini}, A. and {Maris}, M. and {Martin}, P.~G. and {Martinelli}, M. and {Mart{\'\i}nez-Gonz{\'a}lez}, E. and {Masi}, S. and {Matarrese}, S. and {McGehee}, P. and {Meinhold}, P.~R. and {Melchiorri}, A. and {Melin}, J.-B. and {Mendes}, L. and {Mennella}, A. and {Migliaccio}, M. and {Millea}, M. and {Mitra}, S. and {Miville-Desch{\^e}nes}, M.-A. and {Moneti}, A. and {Montier}, L. and {Morgante}, G. and {Mortlock}, D. and {Moss}, A. and {Munshi}, D. and {Murphy}, J.~A. and {Naselsky}, P. and {Nati}, F. and {Natoli}, P. and {Netterfield}, C.~B. and {N{\o}rgaard-Nielsen}, H.~U. and {Noviello}, F. and {Novikov}, D. and {Novikov}, I. and {Oxborrow}, C.~A. and {Paci}, F. and {Pagano}, L. and {Pajot}, F. and {Paladini}, R. and {Paoletti}, D. and {Partridge}, B. and {Pasian}, F. and {Patanchon}, G. and {Pearson}, T.~J. and {Perdereau}, O. and {Perotto}, L. and {Perrotta}, F. and {Pettorino}, V. and {Piacentini}, F. and {Piat}, M. and {Pierpaoli}, E. and {Pietrobon}, D. and {Plaszczynski}, S. and {Pointecouteau}, E. and {Polenta}, G. and {Popa}, L. and {Pratt}, G.~W. and {Pr{\'e}zeau}, G.},
        title = "{Planck 2015 results. XIII. Cosmological parameters}",
      journal = {\aap},
         year = 2016,
        month = sep,
       volume = {594},
          eid = {A13},
        pages = {A13},
          doi = {10.1051/0004-6361/201525830},
archivePrefix = {arXiv},
       eprint = {1502.01589},
 primaryClass = {astro-ph.CO},
       adsurl = {https://ui.adsabs.harvard.edu/abs/2016A&A...594A..13P}
}

@ARTICLE{Planck2020,
       author = {{Planck Collaboration} and {Aghanim}, N. and {Akrami}, Y. and {Ashdown}, M. and {Aumont}, J. and {Baccigalupi}, C. and {Ballardini}, M. and {Banday}, A.~J. and {Barreiro}, R.~B. and {Bartolo}, N. and {Basak}, S. and {Battye}, R. and {Benabed}, K. and {Bernard}, J.-P. and {Bersanelli}, M. and {Bielewicz}, P. and {Bock}, J.~J. and {Bond}, J.~R. and {Borrill}, J. and {Bouchet}, F.~R. and {Boulanger}, F. and {Bucher}, M. and {Burigana}, C. and {Butler}, R.~C. and {Calabrese}, E. and {Cardoso}, J.-F. and {Carron}, J. and {Challinor}, A. and {Chiang}, H.~C. and {Chluba}, J. and {Colombo}, L.~P.~L. and {Combet}, C. and {Contreras}, D. and {Crill}, B.~P. and {Cuttaia}, F. and {de Bernardis}, P. and {de Zotti}, G. and {Delabrouille}, J. and {Delouis}, J.-M. and {Di Valentino}, E. and {Diego}, J.~M. and {Dor{\'e}}, O. and {Douspis}, M. and {Ducout}, A. and {Dupac}, X. and {Dusini}, S. and {Efstathiou}, G. and {Elsner}, F. and {En{\ss}lin}, T.~A. and {Eriksen}, H.~K. and {Fantaye}, Y. and {Farhang}, M. and {Fergusson}, J. and {Fernandez-Cobos}, R. and {Finelli}, F. and {Forastieri}, F. and {Frailis}, M. and {Fraisse}, A.~A. and {Franceschi}, E. and {Frolov}, A. and {Galeotta}, S. and {Galli}, S. and {Ganga}, K. and {G{\'e}nova-Santos}, R.~T. and {Gerbino}, M. and {Ghosh}, T. and {Gonz{\'a}lez-Nuevo}, J. and {G{\'o}rski}, K.~M. and {Gratton}, S. and {Gruppuso}, A. and {Gudmundsson}, J.~E. and {Hamann}, J. and {Handley}, W. and {Hansen}, F.~K. and {Herranz}, D. and {Hildebrandt}, S.~R. and {Hivon}, E. and {Huang}, Z. and {Jaffe}, A.~H. and {Jones}, W.~C. and {Karakci}, A. and {Keih{\"a}nen}, E. and {Keskitalo}, R. and {Kiiveri}, K. and {Kim}, J. and {Kisner}, T.~S. and {Knox}, L. and {Krachmalnicoff}, N. and {Kunz}, M. and {Kurki-Suonio}, H. and {Lagache}, G. and {Lamarre}, J.-M. and {Lasenby}, A. and {Lattanzi}, M. and {Lawrence}, C.~R. and {Le Jeune}, M. and {Lemos}, P. and {Lesgourgues}, J. and {Levrier}, F. and {Lewis}, A. and {Liguori}, M. and {Lilje}, P.~B. and {Lilley}, M. and {Lindholm}, V. and {L{\'o}pez-Caniego}, M. and {Lubin}, P.~M. and {Ma}, Y.-Z. and {Mac{\'\i}as-P{\'e}rez}, J.~F. and {Maggio}, G. and {Maino}, D. and {Mandolesi}, N. and {Mangilli}, A. and {Marcos-Caballero}, A. and {Maris}, M. and {Martin}, P.~G. and {Martinelli}, M. and {Mart{\'\i}nez-Gonz{\'a}lez}, E. and {Matarrese}, S. and {Mauri}, N. and {McEwen}, J.~D. and {Meinhold}, P.~R. and {Melchiorri}, A. and {Mennella}, A. and {Migliaccio}, M. and {Millea}, M. and {Mitra}, S. and {Miville-Desch{\^e}nes}, M.-A. and {Molinari}, D. and {Montier}, L. and {Morgante}, G. and {Moss}, A. and {Natoli}, P. and {N{\o}rgaard-Nielsen}, H.~U. and {Pagano}, L. and {Paoletti}, D. and {Partridge}, B. and {Patanchon}, G. and {Peiris}, H.~V. and {Perrotta}, F. and {Pettorino}, V. and {Piacentini}, F. and {Polastri}, L. and {Polenta}, G. and {Puget}, J.-L. and {Rachen}, J.~P. and {Reinecke}, M. and {Remazeilles}, M. and {Renzi}, A. and {Rocha}, G. and {Rosset}, C. and {Roudier}, G. and {Rubi{\~n}o-Mart{\'\i}n}, J.~A. and {Ruiz-Granados}, B. and {Salvati}, L. and {Sandri}, M. and {Savelainen}, M. and {Scott}, D. and {Shellard}, E.~P.~S. and {Sirignano}, C. and {Sirri}, G. and {Spencer}, L.~D. and {Sunyaev}, R. and {Suur-Uski}, A.-S. and {Tauber}, J.~A. and {Tavagnacco}, D. and {Tenti}, M. and {Toffolatti}, L. and {Tomasi}, M. and {Trombetti}, T. and {Valenziano}, L. and {Valiviita}, J. and {Van Tent}, B. and {Vibert}, L. and {Vielva}, P. and {Villa}, F. and {Vittorio}, N. and {Wandelt}, B.~D. and {Wehus}, I.~K. and {White}, M. and {White}, S.~D.~M. and {Zacchei}, A. and {Zonca}, A.},
        title = "{Planck 2018 results. VI. Cosmological parameters}",
      journal = {\aap},
         year = 2020,
        month = sep,
       volume = {641},
          eid = {A6},
        pages = {A6},
          doi = {10.1051/0004-6361/201833910},
archivePrefix = {arXiv},
       eprint = {1807.06209},
 primaryClass = {astro-ph.CO},
       adsurl = {https://ui.adsabs.harvard.edu/abs/2020A&A...641A...6P}
}

@ARTICLE{Boss,
       author = {{Dawson}, Kyle S. and {Schlegel}, David J. and {Ahn}, Christopher P. and {Anderson}, Scott F. and {Aubourg}, {\'E}ric and {Bailey}, Stephen and {Barkhouser}, Robert H. and {Bautista}, Julian E. and {Beifiori}, Alessandra and {Berlind}, Andreas A. and {Bhardwaj}, Vaishali and {Bizyaev}, Dmitry and {Blake}, Cullen H. and {Blanton}, Michael R. and {Blomqvist}, Michael and {Bolton}, Adam S. and {Borde}, Arnaud and {Bovy}, Jo and {Brandt}, W.~N. and {Brewington}, Howard and {Brinkmann}, Jon and {Brown}, Peter J. and {Brownstein}, Joel R. and {Bundy}, Kevin and {Busca}, N.~G. and {Carithers}, William and {Carnero}, Aurelio R. and {Carr}, Michael A. and {Chen}, Yanmei and {Comparat}, Johan and {Connolly}, Natalia and {Cope}, Frances and {Croft}, Rupert A.~C. and {Cuesta}, Antonio J. and {da Costa}, Luiz N. and {Davenport}, James R.~A. and {Delubac}, Timoth{\'e}e and {de Putter}, Roland and {Dhital}, Saurav and {Ealet}, Anne and {Ebelke}, Garrett L. and {Eisenstein}, Daniel J. and {Escoffier}, S. and {Fan}, Xiaohui and {Filiz Ak}, N. and {Finley}, Hayley and {Font-Ribera}, Andreu and {G{\'e}nova-Santos}, R. and {Gunn}, James E. and {Guo}, Hong and {Haggard}, Daryl and {Hall}, Patrick B. and {Hamilton}, Jean-Christophe and {Harris}, Ben and {Harris}, David W. and {Ho}, Shirley and {Hogg}, David W. and {Holder}, Diana and {Honscheid}, Klaus and {Huehnerhoff}, Joe and {Jordan}, Beatrice and {Jordan}, Wendell P. and {Kauffmann}, Guinevere and {Kazin}, Eyal A. and {Kirkby}, David and {Klaene}, Mark A. and {Kneib}, Jean-Paul and {Le Goff}, Jean-Marc and {Lee}, Khee-Gan and {Long}, Daniel C. and {Loomis}, Craig P. and {Lundgren}, Britt and {Lupton}, Robert H. and {Maia}, Marcio A.~G. and {Makler}, Martin and {Malanushenko}, Elena and {Malanushenko}, Viktor and {Mandelbaum}, Rachel and {Manera}, Marc and {Maraston}, Claudia and {Margala}, Daniel and {Masters}, Karen L. and {McBride}, Cameron K. and {McDonald}, Patrick and {McGreer}, Ian D. and {McMahon}, Richard G. and {Mena}, Olga and {Miralda-Escud{\'e}}, Jordi and {Montero-Dorta}, Antonio D. and {Montesano}, Francesco and {Muna}, Demitri and {Myers}, Adam D. and {Naugle}, Tracy and {Nichol}, Robert C. and {Noterdaeme}, Pasquier and {Nuza}, Sebasti{\'a}n E. and {Olmstead}, Matthew D. and {Oravetz}, Audrey and {Oravetz}, Daniel J. and {Owen}, Russell and {Padmanabhan}, Nikhil and {Palanque-Delabrouille}, Nathalie and {Pan}, Kaike and {Parejko}, John K. and {P{\^a}ris}, Isabelle and {Percival}, Will J. and {P{\'e}rez-Fournon}, Ismael and {P{\'e}rez-R{\`a}fols}, Ignasi and {Petitjean}, Patrick and {Pfaffenberger}, Robert and {Pforr}, Janine and {Pieri}, Matthew M. and {Prada}, Francisco and {Price-Whelan}, Adrian M. and {Raddick}, M. Jordan and {Rebolo}, Rafael and {Rich}, James and {Richards}, Gordon T. and {Rockosi}, Constance M. and {Roe}, Natalie A. and {Ross}, Ashley J. and {Ross}, Nicholas P. and {Rossi}, Graziano and {Rubi{\~n}o-Martin}, J.~A. and {Samushia}, Lado and {S{\'a}nchez}, Ariel G. and {Sayres}, Conor and {Schmidt}, Sarah J. and {Schneider}, Donald P. and {Sc{\'o}ccola}, C.~G. and {Seo}, Hee-Jong and {Shelden}, Alaina and {Sheldon}, Erin and {Shen}, Yue and {Shu}, Yiping and {Slosar}, An{\v{z}}e and {Smee}, Stephen A. and {Snedden}, Stephanie A. and {Stauffer}, Fritz and {Steele}, Oliver and {Strauss}, Michael A. and {Streblyanska}, Alina and {Suzuki}, Nao and {Swanson}, Molly E.~C. and {Tal}, Tomer and {Tanaka}, Masayuki and {Thomas}, Daniel and {Tinker}, Jeremy L. and {Tojeiro}, Rita and {Tremonti}, Christy A. and {Vargas Maga{\~n}a}, M. and {Verde}, Licia and {Viel}, Matteo and {Wake}, David A. and {Watson}, Mike and {Weaver}, Benjamin A. and {Weinberg}, David H. and {Weiner}, Benjamin J. and {West}, Andrew A. and {White}, Martin and {Wood-Vasey}, W.~M. and {Yeche}, Christophe and {Zehavi}, Idit and {Zhao}, Gong-Bo and {Zheng}, Zheng},
        title = "{The Baryon Oscillation Spectroscopic Survey of SDSS-III}",
      journal = {\aj},
         year = 2013,
        month = jan,
       volume = {145},
       number = {1},
          eid = {10},
        pages = {10},
          doi = {10.1088/0004-6256/145/1/10},
archivePrefix = {arXiv},
       eprint = {1208.0022},
 primaryClass = {astro-ph.CO},
       adsurl = {https://ui.adsabs.harvard.edu/abs/2013AJ....145...10D}
}

@article{DESI,
doi = {10.1088/1475-7516/2025/02/021},
url = {https://doi.org/10.1088/1475-7516/2025/02/021},
year = {2025},
month = {feb},
publisher = {IOP Publishing},
volume = {2025},
number = {02},
pages = {021},
author = {Adame, A.G. and Aguilar, J. and Ahlen, S. and Alam, S. and Alexander, D.M. and Alvarez, M. and Alves, O. and Anand, A. and Andrade, U. and Armengaud, E. and Avila, S. and Aviles, A. and Awan, H. and Bahr-Kalus, B. and Bailey, S. and Baltay, C. and Bault, A. and Behera, J. and BenZvi, S. and Bera, A. and Beutler, F. and Bianchi, D. and Blake, C. and Blum, R. and Brieden, S. and Brodzeller, A. and Brooks, D. and Buckley-Geer, E. and Burtin, E. and Calderon, R. and Canning, R. and Carnero Rosell, A. and Cereskaite, R. and Cervantes-Cota, J.L. and Chabanier, S. and Chaussidon, E. and Chaves-Montero, J. and Chen, S. and Chen, X. and Claybaugh, T. and Cole, S. and Cuceu, A. and Davis, T.M. and Dawson, K. and de la Macorra, A. and de Mattia, A. and Deiosso, N. and Dey, A. and Dey, B. and Ding, Z. and Doel, P. and Edelstein, J. and Eftekharzadeh, S. and Eisenstein, D.J. and Elliott, A. and Fagrelius, P. and Fanning, K. and Ferraro, S. and Ereza, J. and Findlay, N. and Flaugher, B. and Font-Ribera, A. and Forero-Sánchez, D. and Forero-Romero, J.E. and Frenk, C.S. and Garcia-Quintero, C. and Gaztañaga, E. and Gil-Marín, H. and Gontcho, S.Gontcho A. and Gonzalez-Morales, A.X. and Gonzalez-Perez, V. and Gordon, C. and Green, D. and Gruen, D. and Gsponer, R. and Gutierrez, G. and Guy, J. and Hadzhiyska, B. and Hahn, C. and Hanif, M.M.S. and Herrera-Alcantar, H.K. and Honscheid, K. and Howlett, C. and Huterer, D. and Iršič, V. and Ishak, M. and Juneau, S. and Karaçaylı, N.G. and Kehoe, R. and Kent, S. and Kirkby, D. and Kremin, A. and Krolewski, A. and Lai, Y. and Lan, T.-W. and Landriau, M. and Lang, D. and Lasker, J. and Le Goff, J.M. and Le Guillou, L. and Leauthaud, A. and Levi, M.E. and Li, T.S. and Linder, E. and Lodha, K. and Magneville, C. and Manera, M. and Margala, D. and Martini, P. and Maus, M. and McDonald, P. and Medina-Varela, L. and Meisner, A. and Mena-Fernández, J. and Miquel, R. and Moon, J. and Moore, S. and Moustakas, J. and Mueller, E. and Muñoz-Gutiérrez, A. and Myers, A.D. and Nadathur, S. and Napolitano, L. and Neveux, R. and Newman, J.A. and Nguyen, N.M. and Nie, J. and Niz, G. and Noriega, H.E. and Padmanabhan, N. and Paillas, E. and Palanque-Delabrouille, N. and Pan, J. and Penmetsa, S. and Percival, W.J. and Pieri, M.M. and Pinon, M. and Poppett, C. and Porredon, A. and Prada, F. and Pérez-Fernández, A. and Pérez-Ràfols, I. and Rabinowitz, D. and Raichoor, A. and Ramírez-Pérez, C. and Ramirez-Solano, S. and Rashkovetskyi, M. and Ravoux, C. and Rezaie, M. and Rich, J. and Rocher, A. and Rockosi, C. and Roe, N.A. and Rosado-Marin, A. and Ross, A.J. and Rossi, G. and Ruggeri, R. and Ruhlmann-Kleider, V. and Samushia, L. and Sanchez, E. and Saulder, C. and Schlafly, E.F. and Schlegel, D. and Schubnell, M. and Seo, H. and Shafieloo, A. and Sharples, R. and Silber, J. and Slosar, A. and Smith, A. and Sprayberry, D. and Tan, T. and Tarlé, G. and Taylor, P. and Trusov, S. and Ureña-López, L.A. and Vaisakh, R. and Valcin, D. and Valdes, F. and Vargas-Magaña, M. and Verde, L. and Walther, M. and Wang, B. and Wang, M.S. and Weaver, B.A. and Weaverdyck, N. and Wechsler, R.H. and Weinberg, D.H. and White, M. and Yu, J. and Yu, Y. and Yuan, S. and Yèche, C. and Zaborowski, E.A. and Zarrouk, P. and Zhang, H. and Zhao, C. and Zhao, R. and Zhou, R. and Zhuang, T. and Zou, H. and The DESI collaboration},
title = {DESI 2024 VI:  cosmological constraints from the measurements of baryon acoustic oscillations},
journal = {Journal of Cosmology and Astroparticle Physics}
}

@article{mismatchuv,
      title={The $z \gtrsim 9$ galaxy UV luminosity function from the JWST Advanced Deep Extragalactic Survey: insights into early galaxy evolution and reionization}, 
      author={Lily Whitler and Daniel P. Stark and Michael W. Topping and Brant Robertson and Marcia Rieke and Kevin N. Hainline and Ryan Endsley and Zuyi Chen and William M. Baker and Rachana Bhatawdekar and Andrew J. Bunker and Stefano Carniani and Stéphane Charlot and Jacopo Chevallard and Emma Curtis-Lake and Eiichi Egami and Daniel J. Eisenstein and Jakob M. Helton and Zhiyuan Ji and Benjamin D. Johnson and Pablo G. Pérez-González and Pierluigi Rinaldi and Sandro Tacchella and Christina C. Williams and Christopher N. A. Willmer and Chris Willott and Joris Witstok},
      journal={The Astrophysical Journal},
      volume={992},
      number={1},
      pages={63},
      year={2025},
      doi={10.3847/1538-4357/adfddc},
}

@article{JWSTIMF,
	author = {{Hutter, Anne} and {Cueto, Elie R.} and {Dayal, Pratika} and {Gottlöber, Stefan} and {Trebitsch, Maxime} and {Yepes, Gustavo}},
	title = {ASTRAEUS
 - X. Indications of a top-heavy initial mass function in highly star-forming galaxies from JWST observations at z  >  10},
	DOI= "10.1051/0004-6361/202452460",
	url= "https://doi.org/10.1051/0004-6361/202452460",
	journal = {A&A},
	year = 2025,
	volume = 694,
	pages = "A254",
}

@misc{jwststochasticstar,
      title={Stochastic star formation and the abundance of $z>10$ UV-bright galaxies}, 
      author={Andrey Kravtsov and Vasily Belokurov},
      year={2024},
      eprint={2405.04578},
      archivePrefix={arXiv},
      primaryClass={astro-ph.GA},
      url={https://arxiv.org/abs/2405.04578}, 
}

@article{jwstsfe,
	author = {{Li, Zhaozhou} and {Dekel, Avishai} and {Sarkar, Kartick C.} and {Aung, Han} and {Giavalisco, Mauro} and {Mandelker, Nir} and {Tacchella, Sandro}},
	title = {Feedback-free starbursts at cosmic dawn - Observable predictions for JWST},
	DOI= "10.1051/0004-6361/202348727",
	url= "https://doi.org/10.1051/0004-6361/202348727",
	journal = {A&A},
	year = 2024,
	volume = 690,
	pages = "A108",
}

@article{Furlanetto_2006,
   title={Cosmology at low frequencies: The 21cm transition and the high-redshift Universe},
   volume={433},
   ISSN={0370-1573},
   url={http://dx.doi.org/10.1016/j.physrep.2006.08.002},
   DOI={10.1016/j.physrep.2006.08.002},
   number={4–6},
   journal={Physics Reports},
   publisher={Elsevier BV},
   author={Furlanetto, Steven R. and Peng Oh, S. and Briggs, Frank H.},
   year={2006},
   month=oct, pages={181–301} }

@article{Pritchard_2012,
   title={21 cm cosmology in the 21st century},
   volume={75},
   ISSN={1361-6633},
   url={http://dx.doi.org/10.1088/0034-4885/75/8/086901},
   DOI={10.1088/0034-4885/75/8/086901},
   number={8},
   journal={Reports on Progress in Physics},
   publisher={IOP Publishing},
   author={Pritchard, Jonathan R and Loeb, Abraham},
   year={2012},
   month=jul, pages={086901} }

@ARTICLE{Barkana_2016,
       author = {{Barkana}, Rennan},
        title = "{The rise of the first stars: Supersonic streaming, radiative feedback, and 21-cm cosmology}",
      journal = {\physrep},
         year = 2016,
        month = jul,
       volume = {645},
        pages = {1-59},
          doi = {10.1016/j.physrep.2016.06.006},
       adsurl = {https://ui.adsabs.harvard.edu/abs/2016PhR...645....1B}
}

@BOOK{Mesinger_2019,
       author = {{Mesinger}, Andrei},
        title = "{The Cosmic 21-cm Revolution; Charting the first billion years of our universe}",
         year = 2019,
    publisher = {IOP Publishing},
          doi = {10.1088/2514-3433/ab4a73},
       adsurl = {https://ui.adsabs.harvard.edu/abs/2019cosm.book.....M}
}

@article{gesseyjones2024,
    author = {Gessey-Jones, T and Pochinda, S and Bevins, H T J and Fialkov, A and Handley, W J and de Lera Acedo, E and Singh, S and Barkana, R},
    title = {On the constraints on superconducting cosmic strings from 21-cm cosmology},
    journal = {Monthly Notices of the Royal Astronomical Society},
    volume = {529},
    number = {1},
    pages = {519-536},
    year = {2024},
    month = {02},
    issn = {0035-8711},
    doi = {10.1093/mnras/stae512},
    url = {https://doi.org/10.1093/mnras/stae512},
    eprint = {https://academic.oup.com/mnras/article-pdf/529/1/519/56814830/stae512.pdf},
}

@article{gesseyjones2022,
    author = {Gessey-Jones, T and Sartorio, N S and Fialkov, A and Mirouh, G M and Magg, M and Izzard, R G and de Lera Acedo, E and Handley, W J and Barkana, R},
    title = {Impact of the primordial stellar initial mass function on the 21-cm signal},
    journal = {Monthly Notices of the Royal Astronomical Society},
    volume = {516},
    number = {1},
    pages = {841-860},
    year = {2022},
    month = {07},
    issn = {0035-8711},
    doi = {10.1093/mnras/stac2049},
    url = {https://doi.org/10.1093/mnras/stac2049},
    eprint = {https://academic.oup.com/mnras/article-pdf/516/1/841/45633596/stac2049.pdf},
}

@ARTICLE{Schauer2019,
       author = {{Schauer}, Anna T.~P. and {Liu}, Boyuan and {Bromm}, Volker},
        title = "{Constraining First Star Formation with 21 cm Cosmology}",
      journal = {\apjl},
         year = 2019,
        month = may,
       volume = {877},
       number = {1},
          eid = {L5},
        pages = {L5},
          doi = {10.3847/2041-8213/ab1e51},
archivePrefix = {arXiv},
       eprint = {1901.03344},
 primaryClass = {astro-ph.GA},
       adsurl = {https://ui.adsabs.harvard.edu/abs/2019ApJ...877L...5S}
}

@article{Mittal_2022,
doi = {10.1088/1475-7516/2022/03/030},
url = {https://doi.org/10.1088/1475-7516/2022/03/030},
year = {2022},
month = {mar},
publisher = {IOP Publishing},
volume = {2022},
number = {03},
pages = {030},
author = {Mittal, Shikhar and Ray, Anupam and Kulkarni, Girish and Dasgupta, Basudeb},
title = {Constraining primordial black holes as dark matter using the global 21-cm signal with X-ray heating and excess radio background},
journal = {Journal of Cosmology and Astroparticle Physics}
}

@article{Barkana_2018,
author = {Barkana, Rennan},
year = {2018},
month = {03},
pages = {71-74},
title = {Possible interaction between baryons and dark-matter particles revealed by the first stars},
volume = {555},
journal = {Nature},
doi = {10.1038/nature25791}
}

@ARTICLE{HERA,
       author = {{DeBoer}, David R. and {Parsons}, Aaron R. and {Aguirre}, James E. and {Alexander}, Paul and {Ali}, Zaki S. and {Beardsley}, Adam P. and {Bernardi}, Gianni and {Bowman}, Judd D. and {Bradley}, Richard F. and {Carilli}, Chris L. and {Cheng}, Carina and {de Lera Acedo}, Eloy and {Dillon}, Joshua S. and {Ewall-Wice}, Aaron and {Fadana}, Gcobisa and {Fagnoni}, Nicolas and {Fritz}, Randall and {Furlanetto}, Steve R. and {Glendenning}, Brian and {Greig}, Bradley and {Grobbelaar}, Jasper and {Hazelton}, Bryna J. and {Hewitt}, Jacqueline N. and {Hickish}, Jack and {Jacobs}, Daniel C. and {Julius}, Austin and {Kariseb}, MacCalvin and {Kohn}, Saul A. and {Lekalake}, Telalo and {Liu}, Adrian and {Loots}, Anita and {MacMahon}, David and {Malan}, Lourence and {Malgas}, Cresshim and {Maree}, Matthys and {Martinot}, Zachary and {Mathison}, Nathan and {Matsetela}, Eunice and {Mesinger}, Andrei and {Morales}, Miguel F. and {Neben}, Abraham R. and {Patra}, Nipanjana and {Pieterse}, Samantha and {Pober}, Jonathan C. and {Razavi-Ghods}, Nima and {Ringuette}, Jon and {Robnett}, James and {Rosie}, Kathryn and {Sell}, Raddwine and {Smith}, Craig and {Syce}, Angelo and {Tegmark}, Max and {Thyagarajan}, Nithyanandan and {Williams}, Peter K.~G. and {Zheng}, Haoxuan},
        title = "{Hydrogen Epoch of Reionization Array (HERA)}",
      journal = {\pasp},
         year = 2017,
        month = apr,
       volume = {129},
       number = {974},
        pages = {045001},
          doi = {10.1088/1538-3873/129/974/045001},
archivePrefix = {arXiv},
       eprint = {1606.07473},
 primaryClass = {astro-ph.IM},
       adsurl = {https://ui.adsabs.harvard.edu/abs/2017PASP..129d5001D}
}

@article{LOFAR,
	author = {{van Haarlem, M. P.} and {Wise, M. W.} and {Gunst, A. W.} and {Heald, G.} and {McKean, J. P.} and {Hessels, J. W. T.} and {de Bruyn, A. G.} and {Nijboer, R.} and {Swinbank, J.} and {Fallows, R.} and {Brentjens, M.} and {Nelles, A.} and {Beck, R.} and {Falcke, H.} and {Fender, R.} and {Hörandel, J.} and {Koopmans, L. V. E.} and {Mann, G.} and {Miley, G.} and {Röttgering, H.} and {Stappers, B. W.} and {Wijers, R. A. M. J.} and {Zaroubi, S.} and {van den Akker, M.} and {Alexov, A.} and {Anderson, J.} and {Anderson, K.} and {van Ardenne, A.} and {Arts, M.} and {Asgekar, A.} and {Avruch, I. M.} and {Batejat, F.} and {Bähren, L.} and {Bell, M. E.} and {Bell, M. R.} and {van Bemmel, I.} and {Bennema, P.} and {Bentum, M. J.} and {Bernardi, G.} and {Best, P.} and {Bîrzan, L.} and {Bonafede, A.} and {Boonstra, A.-J.} and {Braun, R.} and {Bregman, J.} and {Breitling, F.} and {van de Brink, R. H.} and {Broderick, J.} and {Broekema, P. C.} and {Brouw, W. N.} and {Brüggen, M.} and {Butcher, H. R.} and {van Cappellen, W.} and {Ciardi, B.} and {Coenen, T.} and {Conway, J.} and {Coolen, A.} and {Corstanje, A.} and {Damstra, S.} and {Davies, O.} and {Deller, A. T.} and {Dettmar, R.-J.} and {van Diepen, G.} and {Dijkstra, K.} and {Donker, P.} and {Doorduin, A.} and {Dromer, J.} and {Drost, M.} and {van Duin, A.} and {Eislöffel, J.} and {van Enst, J.} and {Ferrari, C.} and {Frieswijk, W.} and {Gankema, H.} and {Garrett, M. A.} and {de Gasperin, F.} and {Gerbers, M.} and {de Geus, E.} and {Grießmeier, J.-M.} and {Grit, T.} and {Gruppen, P.} and {Hamaker, J. P.} and {Hassall, T.} and {Hoeft, M.} and {Holties, H. A.} and {Horneffer, A.} and {van der Horst, A.} and {van Houwelingen, A.} and {Huijgen, A.} and {Iacobelli, M.} and {Intema, H.} and {Jackson, N.} and {Jelic, V.} and {de Jong, A.} and {Juette, E.} and {Kant, D.} and {Karastergiou, A.} and {Koers, A.} and {Kollen, H.} and {Kondratiev, V. I.} and {Kooistra, E.} and {Koopman, Y.} and {Koster, A.} and {Kuniyoshi, M.} and {Kramer, M.} and {Kuper, G.} and {Lambropoulos, P.} and {Law, C.} and {van Leeuwen, J.} and {Lemaitre, J.} and {Loose, M.} and {Maat, P.} and {Macario, G.} and {Markoff, S.} and {Masters, J.} and {McFadden, R. A.} and {McKay-Bukowski, D.} and {Meijering, H.} and {Meulman, H.} and {Mevius, M.} and {Middelberg, E.} and {Millenaar, R.} and {Miller-Jones, J. C. A.} and {Mohan, R. N.} and {Mol, J. D.} and {Morawietz, J.} and {Morganti, R.} and {Mulcahy, D. D.} and {Mulder, E.} and {Munk, H.} and {Nieuwenhuis, L.} and {van Nieuwpoort, R.} and {Noordam, J. E.} and {Norden, M.} and {Noutsos, A.} and {Offringa, A. R.} and {Olofsson, H.} and {Omar, A.} and {Orrú, E.} and {Overeem, R.} and {Paas, H.} and {Pandey-Pommier, M.} and {Pandey, V. N.} and {Pizzo, R.} and {Polatidis, A.} and {Rafferty, D.} and {Rawlings, S.} and {Reich, W.} and {de Reijer, J.-P.} and {Reitsma, J.} and {Renting, G. A.} and {Riemers, P.} and {Rol, E.} and {Romein, J. W.} and {Roosjen, J.} and {Ruiter, M.} and {Scaife, A.} and {van der Schaaf, K.} and {Scheers, B.} and {Schellart, P.} and {Schoenmakers, A.} and {Schoonderbeek, G.} and {Serylak, M.} and {Shulevski, A.} and {Sluman, J.} and {Smirnov, O.} and {Sobey, C.} and {Spreeuw, H.} and {Steinmetz, M.} and {Sterks, C. G. M.} and {Stiepel, H.-J.} and {Stuurwold, K.} and {Tagger, M.} and {Tang, Y.} and {Tasse, C.} and {Thomas, I.} and {Thoudam, S.} and {Toribio, M. C.} and {van der Tol, B.} and {Usov, O.} and {van Veelen, M.} and {van der Veen, A.-J.} and {ter Veen, S.} and {Verbiest, J. P. W.} and {Vermeulen, R.} and {Vermaas, N.} and {Vocks, C.} and {Vogt, C.} and {de Vos, M.} and {van der Wal, E.} and {van Weeren, R.} and {Weggemans, H.} and {Weltevrede, P.} and {White, S.} and {Wijnholds, S. J.} and {Wilhelmsson, T.} and {Wucknitz, O.} and {Yatawatta, S.} and {Zarka, P.} and {Zensus, A.} and {van Zwieten, J.}},
	title = {LOFAR: The LOw-Frequency ARray},
	DOI= "10.1051/0004-6361/201220873",
	url= "https://doi.org/10.1051/0004-6361/201220873",
	journal = {A&A},
	year = 2013,
	volume = 556,
	pages = "A2",
	month = "",
}

@ARTICLE{MWA,
       author = {{Tingay}, S.~J. and {Goeke}, R. and {Bowman}, J.~D. and {Emrich}, D. and {Ord}, S.~M. and {Mitchell}, D.~A. and {Morales}, M.~F. and {Booler}, T. and {Crosse}, B. and {Wayth}, R.~B. and {Lonsdale}, C.~J. and {Tremblay}, S. and {Pallot}, D. and {Colegate}, T. and {Wicenec}, A. and {Kudryavtseva}, N. and {Arcus}, W. and {Barnes}, D. and {Bernardi}, G. and {Briggs}, F. and {Burns}, S. and {Bunton}, J.~D. and {Cappallo}, R.~J. and {Corey}, B.~E. and {Deshpande}, A. and {Desouza}, L. and {Gaensler}, B.~M. and {Greenhill}, L.~J. and {Hall}, P.~J. and {Hazelton}, B.~J. and {Herne}, D. and {Hewitt}, J.~N. and {Johnston-Hollitt}, M. and {Kaplan}, D.~L. and {Kasper}, J.~C. and {Kincaid}, B.~B. and {Koenig}, R. and {Kratzenberg}, E. and {Lynch}, M.~J. and {Mckinley}, B. and {Mcwhirter}, S.~R. and {Morgan}, E. and {Oberoi}, D. and {Pathikulangara}, J. and {Prabu}, T. and {Remillard}, R.~A. and {Rogers}, A.~E.~E. and {Roshi}, A. and {Salah}, J.~E. and {Sault}, R.~J. and {Udaya-Shankar}, N. and {Schlagenhaufer}, F. and {Srivani}, K.~S. and {Stevens}, J. and {Subrahmanyan}, R. and {Waterson}, M. and {Webster}, R.~L. and {Whitney}, A.~R. and {Williams}, A. and {Williams}, C.~L. and {Wyithe}, J.~S.~B.},
        title = "{The Murchison Widefield Array: The Square Kilometre Array Precursor at Low Radio Frequencies}",
      journal = {\pasa},
         year = 2013,
        month = jan,
       volume = {30},
          eid = {e007},
        pages = {e007},
          doi = {10.1017/pasa.2012.007},
archivePrefix = {arXiv},
       eprint = {1206.6945},
 primaryClass = {astro-ph.IM},
       adsurl = {https://ui.adsabs.harvard.edu/abs/2013PASA...30....7T}
}

@INPROCEEDINGS{SKA,
       author = {{Mellema}, G. and {Koopmans}, L. and {Shukla}, H. and {Datta}, K.~K. and {Mesinger}, A. and {Majumdar}, S.},
        title = "{HI tomographic imaging of the Cosmic Dawn and Epoch of Reionization with SKA}",
    booktitle = {Advancing Astrophysics with the Square Kilometre Array (AASKA14)},
         year = 2015,
        month = apr,
          eid = {10},
        pages = {10},
          doi = {10.22323/1.215.0010},
archivePrefix = {arXiv},
       eprint = {1501.04203},
 primaryClass = {astro-ph.CO},
       adsurl = {https://ui.adsabs.harvard.edu/abs/2015aska.confE..10M}
}

@article{Bowman_2018,
   title={An absorption profile centred at 78 megahertz in the sky-averaged spectrum},
   volume={555},
   ISSN={1476-4687},
   url={http://dx.doi.org/10.1038/nature25792},
   DOI={10.1038/nature25792},
   number={7694},
   journal={Nature},
   publisher={Springer Science and Business Media LLC},
   author={Bowman, Judd D. and Rogers, Alan E. E. and Monsalve, Raul A. and Mozdzen, Thomas J. and Mahesh, Nivedita},
   year={2018},
   month=mar, pages={67–70} }

@article{Philip_2018,
author = {Philip, Liju and Abdurashidova, Z. and Chiang, H. and Ghazi, N. and Gumba, A. and Heilgendorff, H. and Jáuregui Garcia, Jose and Malepe, Kagiso and Nunhokee, C. and Peterson, Jaron and Sievers, J. and Simes, V. and Spann, R.},
year = {2018},
month = {12},
pages = {},
title = {Probing Radio Intensity at High-Z from Marion: 2017 Instrument},
volume = {08},
journal = {Journal of Astronomical Instrumentation},
doi = {10.1142/S2251171719500041}
}

@article{Acedo_2022,
   title={The REACH radiometer for detecting the 21-cm hydrogen signal from redshift z ≈ 7.5–28},
   volume={6},
   ISSN={2397-3366},
   url={http://dx.doi.org/10.1038/s41550-022-01709-9},
   DOI={10.1038/s41550-022-01709-9},
   number={8},
   journal={Nature Astronomy},
   publisher={Springer Science and Business Media LLC},
   author={de Lera Acedo, E. and de Villiers, D. I. L. and Razavi-Ghods, N. and Handley, W. and Fialkov, A. and Magro, A. and Anstey, D. and Bevins, H. T. J. and Chiello, R. and Cumner, J. and Josaitis, A. T. and Roque, I. L. V. and Sims, P. H. and Scheutwinkel, K. H. and Alexander, P. and Bernardi, G. and Carey, S. and Cavillot, J. and Croukamp, W. and Ely, J. A. and Gessey-Jones, T. and Gueuning, Q. and Hills, R. and Kulkarni, G. and Maiolino, R. and Meerburg, P. D. and Mittal, S. and Pritchard, J. R. and Puchwein, E. and Saxena, A. and Shen, E. and Smirnov, O. and Spinelli, M. and Zarb-Adami, K.},
   year={2022},
   month=jul, pages={984–998} }

@ARTICLE{Singh_2018,
       author = {{Singh}, Saurabh and {Subrahmanyan}, Ravi and {Udaya Shankar}, N. and {Sathyanarayana Rao}, Mayuri and {Fialkov}, Anastasia and {Cohen}, Aviad and {Barkana}, Rennan and {Girish}, B.~S. and {Raghunathan}, A. and {Somashekar}, R. and {Srivani}, K.~S.},
        title = "{SARAS 2 Constraints on Global 21 cm Signals from the Epoch of Reionization}",
      journal = {\apj},
         year = 2018,
        month = may,
       volume = {858},
       number = {1},
          eid = {54},
        pages = {54},
          doi = {10.3847/1538-4357/aabae1},
archivePrefix = {arXiv},
       eprint = {1711.11281},
 primaryClass = {astro-ph.CO},
       adsurl = {https://ui.adsabs.harvard.edu/abs/2018ApJ...858...54S}
}

@ARTICLE{Singh_2022,
       author = {{Singh}, Saurabh and {Jishnu}, Nambissan T. and {Subrahmanyan}, Ravi and {Udaya Shankar}, N. and {Girish}, B.~S. and {Raghunathan}, A. and {Somashekar}, R. and {Srivani}, K.~S. and {Sathyanarayana Rao}, Mayuri},
        title = "{On the detection of a cosmic dawn signal in the radio background}",
      journal = {Nature Astronomy},
         year = 2022,
        month = feb,
       volume = {6},
        pages = {607-617},
          doi = {10.1038/s41550-022-01610-5},
archivePrefix = {arXiv},
       eprint = {2112.06778},
 primaryClass = {astro-ph.CO},
       adsurl = {https://ui.adsabs.harvard.edu/abs/2022NatAs...6..607S}
}

@article{Sartorio_2023,
    author = {Sartorio, Nina S and Fialkov, A and Hartwig, T and Mirouh, G M and Izzard, R G and Magg, M and Klessen, R S and Glover, S C O and Chen, L and Tarumi, Y and Hendriks, D D},
    title = {Population III X-ray binaries and their impact on the early universe},
    journal = {Monthly Notices of the Royal Astronomical Society},
    volume = {521},
    number = {3},
    pages = {4039-4055},
    year = {2023},
    month = {03},
    issn = {0035-8711},
    doi = {10.1093/mnras/stad697},
    url = {https://doi.org/10.1093/mnras/stad697},
    eprint = {https://academic.oup.com/mnras/article-pdf/521/3/4039/56454135/stad697.pdf},
}

@article{Bevins_2021,
   title={<scp>maxsmooth</scp>: rapid maximally smooth function fitting with applications in Global 21-cm cosmology},
   volume={502},
   ISSN={1365-2966},
   url={http://dx.doi.org/10.1093/mnras/stab152},
   DOI={10.1093/mnras/stab152},
   number={3},
   journal={Monthly Notices of the Royal Astronomical Society},
   publisher={Oxford University Press (OUP)},
   author={Bevins, H T J and Handley, W J and Fialkov, A and de Lera Acedo, E and Greenhill, L J and Price, D C},
   year={2021},
   month=jan, pages={4405–4425} }

@article{Monsalve_2024,
   title={Simulating the Detection of the Global 21 cm Signal with MIST for Different Models of the Soil and Beam Directivity},
   volume={961},
   ISSN={1538-4357},
   url={http://dx.doi.org/10.3847/1538-4357/ad0f1b},
   DOI={10.3847/1538-4357/ad0f1b},
   number={1},
   journal={The Astrophysical Journal},
   publisher={American Astronomical Society},
   author={Monsalve, Raul A. and Bye, Christian H. and Sievers, Jonathan L. and Bidula, Vadym and Bustos, Ricardo and Chiang, H. Cynthia and Guo, Xinze and Hendricksen, Ian and McGee, Francis and Mena, F. Patricio and Prabhakar, Garima and Restrepo, Oscar and Thyagarajan, Nithyanandan},
   year={2024},
   month=jan, pages={56} }

@article{Shen_2021,
    author = {Shen, Emma and Anstey, Dominic and de Lera Acedo, Eloy and Fialkov, Anastasia and Handley, Will},
    title = {Quantifying ionospheric effects on global 21-cm observations},
    journal = {Monthly Notices of the Royal Astronomical Society},
    volume = {503},
    number = {1},
    pages = {344-353},
    year = {2021},
    month = {02},
    issn = {0035-8711},
    doi = {10.1093/mnras/stab429},
    url = {https://doi.org/10.1093/mnras/stab429},
    eprint = {https://academic.oup.com/mnras/article-pdf/503/1/344/38845082/stab429.pdf},
}

@ARTICLE{Shaver_1999,
       author = {{Shaver}, P.~A. and {Windhorst}, R.~A. and {Madau}, P. and {de Bruyn}, A.~G.},
        title = "{Can the reionization epoch be detected as a global signature in the cosmic background?}",
      journal = {\aap},
         year = 1999,
        month = may,
       volume = {345},
        pages = {380-390},
          doi = {10.48550/arXiv.astro-ph/9901320},
archivePrefix = {arXiv},
       eprint = {astro-ph/9901320},
 primaryClass = {astro-ph},
       adsurl = {https://ui.adsabs.harvard.edu/abs/1999A&A...345..380S}
}

@article{Anstey_2021,
    author = {Anstey, Dominic and de Lera Acedo, Eloy and Handley, Will},
    title = {A general Bayesian framework for foreground modelling and chromaticity correction for global 21 cm experiments},
    journal = {Monthly Notices of the Royal Astronomical Society},
    volume = {506},
    number = {2},
    pages = {2041-2058},
    year = {2021},
    month = {06},
    issn = {0035-8711},
    doi = {10.1093/mnras/stab1765},
    url = {https://doi.org/10.1093/mnras/stab1765},
    eprint = {https://academic.oup.com/mnras/article-pdf/506/2/2041/42467147/stab1765.pdf},
}

@article{Anstey_2022,
    author = {Anstey, Dominic and de Lera Acedo, Eloy and Handley, Will},
    title = {Use of time dependent data in Bayesian global 21-cm foreground and signal modelling},
    journal = {Monthly Notices of the Royal Astronomical Society},
    volume = {520},
    number = {1},
    pages = {850-865},
    year = {2023},
    month = {01},
    issn = {0035-8711},
    doi = {10.1093/mnras/stad156},
    url = {https://doi.org/10.1093/mnras/stad156},
    eprint = {https://academic.oup.com/mnras/article-pdf/520/1/850/49057286/stad156.pdf},
}

@article{Carter_2025,
    author = {Carter, George and Handley, Will and Ashdown, Mark and Razavi-Ghods, Nima},
    title = {The Bayesian Global Sky Model (B-GSM): validation of a data-driven Bayesian simultaneous component separation and calibration algorithm for EoR foreground modelling},
    journal = {Monthly Notices of the Royal Astronomical Society},
    volume = {544},
    number = {2},
    pages = {1463-1487},
    year = {2025},
    month = {10},
    issn = {0035-8711},
    doi = {10.1093/mnras/staf1743},
    url = {https://doi.org/10.1093/mnras/staf1743},
    eprint = {https://academic.oup.com/mnras/article-pdf/544/2/1463/64612557/staf1743.pdf},
}

@ARTICLE{deOliveiraCosta2008,
       author = {{de Oliveira-Costa}, Ang{\'e}lica and {Tegmark}, Max and {Gaensler}, B.~M. and {Jonas}, Justin and {Landecker}, T.~L. and {Reich}, Patricia},
        title = "{A model of diffuse Galactic radio emission from 10 MHz to 100 GHz}",
      journal = {\mnras},
         year = 2008,
        month = jul,
       volume = {388},
       number = {1},
        pages = {247-260},
          doi = {10.1111/j.1365-2966.2008.13376.x},
archivePrefix = {arXiv},
       eprint = {0802.1525},
 primaryClass = {astro-ph},
       adsurl = {https://ui.adsabs.harvard.edu/abs/2008MNRAS.388..247D}
}

@article{Sims_2025,
   title={A general Bayesian model-validation framework based on null-test evidence ratios, with an example application to global 21-cm cosmology},
   volume={541},
   ISSN={1365-2966},
   url={http://dx.doi.org/10.1093/mnras/staf1109},
   DOI={10.1093/mnras/staf1109},
   number={3},
   journal={Monthly Notices of the Royal Astronomical Society},
   publisher={Oxford University Press (OUP)},
   author={Sims, Peter H and Bowman, Judd D and Murray, Steven G and Barrett, John P and Cappallo, Rigel C and Lonsdale, Colin J and Mahesh, Nivedita and Monsalve, Raul A and Rogers, Alan E E and Samson, Titu and Vydula, Akshatha K},
   year={2025},
   month=jul, pages={2262–2281} }

@article{Cumner_2022,
author = {Cumner, J. and de Lera Acedo, E. and de Villiers, D. I. L. and Anstey, D. and Kolitsidas, C. I. and Gurdon, B. and Fagnoni, N. and Alexander, P. and Bernardi, G. and Bevins, H. T. J. and Carey, S. and Cavillot, J. and Chiello, R. and Craeye, C. and Croukamp, W. and Ely, J. A. and Fialkov, A. and Gessey-Jones, T. and Gueuning, Q. and Handley, W. and Hills, R. and Josaitis, A. T. and Kulkarni, G. and Magro, A. and Maiolino, R. and Meerburg, P. D. and Mittal, S. and Pritchard, J. R. and Puchwein, E. and Razavi-Ghods, N. and Roque, I. L. V. and Saxena, A. and Scheutwinkel, K. H. and Shen, E. and Sims, P. H. and Smirnov, O. and Spinelli, M. and Zarb-Adami, K.},
title = {Radio Antenna Design for Sky-Averaged 21cm Cosmology Experiments: The REACH Case},
journal = {Journal of Astronomical Instrumentation},
volume = {11},
number = {01},
pages = {2250001},
year = {2022},
doi = {10.1142/S2251171722500015},

URL = { 
    
        https://doi.org/10.1142/S2251171722500015
    
    

},
eprint = { 
    
        https://doi.org/10.1142/S2251171722500015
    
    

}
}

@article{Hu1999,
  author  = {Hu, B. and Chew, W. C. and Michielssen, E. and Zhao, J.},
  title   = {Fast Inhomogeneous Plane Wave Algorithm for the Fast Analysis of Two-Dimensional Scattering Problems},
  journal = {Radio Science},
  volume  = {34},
  number  = {4},
  pages   = {759--772},
  year    = {1999},
  doi     = {10.1029/1999RS900038}
}

@article{Alkhalifeh2016,
  author  = {Alkhalifeh, K. and Hislop, G. and Ozdemir, N. A. and Craeye, C.},
  title   = {Efficient MoM Simulation of 3-D Antennas in the Vicinity of the Ground},
  journal = {IEEE Transactions on Antennas and Propagation},
  volume  = {64},
  number  = {12},
  pages   = {5335--5344},
  year    = {2016},
  doi     = {10.1109/TAP.2016.2618482}
}

@article{Cavillot2020,
  author  = {Jean Cavillot and Denis Tihon and Francisco Mesa and Eloy de Lera Acedo and Christophe Craeye},
  title   = {Efficient simulation of large irregular arrays on a finite ground plane},
  journal = {IEEE Transactions on Antennas and Propagation},
  volume  = {68},
  pages   = {2753--2764},
  year    = {2020},
  doi     = {10.1109/TAP.2019.2955180}
}

@article{Cavillot2024,
  author  = {Jean Cavillot and Denis Tihon and Quentin Gueuning and Eloy de Lera Acedo and Christophe Craeye},
  title   = {Full-Wave Analysis of Thermal Noise in Antenna Arrays on Top of Layered Medium},
  journal = {IEEE Transactions on Antennas and Propagation},
  volume  = {72},
  pages   = {7560--7573},
  year    = {2024},
  doi     = {10.1109/TAP.2024.3445132}
}

@book{Harrington1993,
  author    = {Harrington, R. F.},
  title     = {Field Computation by Moment Methods},
  publisher = {Wiley--IEEE Press},
  address   = {Hoboken, New Jersey, US},
  year      = {1993}
}

@article{Shen_2022,
    author = {Shen, Emma and Anstey, Dominic and de Lera Acedo, Eloy and Fialkov, Anastasia},
    title = {Bayesian data analysis for sky-averaged 21-cm experiments in the presence of ionospheric effects},
    journal = {Monthly Notices of the Royal Astronomical Society},
    volume = {515},
    number = {3},
    pages = {4565-4573},
    year = {2022},
    month = {07},
    issn = {0035-8711},
    doi = {10.1093/mnras/stac1900},
    url = {https://doi.org/10.1093/mnras/stac1900},
    eprint = {https://academic.oup.com/mnras/article-pdf/515/3/4565/45475179/stac1900.pdf},
}

@article{Leeney_2023,
  title = {Bayesian approach to radio frequency interference mitigation},
  author = {Leeney, S. A. K. and Handley, W. J. and Acedo, E. de Lera},
  journal = {Phys. Rev. D},
  volume = {108},
  issue = {6},
  pages = {062006},
  numpages = {6},
  year = {2023},
  month = {Sep},
  publisher = {American Physical Society},
  doi = {10.1103/PhysRevD.108.062006},
  url = {https://link.aps.org/doi/10.1103/PhysRevD.108.062006}
}

@article{Anstey_2024,
    author = {Anstey, Dominic and Leeney, Samuel A K},
    title = {Enhanced Bayesian RFI mitigation and transient flagging using likelihood reweighting},
    journal = {RAS Techniques and Instruments},
    volume = {3},
    number = {1},
    pages = {372-384},
    year = {2024},
    month = {07},
    issn = {2752-8200},
    doi = {10.1093/rasti/rzae025},
    url = {https://doi.org/10.1093/rasti/rzae025},
    eprint = {https://academic.oup.com/rasti/article-pdf/3/1/372/61224726/rzae025.pdf},
}

@article{Mittal_2024,
    author = {Mittal, Shikhar and Kulkarni, Girish and Anstey, Dominic and de Lera Acedo, Eloy},
    title = {Impact of extragalactic point sources on the low-frequency sky spectrum and cosmic dawn global 21-cm measurements},
    journal = {Monthly Notices of the Royal Astronomical Society},
    volume = {534},
    number = {2},
    pages = {1317-1328},
    year = {2024},
    month = {09},
    issn = {0035-8711},
    doi = {10.1093/mnras/stae2111},
    url = {https://doi.org/10.1093/mnras/stae2111},
    eprint = {https://academic.oup.com/mnras/article-pdf/534/2/1317/59401684/stae2111.pdf},
}

@article{Pattison_2023,
    author = {Pattison, Joe H N and Anstey, Dominic J and de Lera Acedo, Eloy},
    title = {Modelling a hot horizon in global 21-cm experimental foregrounds},
    journal = {Monthly Notices of the Royal Astronomical Society},
    volume = {527},
    number = {2},
    pages = {2413-2425},
    year = {2023},
    month = {11},
    issn = {0035-8711},
    doi = {10.1093/mnras/stad3378},
    url = {https://doi.org/10.1093/mnras/stad3378},
    eprint = {https://academic.oup.com/mnras/article-pdf/527/2/2413/53404166/stad3378.pdf},
}

@article{pattison_2025a,
    author = {Pattison, Joe H N and Cavillot, Jean and Bevins, Harry T J and Anstey, Dominic J and Cumner, John M and de Lera Acedo, Eloy},
    title = {Global 21 cm signal recovery under changing environmental conditions},
    journal = {Monthly Notices of the Royal Astronomical Society},
    volume = {538},
    number = {3},
    pages = {1301-1313},
    year = {2025},
    month = {02},
    issn = {0035-8711},
    doi = {10.1093/mnras/staf315},
    url = {https://doi.org/10.1093/mnras/staf315},
    eprint = {https://academic.oup.com/mnras/article-pdf/538/3/1301/62178753/staf315.pdf},
}

@article{pattison_2025b,
      title={Quantifying the Impact of Lunar and Planetary Occultation on Experimental Global 21 cm Cosmology}, 
      author={Joe H. N. Pattison and Dominic J. Anstey and Eloy de Lera Acedo},
      journal={Monthly Notices of the Royal Astronomical Society},
      volume={546},
      number={4},
      pages={stag258},
      year={2026},
      doi={10.1093/mnras/stag258},
}

@article{Pagano_2023,
   title={A general Bayesian framework to account for foreground map errors in global 21-cm experiments},
   volume={527},
   ISSN={1365-2966},
   url={http://dx.doi.org/10.1093/mnras/stad3392},
   DOI={10.1093/mnras/stad3392},
   number={3},
   journal={Monthly Notices of the Royal Astronomical Society},
   publisher={Oxford University Press (OUP)},
   author={Pagano, Michael and Sims, Peter and Liu, Adrian and Anstey, Dominic and Handley, Will and de Lera Acedo, Eloy},
   year={2023},
   month=nov, pages={5649–5667} }

@article{polychord,
    author = {Handley, W. J. and Hobson, M. P. and Lasenby, A. N.},
    title = {polychord: next-generation nested sampling},
    journal = {Monthly Notices of the Royal Astronomical Society},
    volume = {453},
    number = {4},
    pages = {4384-4398},
    year = {2015},
    month = {09},
    issn = {0035-8711},
    doi = {10.1093/mnras/stv1911},
    url = {https://doi.org/10.1093/mnras/stv1911},
    eprint = {https://academic.oup.com/mnras/article-pdf/453/4/4384/8034904/stv1911.pdf},
}

@article{Skilling_2006,
    author = "Skilling, John",
    title = "{Nested sampling for general Bayesian computation}",
    doi = "10.1214/06-BA127",
    journal = "Bayesian Analysis",
    volume = "1",
    number = "4",
    pages = "833--859",
    year = "2006"
}

@article{Hoffman_2014,
  title   = {The No-U-Turn Sampler: Adaptively Setting Path Lengths in Hamiltonian Monte Carlo},
  author  = {Hoffman, Matthew D. and Gelman, Andrew},
  journal = {Journal of Machine Learning Research},
  volume  = {15},
  pages   = {1593--1623},
  year    = {2014}
}

@misc{lovick_2025,
      title={High-Dimensional Bayesian Model Comparison in Cosmology with GPU-accelerated Nested Sampling and Neural Emulators}, 
      author={Toby Lovick and David Yallup and Davide Piras and Alessio Spurio Mancini and Will Handley},
      year={2025},
      eprint={2509.13307},
      archivePrefix={arXiv},
      primaryClass={astro-ph.CO},
      url={https://arxiv.org/abs/2509.13307}, 
}

@article{Saxena_2023,
   title={Sky-averaged 21-cm signal extraction using multiple antennas with an SVD framework: the REACH case},
   volume={522},
   ISSN={1365-2966},
   url={http://dx.doi.org/10.1093/mnras/stad1047},
   DOI={10.1093/mnras/stad1047},
   number={1},
   journal={Monthly Notices of the Royal Astronomical Society},
   publisher={Oxford University Press (OUP)},
   author={Saxena, Anchal and Meerburg, P Daniel and de Lera Acedo, Eloy and Handley, Will and Koopmans, Léon V E},
   year={2023},
   month=Apr, pages={1022–1032} }

@ARTICLE{Sims_2025b ,
       author = {{Sims}, Peter H. and {Bowman}, Judd D. and {Murray}, Steven G. and {Barrett}, John P. and {Cappallo}, Rigel C. and {Lonsdale}, Colin J. and {Mahesh}, Nivedita and {Monsalve}, Raul A. and {Rogers}, Alan E.~E. and {Samson}, Titu and et al.},
        title = "{A Bayesian approach to modelling spectrometer data chromaticity corrected using beam factors ─ II. Model priors and posterior odds}",
      journal = {\mnras},
         year = 2025,
        month = dec,
       volume = {544},
       number = {2},
        pages = {2340-2364},
          doi = {10.1093/mnras/staf1767},
archivePrefix = {arXiv},
       eprint = {2506.20042},
 primaryClass = {astro-ph.IM},
       adsurl = {https://ui.adsabs.harvard.edu/abs/2025MNRAS.544.2340S}
}

@article{Jeffrey_2024,
   title={Evidence Networks: simple losses for fast, amortized, neural Bayesian model comparison},
   volume={5},
   ISSN={2632-2153},
   url={http://dx.doi.org/10.1088/2632-2153/ad1a4d},
   DOI={10.1088/2632-2153/ad1a4d},
   number={1},
   journal={Machine Learning: Science and Technology},
   publisher={IOP Publishing},
   author={Jeffrey, Niall and Wandelt, Benjamin D},
   year={2024},
   month=jan, pages={015008} }

@article{kern_2025,
    author = {Kern, Nicholas},
    title = {A differentiable, end-to-end forward model for 21 cm cosmology: estimating the foreground, instrument, and signal joint posterior},
    journal = {Monthly Notices of the Royal Astronomical Society},
    volume = {541},
    number = {2},
    pages = {687-713},
    year = {2025},
    month = {06},
    issn = {0035-8711},
    doi = {10.1093/mnras/staf1007},
    url = {https://doi.org/10.1093/mnras/staf1007},
    eprint = {https://academic.oup.com/mnras/article-pdf/541/2/687/63527997/staf1007.pdf},
}

@article{Wilensky_2025,
    author = {Wilensky, Michael J and Bull, Philip and Fagnoni, Nicolas},
    title = {High dimensional beam inference II: inference of a perturbed HERA beam from simulated visibility data},
    journal = {RAS Techniques and Instruments},
    volume = {4},
    pages = {rzaf042},
    year = {2025},
    month = {09},
    issn = {2752-8200},
    doi = {10.1093/rasti/rzaf042},
    url = {https://doi.org/10.1093/rasti/rzaf042},
    eprint = {https://academic.oup.com/rasti/article-pdf/doi/10.1093/rasti/rzaf042/64303441/rzaf042.pdf},
}

@ARTICLE{Gunn_1965,
       author = {{Gunn}, James E. and {Peterson}, Bruce A.},
        title = "{On the Density of Neutral Hydrogen in Intergalactic Space.}",
      journal = {\apj},
         year = 1965,
        month = nov,
       volume = {142},
        pages = {1633-1636},
          doi = {10.1086/148444},
       adsurl = {https://ui.adsabs.harvard.edu/abs/1965ApJ...142.1633G}
}

@article{Fan_2006,
   title={Constraining the Evolution of the Ionizing Background and the Epoch of Reionization with z ~ 6 Quasars. II. A Sample of 19 Quasars},
   volume={132},
   ISSN={1538-3881},
   url={http://dx.doi.org/10.1086/504836},
   DOI={10.1086/504836},
   number={1},
   journal={The Astronomical Journal},
   publisher={American Astronomical Society},
   author={Fan, Xiaohui and Strauss, Michael A. and Becker, Robert H. and White, Richard L. and Gunn, James E. and Knapp, Gillian R. and Richards, Gordon T. and Schneider, Donald P. and Brinkmann, J. and Fukugita, Masataka},
   year={2006},
   month=jun,
   pages={117-136}
}

@article{Becker_2015,
   title={Evidence of patchy hydrogen reionization from an extreme Ly-alpha trough below redshift six},
   volume={447},
   ISSN={0035-8711},
   url={http://dx.doi.org/10.1093/mnras/stu2646},
   DOI={10.1093/mnras/stu2646},
   number={4},
   journal={Monthly Notices of the Royal Astronomical Society},
   publisher={Oxford University Press (OUP)},
   author={Becker, George D. and Bolton, James S. and Madau, Piero and Pettini, Max and Ryan-Weber, Emma V. and Venemans, Bram P.},
   year={2015},
   month=jan,
   pages={3402-3419}
}

@article{Eilers_2018,
   title={The Opacity of the Intergalactic Medium Measured along Quasar Sightlines at z ~ 6},
   volume={864},
   ISSN={1538-4357},
   url={http://dx.doi.org/10.3847/1538-4357/aad4fd},
   DOI={10.3847/1538-4357/aad4fd},
   number={1},
   journal={The Astrophysical Journal},
   publisher={American Astronomical Society},
   author={Eilers, Anna-Christina and Davies, Frederick B. and Hennawi, Joseph F.},
   year={2018},
   month=aug,
   pages={53}
}

@article{Bosman_2022,
   title={Hydrogen reionization ends by z = 5.3: Lyman-alpha optical depth measured by the XQR-30 sample},
   volume={514},
   ISSN={1365-2966},
   url={http://dx.doi.org/10.1093/mnras/stac1046},
   DOI={10.1093/mnras/stac1046},
   number={1},
   journal={Monthly Notices of the Royal Astronomical Society},
   publisher={Oxford University Press (OUP)},
   author={Bosman, Sarah E. I. and Davies, Frederick B. and Becker, George D. and Keating, Laura C. and Davies, Rebecca L. and Zhu, Yongda and Eilers, Anna-Christina and D'Odorico, Valentina and Bian, Fuyan and Bischetti, Manuela and Cristiani, Stefano V. and Fan, Xiaohui and Farina, Emanuele P. and Haehnelt, Martin G. and Hennawi, Joseph F. and Kulkarni, Girish and Mesinger, Andrei and Meyer, Romain A. and Onoue, Masafusa and Pallottini, Andrea and Qin, Yuxiang and Ryan-Weber, Emma and Schindler, Jan-Torge and Walter, Fabian and Wang, Feige and Yang, Jinyi},
   year={2022},
   month=jun,
   pages={55-76}
}

@article{Qin_2021,
   title={Reionization and galaxy inference from the high-redshift Ly-alpha forest},
   volume={506},
   ISSN={1365-2966},
   url={http://dx.doi.org/10.1093/mnras/stab1833},
   DOI={10.1093/mnras/stab1833},
   number={2},
   journal={Monthly Notices of the Royal Astronomical Society},
   publisher={Oxford University Press (OUP)},
   author={Qin, Yuxiang and Mesinger, Andrei and Bosman, Sarah E. I. and Viel, Matteo},
   year={2021},
   month=jun,
   pages={2390-2407},
   eprint={2101.09033},
   archivePrefix={arXiv},
   primaryClass={astro-ph.CO}
}

@article{Tauscher_2018a,
   title={Global 21 cm Signal Extraction from Foreground and Instrumental Effects. I. Pattern Recognition Framework for Separation Using Training Sets},
   volume={853},
   ISSN={1538-4357},
   url={http://dx.doi.org/10.3847/1538-4357/aaa41f},
   DOI={10.3847/1538-4357/aaa41f},
   number={2},
   journal={The Astrophysical Journal},
   publisher={American Astronomical Society},
   author={Tauscher, Keith and Rapetti, David and Burns, Jack O. and Switzer, Eric},
   year={2018},
   month=feb,
   pages={187},
   adsurl={https://ui.adsabs.harvard.edu/abs/2018ApJ...853..187T/abstract}
}

@article{Rapetti_2020,
   title={Global 21 cm Signal Extraction from Foreground and Instrumental Effects. II. Efficient and Self-consistent Technique for Constraining Nonlinear Signal Models},
   volume={897},
   ISSN={1538-4357},
   url={http://dx.doi.org/10.3847/1538-4357/ab9b29},
   DOI={10.3847/1538-4357/ab9b29},
   number={2},
   journal={The Astrophysical Journal},
   publisher={American Astronomical Society},
   author={Rapetti, David and Tauscher, Keith and Mirocha, Jordan and Burns, Jack O.},
   year={2020},
   month=jul,
   pages={174},
   adsurl={https://ui.adsabs.harvard.edu/abs/2020ApJ...897..174R/abstract}
}

@article{Tauscher_2021,
   title={Global 21 cm Signal Extraction from Foreground and Instrumental Effects. IV. Accounting for Realistic Instrument Uncertainties and Their Overlap with Foreground and Signal Models},
   volume={915},
   ISSN={1538-4357},
   url={http://dx.doi.org/10.3847/1538-4357/ac00af},
   DOI={10.3847/1538-4357/ac00af},
   number={1},
   journal={The Astrophysical Journal},
   publisher={American Astronomical Society},
   author={Tauscher, Keith and Rapetti, David and Nhan, Bang D. and Handy, Alec and Bassett, Neil and Hibbard, Joshua and Bordenave, David and Bradley, Richard F. and Burns, Jack O.},
   year={2021},
   month=jul,
   pages={66},
   adsurl={https://ui.adsabs.harvard.edu/abs/2021ApJ...915...66T/abstract}
}

@article{Tauscher_2020b,
   title={Formulating and Critically Examining the Assumptions of Global 21 cm Signal Analyses: How to Avoid the False Troughs That Can Appear in Single-spectrum Fits},
   volume={897},
   ISSN={1538-4357},
   url={http://dx.doi.org/10.3847/1538-4357/ab9a3f},
   DOI={10.3847/1538-4357/ab9a3f},
   number={2},
   journal={The Astrophysical Journal},
   publisher={American Astronomical Society},
   author={Tauscher, Keith and Rapetti, David and Burns, Jack O.},
   year={2020},
   month=jul,
   pages={132},
   adsurl={https://ui.adsabs.harvard.edu/abs/2020ApJ...897..132T/abstract}
}

@article{Hibbard_2020,
   title={Modeling the Galactic Foreground and Beam Chromaticity for Global 21 cm Cosmology},
   volume={905},
   ISSN={1538-4357},
   url={http://dx.doi.org/10.3847/1538-4357/abc3c5},
   DOI={10.3847/1538-4357/abc3c5},
   number={2},
   journal={The Astrophysical Journal},
   publisher={American Astronomical Society},
   author={Hibbard, Joshua J. and Tauscher, Keith and Rapetti, David and Burns, Jack O.},
   year={2020},
   month=dec,
   pages={113},
   adsurl={https://ui.adsabs.harvard.edu/abs/2020ApJ...905..113H/abstract}
}

@misc{Hibbard_2024,
   title={{\texttt{MEDEA}}: A New Model for Emulating Radio Antenna Beam Patterns for 21-cm Cosmology and Antenna Design Studies},
   author={Hibbard, Joshua J. and Nhan, Bang D. and Rapetti, David and Burns, Jack O.},
   year={2024},
   eprint={2408.16135},
   archivePrefix={arXiv},
   primaryClass={astro-ph.IM},
   doi={10.48550/arXiv.2408.16135},
   url={https://arxiv.org/abs/2408.16135}
}

@article{Bassett_2021a,
   title={Ensuring Robustness in Training-set-based Global 21 cm Cosmology Analysis},
   volume={908},
   ISSN={1538-4357},
   url={http://dx.doi.org/10.3847/1538-4357/abdb29},
   DOI={10.3847/1538-4357/abdb29},
   number={2},
   journal={The Astrophysical Journal},
   publisher={American Astronomical Society},
   author={Bassett, Neil and Rapetti, David and Tauscher, Keith and Burns, Jack O. and Hibbard, Joshua J.},
   year={2021},
   month=feb,
   pages={189},
   adsurl={https://ui.adsabs.harvard.edu/abs/2021ApJ...908..189B/abstract}
}

@article{Bassett_2021b,
   title={Lost Horizon: Quantifying the Effect of Local Topography on Global 21 cm Cosmology Data Analysis},
   volume={923},
   ISSN={1538-4357},
   url={http://dx.doi.org/10.3847/1538-4357/ac1cde},
   DOI={10.3847/1538-4357/ac1cde},
   number={1},
   journal={The Astrophysical Journal},
   publisher={American Astronomical Society},
   author={Bassett, Neil and Rapetti, David and Tauscher, Keith and Nhan, Bang D. and Bordenave, David D. and Hibbard, Joshua J. and Burns, Jack O.},
   year={2021},
   month=dec,
   pages={33},
   adsurl={https://ui.adsabs.harvard.edu/abs/2021ApJ...923...33B/abstract}
}

@article{McQuinn_2005,
   title={The Kinetic Sunyaev-Zeldovich Effect from Reionization},
   volume={630},
   ISSN={1538-4357},
   url={http://dx.doi.org/10.1086/432049},
   DOI={10.1086/432049},
   number={2},
   journal={The Astrophysical Journal},
   publisher={American Astronomical Society},
   author={McQuinn, Matthew and Furlanetto, Steven R. and Hernquist, Lars and Zahn, Oliver and Zaldarriaga, Matias},
   year={2005},
   month=sep,
   pages={643-656}
}

@article{Reichardt_2021,
   title={An Improved Measurement of the Secondary Cosmic Microwave Background Anisotropies from the SPT-SZ + SPTpol Surveys},
   volume={908},
   ISSN={1538-4357},
   url={http://dx.doi.org/10.3847/1538-4357/abd407},
   DOI={10.3847/1538-4357/abd407},
   number={2},
   journal={The Astrophysical Journal},
   publisher={American Astronomical Society},
   author={Reichardt, C. L. and Patil, S. and Ade, P. A. R. and Anderson, A. J. and Austermann, J. E. and Avva, J. S. and Baxter, E. and Beall, J. A. and Bender, A. N. and Benson, B. A. and Bianchini, F. and Bleem, L. E. and Carlstrom, J. E. and Chang, C. L. and Chaubal, P. and Chiang, H. C. and Chou, T. L. and Citron, R. and Moran, C. Corbett and Crawford, T. M. and Crites, A. T. and de Haan, T. and Dobbs, M. A. and Everett, W. and Gallicchio, J. and George, E. M. and Gilbert, A. and Gupta, N. and Halverson, N. W. and Harrington, N. and Henning, J. W. and Hilton, G. C. and Holder, G. P. and Holzapfel, W. L. and Hrubes, J. D. and Huang, N. and Hubmayr, J. and Irwin, K. D. and Knox, L. and Lee, A. T. and Li, D. and Lowitz, A. and Luong-Van, D. and McMahon, J. J. and Mehl, J. and Meyer, S. S. and Millea, M. and Mocanu, L. M. and Mohr, J. J. and Montgomery, J. and Nadolski, A. and Natoli, T. and Nibarger, J. P. and Noble, G. and Novosad, V. and Omori, Y. and Padin, S. and Pryke, C. and Ruhl, J. E. and Saliwanchik, B. R. and Sayre, J. T. and Schaffer, K. K. and Shirokoff, E. and Sievers, C. and Smecher, G. and Spieler, H. G. and Staniszewski, Z. and Stark, A. A. and Tucker, C. and Vanderlinde, K. and Veach, T. and Vieira, J. D. and Wang, G. and Whitehorn, N. and Williamson, R. and Wu, W. L. K. and Yefremenko, V.},
   year={2021},
   month=feb,
   pages={199}
}

@misc{iliev_2024,
   title={The Thermal Sunyaev-Zeldovich Effect from the Epoch of Reionization},
   author={Iliev, Ilian T. and Hosein, Azizah R. and Chluba, Jens and Conaboy, Luke and Attard, David and Mondal, Rajesh and Ahn, Kyungjin and Gottlober, Stefan and Lewis, Joseph and Ocvirk, Pierre and Park, Hyunbae and Shapiro, Paul R. and Sorce, Jenny G. and Yepes, Gustavo},
   year={2024},
   eprint={2412.04385},
   archivePrefix={arXiv},
   primaryClass={astro-ph.CO},
   url={https://arxiv.org/abs/2412.04385}
}

@article{Liu_2014a,
   title={Epoch of reionization window. I. Mathematical formalism},
   volume={90},
   number={2},
   journal={Physical Review D},
   publisher={American Physical Society},
   author={Liu, Adrian and Parsons, Aaron R. and Trott, Cathryn M.},
   year={2014},
   month=jul,
   doi={10.1103/PhysRevD.90.023018},
   url={https://doi.org/10.1103/PhysRevD.90.023018}
}

@article{Liu_2014b,
   title={Epoch of reionization window. II. Statistical methods for foreground wedge reduction},
   volume={90},
   number={2},
   journal={Physical Review D},
   publisher={American Physical Society},
   author={Liu, Adrian and Parsons, Aaron R. and Trott, Cathryn M.},
   year={2014},
   month=jul,
   doi={10.1103/PhysRevD.90.023019},
   url={https://doi.org/10.1103/PhysRevD.90.023019}
}

@article{Thyagarajan_2016,
   title={Effects of antenna beam chromaticity on redshifted 21 cm power spectrum and implications for Hydrogen Epoch of Reionization Array},
   volume={825},
   number={1},
   journal={The Astrophysical Journal},
   publisher={American Astronomical Society},
   author={Thyagarajan, Nithyanandan and Parsons, Aaron R. and DeBoer, David R. and Bowman, Judd D. and Ewall-Wice, Aaron M. and Neben, Abraham R. and Patra, Nipanjana},
   year={2016},
   month=jun,
   pages={9},
   doi={10.3847/0004-637X/825/1/9},
   url={https://doi.org/10.3847/0004-637X/825/1/9}
}

@article{Neben_2016,
   title={The Hydrogen Epoch of Reionization Array dish. I. Beam pattern measurements and science implications},
   volume={826},
   number={2},
   journal={The Astrophysical Journal},
   publisher={American Astronomical Society},
   author={Neben, Abraham R. and Bradley, Richard F. and Hewitt, Jacqueline N. and DeBoer, David R. and Parsons, Aaron R. and Aguirre, James E. and Ali, Zaki S. and others},
   year={2016},
   month=jul,
   pages={199},
   doi={10.3847/0004-637X/826/2/199},
   url={https://doi.org/10.3847/0004-637X/826/2/199}
}

@article{Fagnoni_2020,
   title={Understanding the HERA Phase I receiver system with simulations and its impact on the detectability of the EoR delay power spectrum},
   volume={500},
   number={1},
   journal={Monthly Notices of the Royal Astronomical Society},
   publisher={Oxford University Press},
   author={Fagnoni, Nicolas and de Lera Acedo, Eloy and DeBoer, David R. and Abdurashidova, Zara and Aguirre, James E. and Alexander, Paul and Ali, Zaki S. and others},
   year={2020},
   month=oct,
   pages={1232--1242},
   doi={10.1093/mnras/staa3268},
   url={https://doi.org/10.1093/mnras/staa3268}
}

@article{O_Hara_2025,
   title={Uncovering the effects of array mutual coupling in 21-cm experiments with the SKA-Low radio telescope},
   volume={538},
   number={1},
   journal={Monthly Notices of the Royal Astronomical Society},
   publisher={Oxford University Press},
   author={O'Hara, Oscar S. D. and Gueuning, Quentin and de Lera Acedo, Eloy and Dulwich, Fred and Cumner, John and Anstey, Dominic and Brown, Anthony and Fialkov, Anastasia and Dhandha, Jiten and Faulkner, Andrew and Liu, Yuchen},
   year={2025},
   month=feb,
   pages={31--48},
   doi={10.1093/mnras/staf264},
   url={https://doi.org/10.1093/mnras/staf264}
}

@article{Vedantham_2014,
   title={Chromatic effects in the 21 cm global signal from the cosmic dawn},
   volume={437},
   number={2},
   journal={Monthly Notices of the Royal Astronomical Society},
   publisher={Oxford University Press},
   author={Vedantham, H. K. and Koopmans, L. V. E. and de Bruyn, A. G. and Wijnholds, S. J. and Ciardi, B. and Brentjens, M. A.},
   year={2014},
   pages={1056--1069},
   doi={10.1093/mnras/stt1878},
   url={https://doi.org/10.1093/mnras/stt1878}
}

@article{Monsalve_2017,
   title={Results from EDGES High-band. I. Constraints on phenomenological models for the global 21 cm signal},
   volume={847},
   number={1},
   journal={The Astrophysical Journal},
   publisher={American Astronomical Society},
   author={Monsalve, Raul A. and Rogers, Alan E. E. and Bowman, Judd D. and Mozdzen, Thomas J.},
   year={2017},
   month=sep,
   pages={64},
   doi={10.3847/1538-4357/aa88d1},
   url={https://doi.org/10.3847/1538-4357/aa88d1}
}

@article{Cumner_2024,
   title={The effects of the antenna power pattern uncertainty within a global 21 cm experiment},
   volume={531},
   number={4},
   journal={Monthly Notices of the Royal Astronomical Society},
   publisher={Oxford University Press},
   author={Cumner, John and Pieterse, Carla and de Villiers, Dirk and de Lera Acedo, Eloy},
   year={2024},
   month=jun,
   pages={4734--4745},
   doi={10.1093/mnras/stae1475},
   url={https://doi.org/10.1093/mnras/stae1475}
}

@inproceedings{Pieterse_2024,
   title={Investigating Frequency Dependency in Characteristic Basis Function Pattern Modelling with Geometric Perturbations},
   author={Pieterse, Carla M. and Venter, Mariet and de Villiers, Dirk I. L.},
   booktitle={2024 International Conference on Electromagnetics in Advanced Applications (ICEAA)},
   year={2024},
   pages={407--412},
   doi={10.1109/ICEAA61917.2024.10701964},
   url={https://doi.org/10.1109/ICEAA61917.2024.10701964}
}

@misc{Pattison_2026,
   title={Impact of Antenna Structure and Orientation on Forward-Modelled Global 21 cm Signal Recovery},
   author={Pattison, Joe H. N. and Cumner, John M. and Anstey, Dominic J. and Pegwal, Saurabh and Croukamp, Wessel and de Villiers, Dirk I. L. and de Lera Acedo, Eloy},
   year={2026},
   eprint={2603.24378},
   archivePrefix={arXiv},
   primaryClass={astro-ph.CO},
   url={https://arxiv.org/abs/2603.24378}
}

@article{GAIAOG,
   title={The Gaia mission},
   volume={595},
   journal={Astronomy \& Astrophysics},
   publisher={EDP Sciences},
   author={{Gaia Collaboration} and Prusti, T. and de Bruijne, J. H. J. and Brown, A. G. A. and Vallenari, A. and Babusiaux, C. and others},
   year={2016},
   month=nov,
   pages={A1},
   doi={10.1051/0004-6361/201629272},
   url={https://doi.org/10.1051/0004-6361/201629272}
}

@article{WEAVE,
   title={The wide-field, multiplexed, spectroscopic facility WEAVE: Survey design, overview, and simulated implementation},
   volume={530},
   number={3},
   journal={Monthly Notices of the Royal Astronomical Society},
   publisher={Oxford University Press},
   author={Jin, Shoko and Trager, Scott C. and Dalton, Gavin B. and Aguerri, J. Alfonso L. and Drew, J. E. and others},
   year={2023},
   month=mar,
   pages={2688--2730},
   doi={10.1093/mnras/stad557},
   url={https://doi.org/10.1093/mnras/stad557}
}

@inproceedings{4MOST,
   title={4MOST: 4-metre multi-object spectroscopic telescope},
   volume={8446},
   series={Society of Photo-Optical Instrumentation Engineers (SPIE) Conference Series},
   booktitle={Ground-based and Airborne Instrumentation for Astronomy IV},
   publisher={SPIE},
   author={de Jong, Roelof S. and Bellido-Tirado, Olga and Chiappini, Cristina and Depagne, Eric and Haynes, Roger and others},
   editor={McLean, Ian S. and Ramsay, Suzanne K. and Takami, Hideki},
   year={2012},
   month=oct,
   pages={84460T},
   doi={10.1117/12.926239},
   url={https://doi.org/10.1117/12.926239}
}

@inproceedings{nenufar,
   title={{NenuFAR}: Instrument description and science case},
   author={Zarka, Philippe and Tagger, Michel and Denis, L. and Girard, J. N. and Konovalenko, A. and others},
   booktitle={2015 International Conference on Antenna Theory and Techniques (ICATT)},
   address={Kharkiv, Ukraine},
   year={2015},
   month=apr,
   doi={10.1109/ICATT.2015.7136773},
   url={https://hal.science/hal-01196457},
   eprint={hal-01196457}
}

@article{wilensky_2024,
      title={High-dimensional Inference of Radio Interferometer Beam Patterns I: Parametric Model of the {HERA} Beams},
      author={Wilensky, Michael J. and Burba, Jacob and Bull, Philip and Garsden, Hugh and Glasscock, Katrine A. and Fagnoni, Nicolas and de Lera Acedo, Eloy and DeBoer, David R. and Razavi-Ghods, Nima},
      journal={RAS Techniques and Instruments},
      volume={3},
      number={1},
      pages={400--414},
      year={2024},
      doi={10.1093/rasti/rzae029},
}

@article{sims_2023,
    author = {Sims, Peter H and Bowman, Judd D and Murray, Steven G and Barrett, John P and Cappallo, Rigel C and Lonsdale, Colin J and Mahesh, Nivedita and Monsalve, Raul A and Rogers, Alan E E and Samson, Titu and Vydula, Akshatha K},
    title = {A Bayesian approach to modelling spectrometer data chromaticity corrected using beam factors – II. Model priors and posterior odds},
    journal = {Monthly Notices of the Royal Astronomical Society},
    volume = {544},
    number = {2},
    pages = {2340-2364},
    year = {2025},
    month = {12},
    issn = {0035-8711},
    doi = {10.1093/mnras/staf1767},
    url = {https://doi.org/10.1093/mnras/staf1767},
    eprint = {https://academic.oup.com/mnras/article-pdf/544/2/2340/64711504/staf1767.pdf},
}

@article{Tutt_2026,
   title={Optimizing foreground modelling for global 21-cm cosmology with GPU-accelerated nested sampling},
   volume={550},
   ISSN={1365-2966},
   url={http://dx.doi.org/10.1093/mnras/stag1101},
   DOI={10.1093/mnras/stag1101},
   number={1},
   journal={Monthly Notices of the Royal Astronomical Society},
   publisher={Oxford University Press (OUP)},
   author={Tutt, Jacob L and Sims, Peter H and Pattison, Joe H N and Anstey, Dominic J and Leeney, Samuel A K and de Lera Acedo, Eloy},
   year={2026},
   month=June }

@misc{Robins_2026,
      title={Synchrotron and free-free mapping with simulated REACH observations between 50-170 MHz}, 
      author={Daniel Robins and Dominic Anstey and Harry Bevins and Eloy de Lera Acedo and Melis O. Irfan},
      year={2026},
      eprint={2607.00299},
      archivePrefix={arXiv},
      primaryClass={astro-ph.CO},
      url={https://arxiv.org/abs/2607.00299}, 
}

@article{Bull_2025,
   title={RHINO: a large horn antenna for detecting the 21 cm global signal},
   volume={4},
   ISSN={2752-8200},
   url={http://dx.doi.org/10.1093/rasti/rzaf046},
   DOI={10.1093/rasti/rzaf046},
   journal={RAS Techniques and Instruments},
   publisher={Oxford University Press (OUP)},
   author={Bull, Philip and El-Makadema, Ahmed and Garsden, Hugh and Edgley, John and Roddis, Neil and Chluba, Jens and Conselice, Christopher J and Dutta, Sohini and Glasscock, Katrine A and Nasirudin, Ainulnabilah and Norris, Jordan and Wilensky, Michael J and Ye, Isabelle and Zhang, Zheng},
   year={2025} }

@misc{Wiersema_2026,
Author = {Roeland Wiersema},
Title = {JAXMg: A multi-GPU linear solver in JAX},
Year = {2026},
Eprint = {arXiv:2601.14466},
}

@article{Gunapati_2022,
   title={Variational inference as an alternative to MCMC for parameter estimation and model selection},
   volume={39},
   ISSN={1448-6083},
   url={http://dx.doi.org/10.1017/pasa.2021.64},
   DOI={10.1017/pasa.2021.64},
   journal={Publications of the Astronomical Society of Australia},
   publisher={Cambridge University Press (CUP)},
   author={Gunapati, Geetakrishnasai and Jain, Anirudh and Srijith, P. K. and Desai, Shantanu},
   year={2022} }

@misc{lovick_2026,
      title={Automatic Laplace Collapsed Sampling: Scalable Marginalisation of Latent Parameters via Automatic Differentiation}, 
      author={Toby Lovick and David Yallup and Will Handley},
      year={2026},
      eprint={2603.26644},
      archivePrefix={arXiv},
      primaryClass={cs.LG},
      url={https://arxiv.org/abs/2603.26644}, 
}

@misc{leeney_2026,
      title={Conditional Neural Bayes Ratio Estimation for Experimental Design Optimisation}, 
      author={S. A. K. Leeney and T. Gessey-Jones and W. J. Handley and E. de Lera Acedo and H. T. J. Bevins and J. L. Tutt},
      year={2026},
      eprint={2603.26489},
      archivePrefix={arXiv},
      primaryClass={astro-ph.IM},
      url={https://arxiv.org/abs/2603.26489}, 
}

@inproceedings{Isamabrd_2024,
      title={Isambard-{AI}: A Leadership-Class Supercomputer Optimised Specifically for Artificial Intelligence},
      author={McIntosh-Smith, Simon and Alam, Sadaf R. and Woods, Christopher J.},
      booktitle={CUG 2024 Proceedings},
      pages={44--54},
      year={2024},
}

@article{Polanska_2025,
   title={Learned harmonic mean estimation of the Bayesian evidence with normalizing flows},
   volume={8},
   ISSN={2565-6120},
   url={http://dx.doi.org/10.33232/001c.146026},
   DOI={10.33232/001c.146026},
   journal={The Open Journal of Astrophysics},
   publisher={Maynooth University},
   author={Polanska, Alicja and Price, Matthew A. and Piras, Davide and Spurio Mancini, Alessio and McEwen, Jason D.},
   year={2025},
   month=Oct }

\appendix

\section{Marginalising directly Over SVD Weights}
\label{app:posterior_grids}

In Section~\ref{sec:basis_functions}, we argued that the core limitation of using the SVD weights as a direct parameterisation of the beam uncertainty is not the dimensionality but that it does not inform the inference how the weights should evolve coherently across frequency and bases. In other words, it is statistically too flexible for cosmological recovery. This appendix makes that point explicit by attempting to perform signal recovery directly in this space.

The equivalent forward model and analytical marginalisation derivation are first outlined in Section~\ref{app:Equiv_Formulation}, before presenting the resulting inference in Section~\ref{app:Equiv_Results} and comparing it to the PCA-based beam marginalisation results in Section~\ref{sec:beam_marginalisation_results}.

\subsection{Mathematical Formulation}
\label{app:Equiv_Formulation}

The direct SVD-weight model follows a parrallel mathematical derivation to that presented in Sections~\ref{sec:forward_model} and \ref{sec:beam_marginalisation}, to keep this equivalent formulation distinct from the main analysis, we denote equations with a $\star$ superscript. The foreground model is therefore defined as
\begin{equation}
\label{eqn:equiv_direct_svd_forward_model}
\begin{split}
T_{\mathrm{FG}}^{\star}(\nu_{\ell},t)
=
\sum_{k=1}^{R}
w_{k,\ell}
\sum_{i=1}^{N_{\alpha}}
\sum_{j=1}^{N_{\beta}}
\alpha_i\,
\mathcal{B}^{\star}_{i,j,k}(t)
\left( \frac{\nu_{\ell}}{\nu_0} \right)^{-\beta_j}
+
T_{\mathrm{CMB}},
\end{split}
\end{equation}
\noindent where $w_{k,\ell}\equiv w_k(\nu_{\ell})$ is the SVD weight associated with angular basis function $B_k(\theta,\phi)$ at frequency $\nu_{\ell}$. Since the additional propogating through PCA basis functions and weights has been removed, the basis response functions are simplifed to: 
\begin{equation}
\label{eqn:equiv_basis_response_functions}
\begin{split}
\mathcal{B}^{\star}_{i,j,k}(t)
=
\frac{1}{4\pi}
\int_{4\pi}
M_{\alpha,i}&(\theta,\phi,t)\,
M_{\beta,j}(\theta,\phi,t)\,
 \\\times
& \left[
T_{230}(\theta,\phi,t)-T_{\mathrm{CMB}}
\right] 
B_k(\theta,\phi)\,
\mathrm{d}\Omega .
\end{split}
\end{equation}

\noindent Consequentely each frequency channel has its own independent set of SVD weights. Defining the per-frequency weight vector $\mathbf{w}_{\ell}=(w_{1,\ell},\ldots,w_{R,\ell})^{\mathrm{T}}$, the model for the time-binned data vector at frequency $\nu_{\ell}$, conditioned on the astrophysical parameters $\boldsymbol{\phi}\equiv(\boldsymbol{\alpha},\boldsymbol{\beta},\boldsymbol{\theta}_{21})$, can be written as
\begin{equation}
\label{eqn:equiv_frequency_linear_model}
\mathbf{m}^{\star}_{\ell}(\boldsymbol{\phi})
=
\mathbf{b}_{\ell}(\boldsymbol{\theta}_{21})
+
\mathbf{H}^{\star}_{\ell}(\boldsymbol{\alpha},\boldsymbol{\beta})
\mathbf{w}_{\ell},
\end{equation}
\noindent where $\mathbf{b}_{\ell}=T_{\mathrm{CMB}}\mathbf{1}+T_{21}(\nu_{\ell}\mid\boldsymbol{\theta}_{21})\mathbf{1}$ and the $k$th column of $\mathbf{H}^{\star}_{\ell}$ is
\begin{equation}
\label{eqn:equiv_frequency_response_columns}
h^{\star}_{k,\ell}(t\mid\boldsymbol{\alpha},\boldsymbol{\beta})
=
\sum_{i=1}^{N_{\alpha}}
\sum_{j=1}^{N_{\beta}}
\alpha_i\,
\mathcal{B}^{\star}_{i,j,k}(t)
\left( \frac{\nu_{\ell}}{\nu_0} \right)^{-\beta_j}.
\end{equation}

A Gaussian prior can then be placed on the weights using the mean SVD weights across the training set, $\bar{\mathbf{w}}_{\ell}$, and a diagonal covariance, $\mathbf{\Sigma}_{w,\ell}$. Since the prior is defined independently at each frequency, it factorises as
\begin{equation}
\label{eqn:equiv_frequency_factorised_prior}
\pi(\mathbf{w})
=
\prod_{\ell=1}^{N_{\nu}}
\mathcal{N}
\left(
\mathbf{w}_{\ell}
\mid
\bar{\mathbf{w}}_{\ell},
\mathbf{\Sigma}_{w,\ell}
\right).
\end{equation}
\noindent The effective mean and covariance for each channel are then
\begin{equation}
\label{eqn:equiv_frequency_effective_covariance}
\boldsymbol{\mu}^{\star}_{\ell}
=
\mathbf{b}_{\ell}
+
\mathbf{H}^{\star}_{\ell}\bar{\mathbf{w}}_{\ell},
\qquad
\mathbf{C}^{\star}_{\mathrm{eff},\ell}
=
\mathbf{C}_{n,\ell}
+
\mathbf{H}^{\star}_{\ell}
\mathbf{\Sigma}_{w,\ell}
\left(\mathbf{H}^{\star}_{\ell}\right)^{\mathrm{T}},
\end{equation}
\noindent with the full marginalised likelihood given by the sum of the independent channel likelihoods,
\begin{equation}
\label{eqn:equiv_frequency_marginalised_likelihood}
\ln \mathcal{L}^{\star}_{\mathrm{marg}}
=
\sum_{\ell=1}^{N_{\nu}}
\ln
\mathcal{N}
\left(
\mathbf{d}_{\ell}
\mid
\boldsymbol{\mu}^{\star}_{\ell},
\mathbf{C}^{\star}_{\mathrm{eff},\ell}
\right).
\end{equation}

One of the key advantages of this factorisation is that it allows the marginalisation regime to move from a single large linear solve over all $N_{\nu} \times R$ weights to $N_{\nu}$ independent solves, each over only $R$ weights, where $R$ is the number of retained SVD modes defined in Section~\ref{sec:basis_functions}. For each channel, the corresponding linear system is therefore:
\begin{equation}
\label{eqn:equiv_frequency_basis_solve}
\mathbf{A}_{\ell}
=
\mathbf{\Sigma}_{w,\ell}^{-1}
+
\left(\mathbf{H}^{\star}_{\ell}\right)^{\mathrm{T}}
\mathbf{C}_{n,\ell}^{-1}
\mathbf{H}^{\star}_{\ell}.
\end{equation}
\noindent Since dense Cholesky decomposition scales as $\mathcal{O}(N^3)$, this changes the scaling from $\mathcal{O}[(N_{\nu}R)^3]$ to $N_{\nu}\mathcal{O}(R^3)$, corresponding to significantly reduced computational cost for the marginalisation. Additionally, exploting the inference frameworks compatability with modern accelerators, the per-frequency solves can be trivially parallelised further reducing the wall-clock time required for inference.

However, the same factorisation that makes the calculation efficient is also what makes the parameterisation statistically weak. In antenna-temperature space, the prior allows each frequency channel to be fit independently across the full beam-induced temperature range shown in Figure~\ref{fig:beam_temperature_variation}. Without the inter-frequency correlations learned by the PCA parameterisation, the beam model is therefore flexible enough to absorb any cosmological structure into the instrumental uncertainty, as shown in Appendix~\ref{app:Equiv_Results}.

We note that these correlations could, in principle, be imposed directly in the SVD-weight space by replacing the factorised prior with a dense covariance matrix, or by using Gaussian-process style kernels to enforce smoothness across frequency. However, doing so removes the computational advantage described above, requiring a $34{,}000 \times 34{,}000$ covariance matrix, making direct SVD-weight marginalisation substantially less attractive than the compressed PCA-space formulation.

\subsection{Cosmological and Foreground Recovery}
\label{app:Equiv_Results}

\begin{figure}
    \centering
    \includegraphics[width=\columnwidth]{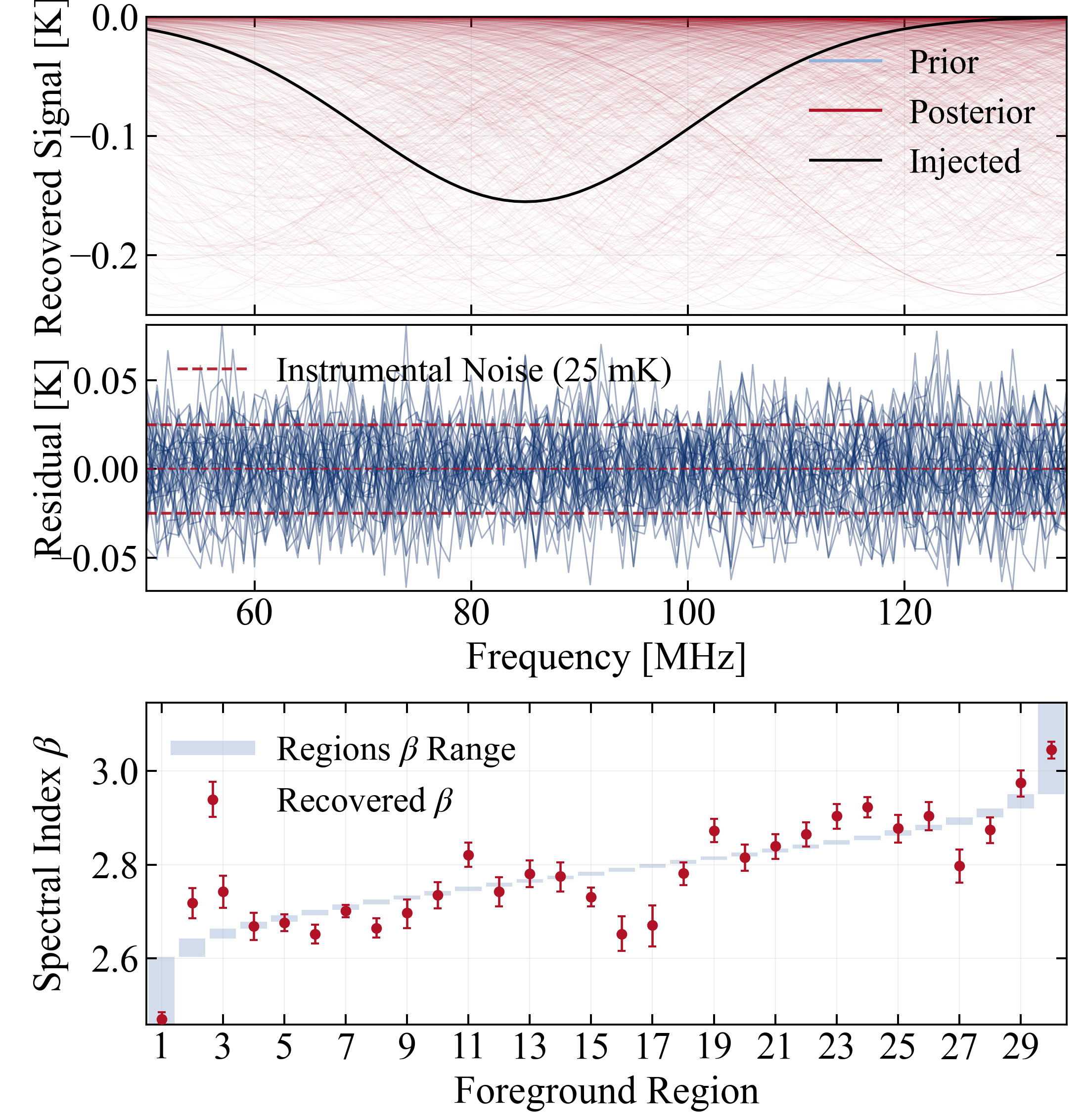}
    \caption{Direct-SVD-weight marginalisation alternative to the beam-marginalised recovery shown in Figure~\ref{fig:400_signal_recovery_plot}. The inference is performed by marginalising independently over the SVD weights at each frequency, rather than over the correlated PCA beam coefficients used in the main analysis. Top panel: recovered signal posterior draws as pencil lines compared to the injected absorption profile. Middle panel: full-model residuals at the conditional maximum-posterior beam parameters. Bottom panel: recovered foreground spectral-index parameters compared to the regional $\beta$ ranges used to generate the data.}
    \label{fig:direct_svd_weight_recovery}
\end{figure}

\begin{table*}
    \centering
    \caption{Comparison between direct SVD-weight marginalisation and the PCA-space marginalisation used in the main analysis. The prior-volume contraction reports the percentage contraction of the sampled signal posterior relative to its prior volume, calculated through the KL divergence, while the bayes factor between the models is quoted as $\log \mathcal{Z}_{\mathrm{PCA}}-\log \mathcal{Z}_{\mathrm{SVD}}$.}
    \label{tab:svd_pca_marginalisation_comparison}
    \setlength{\tabcolsep}{4pt}
    \begin{tabular}{lccccccccccc}
        \toprule
        & Beam 0 & Beam 1 & Beam 2 & Beam 3 & Beam 4 & Beam 5 & Beam 6 & Beam 7 & Beam 8 & Beam 9 & Beam 10 \\
        \midrule
        \begin{tabular}[c]{@{}l@{}}SVD-model\\Prior contraction [\%]\end{tabular} & 0.20 & 0.34 & 0.22 & 0.20 & 0.13 & 0.19 & 0.23 & 0.19 & 0.12 & 0.42 & 0.30 \\
        \begin{tabular}[c]{@{}l@{}}PCA-model\\Prior contraction [\%]\end{tabular} & 99.82 & 99.77 & 99.59 & 99.81 & 99.94 & 99.61 & 99.59 & 99.51 & 99.86 & 99.81 & 99.85 \\
        $\log \mathcal{Z}_{\mathrm{PCA}}-\log \mathcal{Z}_{\mathrm{SVD}}$ & 1868 & 1900 & 1941 & 1929 & 1786 & 1749 & 1838 & 1934 & 2022 & 1841 & 1946 \\
        \bottomrule
    \end{tabular}
\end{table*}

Figure~\ref{fig:direct_svd_weight_recovery} shows the recovered signal posterior, foreground spectral-index parameters and maximum-a-posteriori residuals for Beam 1 with an injected signal at $\nu_{21}=85~\mathrm{MHz}$. This is the same simulated dataset shown in Figure~\ref{fig:400_signal_recovery_plot}, but analysed using the direct SVD-weight marginalisation described in Appendix~\ref{app:Equiv_Formulation}. Compared to the fixed-beam failure in Figure~\ref{fig:fixed_beam_failure}, the residuals are substantially improved and are consistent with the instrumental-noise level. However, this is not evidence of successful signal recovery. Instead, the independent SVD weights are sufficiently flexible to absorb residual chromatic structure, including smooth cosmological structure, into the beam model itself. As a result, the recovered 21-cm signal posterior remains almost completely unconstrained, and the foreground spectral-index recovery is noticeably degraded relative to the PCA-space result in Figure~\ref{fig:400_signal_recovery_plot}.

This behaviour is quantified across all 11 validation beams in Table~\ref{tab:svd_pca_marginalisation_comparison}, which compares the prior to posterior contraction and Bayesian evidence between the direct SVD-weight marginalisation and the PCA-space marginalisation used in the main analysis. To estimate this contraction, we use the \texttt{margarine} package \citep{Bevins_2023_margarine}, which trains masked autoregressive flows to estimate marginal posterior densities for the cosmological signal parameters. From these density estimates, we calculate the marginal Kullback--Leibler (KL) divergence,
\begin{equation}
D_{\mathrm{KL}}
=
\int
P(\boldsymbol{\theta}_{21}\mid \mathbf{d})
\log
\left[
\frac{
P(\boldsymbol{\theta}_{21}\mid \mathbf{d})
}{
\pi(\boldsymbol{\theta}_{21})
}
\right]
\mathrm{d}\boldsymbol{\theta}_{21},
\label{eqn:marginal_kl_divergence}
\end{equation}
\noindent where $P(\boldsymbol{\theta}_{21}\mid \mathbf{d})$ is the marginal posterior on the signal parameters and $\pi(\boldsymbol{\theta}_{21})$ is the corresponding prior. The KL divergence gives the information gain from prior to posterior, and can be related to the fractional posterior volume through $\exp(-D_{\mathrm{KL}}) \simeq V_{\mathrm{post}}/V_{\mathrm{prior}}$. We therefore quote the prior-volume contraction as
\begin{equation}
C_{\mathrm{prior}}
=
100
\left[
1
-
\exp(-D_{\mathrm{KL}})
\right]
\simeq
100
\left[
1
-
\frac{V_{\mathrm{post}}}{V_{\mathrm{prior}}}
\right].
\label{eqn:prior_volume_contraction}
\end{equation}

Table~\ref{tab:svd_pca_marginalisation_comparison} shows that, in every case, the direct SVD-weight posterior contracts by less than $0.5$ per cent of the prior volume, indicating that the signal parameters remain effectively unconstrained. In contrast, the PCA-space marginalisation achieves a minimum contraction of $99.5$ per cent across the held-out beams, with the Bayesian evidence overwhelmingly favouring the correlated PCA parameterisation. This demonstrates that the PCA step is not merely a computational convenience: it is what prevents the beam model from becoming flexible enough to absorb the cosmological signal itself.

\FloatBarrier
\section{Recovery Across Injected Signal Profiles}
\label{app:signal_shape_recovery}

\begin{figure*}
  \centering
  \includegraphics[width=\textwidth,height=0.9\textheight,keepaspectratio]{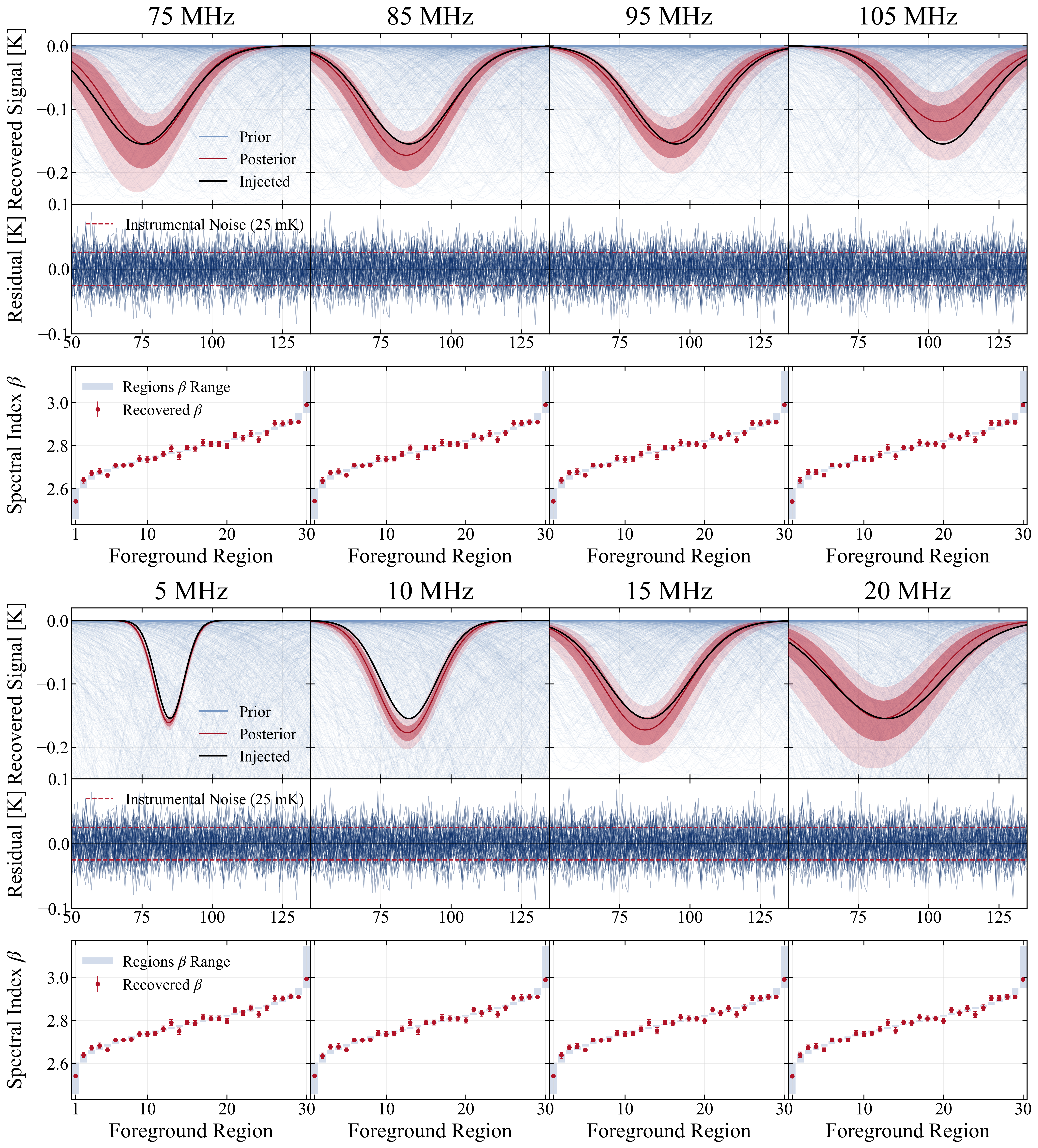}
  \caption{Beam-marginalised inference results for data generated with Beam~1 across variations in the injected 21-cm absorption profile. The upper block varies the central frequency over $\nu_{21} =\{75,85,95,105\}$~MHz at fixed amplitude $A_{21}=155$~mK and width $\sigma_{21}=15$~MHz, while the lower block varies the width over $\sigma_{21}=\{5,10,15,20\}$~MHz at fixed central frequency $\nu_{21} =85$~MHz and amplitude $A_{21}=155$~mK. Within each block, the top row shows the recovered signal posterior, with the 68 and 95 per cent credible regions shaded in red, the central red curve showing the posterior-median signal and the black curve showing the injected signal. The middle row shows the full-model residuals at the conditional maximum-posterior beam parameters, with dashed red lines indicating the $\pm25$~mK instrumental-noise level. The bottom row shows the recovered foreground spectral-index parameters compared with the regional $\beta$ ranges used to generate the data.}
  \label{fig:signal_shape_recovery}
\end{figure*}

In direct comparison with the Beam~1 signal-recovery results shown for the amplitude sweep in Figure~\ref{fig:400_signal_recovery_plot}, Figure~\ref{fig:signal_shape_recovery} presents the corresponding results for variations in central frequency and width. Across the sweep of central-frequency, the recovered posteriors remain consistent with the injected signal, although the uncertainty increases as the absorption profile approaches the edge of the observing band. A similar broadening is seen as the injected width increases, reflecting the greater degeneracy of broad, spectrally smooth profiles with the foreground and instrumental structure. In all cases for Beam~1, the reconstructed residuals remain consistent with the instrumental-noise level and the foreground spectral indices are inline with the corresponding regional ranges.
  
\label{lastpage}
\end{document}